\documentclass[dvipsnames]{jfm}

\usepackage[dvipsnames]{xcolor}

\usepackage{newtxtext}
\usepackage{newtxmath}

\usepackage{graphicx}
\graphicspath{{Pictures/}}

\usepackage{natbib}
\usepackage{siunitx}
\usepackage{mathtools}

\usepackage{tabularx}
\usepackage{multirow}
\usepackage{multicol}
\usepackage{booktabs}
\usepackage{amsmath}
\usepackage{amsfonts}
\usepackage{tikz}
\usepackage{MnSymbol}

\definecolor{mycolor1}{RGB}{0, 114, 189}
\definecolor{mycolor2}{RGB}{218, 83, 25}
\definecolor{mycolor3}{RGB}{238, 178, 32}
\definecolor{mycolor4}{RGB}{126, 47, 142}
\definecolor{mycolor5}{RGB}{119, 173, 48}
\definecolor{mycolor6}{RGB}{77, 191, 239}
\definecolor{mycolor7}{RGB}{163, 20, 47}
\definecolor{mycolor8}{RGB}{192, 192, 192}
\definecolor{mycolor9}{RGB}{0, 0, 255}
\definecolor{mycolor10}{RGB}{0, 0, 0}
\definecolor{matplotlibgreen}{HTML}{2ca02c}
\definecolor{matplotlibbrown}{HTML}{A52A2A}

\usepackage{hyperref}
\hypersetup{
   colorlinks = true,
   urlcolor   = blue,
   citecolor  = blue,
   allcolors  = blue
}

\shorttitle{DNS of complete transition to turbulence with a supercritical fluid}
\shortauthor{P.~C.~Boldini et al.}

\title{Direct numerical simulation of complete transition to turbulence with a fluid at supercritical pressure}

\author{P.C.~Boldini\aff{1}\dagger,
 B.~Bugeat\aff{2}, J.~W.~R.~Peeters\aff{1}, M.~Kloker\aff{3}, R.~Pecnik\aff{1} \corresp{\email{p.c.boldini@tudelft.nl}, r.pecnik@tudelft.nl}}

\affiliation{\aff{1}Process and Energy Department, Delft University of Technology, Leeghwaterstraat 39, 2628 CB Delft, the Netherlands
            \aff{2}School of Engineering, University of Leicester, University Road, Leicester, LE1 7RH, United Kingdom       
            \aff{3}Institute of Aerodynamics and Gas Dynamics, University of Stuttgart, Pfaffenwaldring 21, 70569 Stuttgart, Germany}

\begin{document}
\maketitle

\begin{abstract}
The objective of this work is to investigate the unexplored laminar-to-turbulent transition of a heated flat-plate boundary layer with a fluid at supercritical pressure. Two temperature ranges are considered: a subcritical case, where the fluid remains entirely in the liquid-like regime, and a transcritical case, where the pseudo-critical (Widom) line is crossed and pseudo-boiling occurs. Fully compressible direct numerical simulations are used to study (i) the linear and nonlinear instabilities, (ii) the breakdown to turbulence, and (iii) the fully developed turbulent boundary layer. In the transcritical regime, two-dimensional forcing generates not only a train of billow-like structures around the Widom line, resembling Kelvin-Helmholtz instability, but also near-wall travelling regions of flow reversal. These spanwise-oriented billows dominate the early nonlinear stage. When high subharmonic three-dimensional forcing is applied, staggered $\Lambda$-vortices emerge more abruptly than in the subcritical case. However, unlike the classic H-type breakdown under zero pressure gradient observed in ideal-gas and subcritical regimes, the H-type breakdown is triggered by strong shear layers caused by flow reversals -- similar to that observed in adverse-pressure-gradient boundary layers. Without oblique wave forcing, transition is only slightly delayed and follows a naturally selected fundamental breakdown (K-type) scenario. Hence, in the transcritical regime, it is possible to trigger nonlinearities and achieve transition to turbulence relatively early using only a single two-dimensional wave that strongly amplifies background noise. In the fully turbulent region, we demonstrate that variable-property scaling accurately predicts turbulent skin-friction and heat-transfer coefficients.
\end{abstract}


\section{Introduction} \label{sec:intro}

Fluids at supercritical pressure have enabled the development of more efficient, compact industrial processes and continue to offer promising opportunities for future energy-conversion technologies \citep{Guardone1}. Among different fluids, carbon dioxide ($\text{CO}_2$) has emerged as a promising working medium for power cycles in geothermal and concentrated solar energy systems, as well as for heat pumps in industrial heating applications. In the aerospace sector, supercritical fuel injection improves mixing and combustion efficiency in both air-breathing and liquid rocket engines \citep{Wang1}. Supercritical fluids also occur in nature -- for example, the $\text{CO}_2$/$\text{N}_2$ mixture in the lower atmosphere of Venus \citep{Morellina1}.

At supercritical conditions, i.e. above the thermodynamic critical point, the liquid–vapour boundary vanishes and strong variations in thermophysical properties occur within a narrow temperature range around the pseudo-critical temperature $T_\mathit{pc}$ -- also referred to as pseudo-boiling \citep{Banuti1} -- at which the isobaric heat capacity reaches its maximum. In this non-ideal, single-phase thermodynamic region, the ideal-gas assumption fails.

Supercritical fluids have been extensively investigated in fully developed turbulent flows. \citet{Yoo1} highlighted the complexity of supercritical heat-transfer measurements, in which thermophysical property variations may either enhance or deteriorate heat transfer. Early computational studies on turbulent flows at supercritical pressure mainly focused on closed flow configurations, such as pipes \citep{Bae1,Nemati1,He1,Cao1}, channels \citep{Patel1,Ma1,Kim1,Guo1,Li1,Li2}, and annular flows \citep{Peeters1}. These studies underlined the significant impact of local property variations on large-scale structures and the presence of relaminarisation mechanisms. Semi-local scaling demonstrated to best characterise turbulence in variable-property flows, assuming weak density and viscosity fluctuations. However, in turbulent boundary layers with parahydrogen at supercritical pressure and transcritical temperature conditions, large density fluctuations of order $\sqrt{\smash[b]{\overline{\rho^{\prime}\rho^{\prime}}}}/\overline{\rho} \approx 0.4-1.0$ significantly alter near-wall turbulence and its statistics \citep{Kawai1}. As a result, semi-local scaling failed to collapse the velocity transformation.

Conversely, many industrial applications operating at supercritical conditions frequently involve flows that have not yet reached a fully turbulent state. Even under the ideal-gas assumption, transition to turbulence remains of fundamental importance across subsonic, supersonic, and high-Mach-number flows \citep{Saric1,Fedorov1}. In high-speed flows, real-gas effects have been studied extensively \citep{Candler1}, yet are often erroneously linked to non-ideal gas effects \citep{Guardone1}. Unlike real-gas effects, non-ideal gas effects near the thermodynamic critical point are characterised by strong stratification \citep{Govindarajan1} and abrupt variations in thermophysical properties.

Transition to turbulence can follow different routes depending on the type of flow disturbances \citep{Morkovin1}. In this work, we focus on the transitional sequence typically observed under low levels of free-stream turbulence: (i) receptivity to external disturbances, (ii) primary modal or non-modal disturbance growth, (iii) secondary instability and nonlinear interactions, followed by the eventual breakdown to turbulence. Only recently, linear stability theory (LST) of boundary layers with fluids at supercritical pressure has been investigated \citep{Robinet1}. The first such analysis by \citet{Ren1} for an adiabatic zero-pressure-gradient (ZPG) flat-plate boundary layer with supercritical $\text{CO}_2$ revealed strong sensitivity to the temperature profile relative to the pseudo-critical temperature. As the flow crosses the pseudo-critical (Widom) line, i.e. under transcritical temperature conditions, a new inviscid mode (Mode II) emerges, exhibiting growth rates an order of magnitude greater than those of Mode I, which corresponds to the Tollmien--Schlichting (TS) instability. It is worth noting this dual-mode instability occurs only when heating from a liquid-like free stream. Unlike Mack’s second mode in hypersonics, Mode II is not of acoustic nature. Moreover, the Mode-II instability is largest for two-dimensional (2-D) perturbations, as also confirmed for low-Mach, diabatic boundary layers by \citet{Boldini1}. \citet{Bugeat1} further showed that Mode II is associated with a minimum of kinematic viscosity at the Widom line -- a feature common to all non-polar supercritical fluids at transcritical temperature conditions. Thus, according to the generalised-inflection-point (GIP) criterion, such boundary layers are inviscidly unstable. In plane Couette flow, \citet{Bugeat2} demonstrated that this inviscid instability arises from a localised maximum of density-weighted vorticity and consists of two phase-locked vorticity waves induced by shear and baroclinic effects around the kinematic-viscosity minimum.

Alongside previous modal stability analyses, \citet{Boldini1} investigated transient growth with fluids at supercritical pressure. When heating beyond the Widom line -- where Mode II is unstable -- optimal energy growth arises from an interplay between lift-up and Orr mechanisms. Conversely, wall cooling was found to resemble the effect of an adverse pressure gradient (APG) under the ideal-gas assumption. A similar trend appeared in cross-flow (CF) dominated three-dimensional (3-D) boundary layers \citep{Ren2}, where the inviscid TS mode can be amplified far more than classical CF modes, effectively suppressing them -- akin to imposing strong deceleration under the ideal-gas assumption.

In controlled transition, a linearly amplified 2-D wave grows to finite amplitude, triggering secondary instabilities and nonlinear interactions that lead the growth of 3-D waves with subsequent breakdown. These mechanisms remain unexplored for supercritical fluids, despite their importance for accurate transition prediction. Under the ideal-gas assumption, two canonical breakdown paths -- K-type and H-type -- have been studied extensively via experiments \citep{Klebanoff1,Kachanov1} and Direct Numerical Simulations (DNS) \citep{Fasel1,Rist1,Bake1,Sayadi1} in incompressible boundary layers. These scenarios depend on the choice of initial disturbance wavelengths and spanwise wavenumbers. In K-type fundamental resonance, a primary 2-D TS wave and a steady longitudinal vortex mode -- often forced, as in Klebanoff’s experiment -- nonlinearly generate a symmetric pair of oblique modes at the same TS frequency. Alternatively, these oblique modes can be introduced directly, inducing the steady vortex mode. Their 3-D development leads to aligned $\Lambda$-shaped vortices, each consisting of two elongated legs of streamwise vorticity and a tip of spanwise vorticity. In contrast, H-type subharmonic resonance, following Craik’s model \citep{Craik1}, involves forcing oblique modes at half the TS frequency, forming staggered $\Lambda$-vortices without a steady vortex mode. In both scenarios, $\Lambda$-vortices develop into hairpin structures with a localised high-shear layer atop their heads, eventually evolving into $\Omega$- or ring-like vortices that mark the onset of turbulence. DNS studies of K- and H-type breakdown have since been extended to APG flows \citep{Kloker1,Kloker2} and to supersonic and hypersonic regimes in various geometries \citep{Fezer1,Franko1,Sivasubramanian1,Hader1,Unnikrishnan1}.

DNS remains the most effective tool for isolating specific perturbation waves and their effects on the transition routes \citep{Zhong2}. In the context of non-ideal fluid flows, high-order DNS have recently been employed to study the O-type breakdown in boundary layers with dense vapours, such as PP11-vapour at $M_\infty = 2.25$ and $6$ \citep{Sciacovelli1}, and Novec649-vapour at $M_\infty = 0.9$ \citep{Gloerfelt1}, which exhibit negligible dissipation and heat conduction. In these studies, density fluctuations remain small relative to the mean value. In contrast, under supercritical conditions near the Widom line, the abrupt variation of thermodynamic properties induces large density fluctuations and poses major numerical challenges for accurate and robust DNS analysis \citep{Kawai1}. 

The main objective of this work is to investigate the nonlinear interactions and transition to turbulence in a boundary layer with a supercritical fluid, aiming to improve transition prediction under non-ideal gas conditions. Specifically, we focus on elucidating the role of Mode-II instability in the nonlinear regime, examining the subsequent stages beyond the linear stability analysis of \citet{Ren1}. The strongly nonlinear thermodynamics and abrupt fluid property variations are accounted for by combining the Van der Waals cubic equation of state with non-ideal transport-property models. To numerically investigate supercritical fluids and tackle the related computational challenges, we employ the open-source solver CUBic Equation of state Navier--Stokes (CUBENS) \citep{Boldini2}. We perform DNS with controlled transition scenarios using harmonic disturbance forcing to isolate the most critical nonlinear mechanisms and limit modal interactions. To study the nonlinear regime, only a single fundamental 2-D wave is excited in a 2-D DNS set-up, as Mode-II instability is most unstable for 2-D perturbations. For the breakdown to turbulence, 3-D DNS are performed with 3-D forcing, in line with the aforementioned ideal-gas transitional boundary layer simulations. Building upon the 2-D nonlinear analysis, a pair of subharmonic oblique waves are introduced alongside the large amplitude 2-D wave. The amplitude of the oblique waves is set to either finitely large or infinitesimally small. Finally, the fully turbulent regime is analysed to evaluate the accuracy of mean velocity and enthalpy scaling laws. A predictive tool for the turbulent skin friction and heat transfer in non-ideal fluids is developed.

The work is organised as follows: \S\,\ref{sec:methodology} outlines the governing equations, with a focus on non-ideal thermodynamic models and numerical methods. \S\,\ref{sec:flow} presents two flow cases at supercritical pressure (reduced pressure of 1.10) and describes their respective DNS set-ups. Two temperature profiles for a slightly heated wall are considered -- one below and one crossing the pseudo-critical temperature $T_\mathit{pc}$. \S\,\ref{sec:2D} examines the 2-D linear and nonlinear evolution of Mode-II instability. Selected 3-D transitional cases are then presented and analysed in \S\,\ref{sec:3D}, followed by an assessment of the resulting turbulent boundary layers in \S\,\ref{sec:turbulent}. Finally, the study is concluded in \S\,\ref{sec:conclusions}.

\setcounter{equation}{2}

\section{Methodology} \label{sec:methodology}
This section presents the governing equations for supercritical fluid flows. For details on the numerical methods, performance, and validation of the flow solver, the reader is referred to \citet{Boldini2}.

\subsection{Flow-conservation equations} \label{sec:flow_conservation_equations}
We consider single-phase, non-reacting flows of supercritical fluids, governed by the fully compressible Navier--Stokes (NS) equations, expressed in both differential and dimensionless form as
\begin{subequations}
\begin{gather}
\dfrac{\partial\rho} {\partial t }+\dfrac{\partial ( \rho  u_\mathit{j})}{\partial x_\mathit{j}} =0, \nonumber \\
\dfrac{\partial ( \rho u_\mathit{i})}{\partial t} + \dfrac{\partial ( \rho u_\mathit{i} u_\mathit{j})}{\partial x_\mathit{j}} + \dfrac{\partial p}{\partial x_\mathit{i}} - \dfrac{1}{\Rey}\dfrac{\partial \tau_\mathit{ij}}{\partial x_\mathit{j}} = 0, \tag{\theequation$a$--$c$} \\ 
\dfrac{\partial ( \rho e_0)}{\partial t} + \dfrac{\partial ( ( \rho e_0 + p)u_\mathit{j})}{\partial x_\mathit{j}}   -  \dfrac{1}{\Rey} \dfrac{\partial \left(\tau_\mathit{ij} u_\mathit{i}  \right)}{\partial x_\mathit{j}} +   \dfrac{1}{\Rey \, \Pran_\infty \Eck_\infty}\dfrac{\partial q_\mathit{j} }{\partial x_\mathit{j}} = 0, \nonumber
\label{eq:ns}
\end{gather}
\end{subequations}
where $x_i=(x,y,z)$ are the Cartesian coordinates in the streamwise, wall-normal, and spanwise directions, respectively, $t$ is the time, $\rho$ is the density, $u_\mathit{i}=(u,v,w)$ are the velocity components, $p$ is the pressure, and $e_0=e+u_\mathit{j} u_\mathit{j}/2$ is the specific total energy, with $e$ as the specific internal energy. Under the Newtonian fluid assumption, the viscous stress tensor $\tau_\mathit{ij}$ is calculated as $\lambda \delta_\mathit{ij} \, \partial u_\mathit{k}/\partial x_\mathit{k} + \mu (\partial u_\mathit{i}/\partial x_\mathit{j} + \partial u_\mathit{j}/\partial x_\mathit{i})$, where $\mu$ is the dynamic viscosity, $\lambda=-2/3 \mu $ is the Lam\'e's constant with zero bulk viscosity (Stokes' hypothesis) in agreement with \citet{Sciacovelli2} and \citet{Ren1}, and $\delta_\mathit{ij}$ is the Kronecker delta. Additionally, buoyancy effects are neglected. The convective heat flux vector $q_\mathit{j}$ follows Fourier's law as $q_\mathit{j}=-\kappa \, \partial T/\partial x_\mathit{j}$, where $\kappa$ is the thermal conductivity and $T$ the fluid temperature. The conservation equations in \eqref{eq:ns} are non-dimensionalised by the following reference values
\begin{subequations}
\begin{gather}
t=\frac{t^* u^*_{\infty}}{\delta^*},	\quad x_\mathit{i}=\frac{x^*_\mathit{i}}{\delta^*}, \quad u_\mathit{i}=\frac{u^*_\mathit{i}}{u^*_{\infty}}, \quad \rho=\frac{\rho^*}{\rho^*_{\infty}}, \quad p=\frac{p^*}{\rho^*_{\infty} {u^{*^{2}}_{\infty}}}, \quad T=\frac{T^*}{T^*_{\infty}}, \nonumber\\ 
\quad e=\frac{e^*}{u^{*^{2}}_{\infty}}, \quad h=\frac{h^*}{u^{*^{2}}_{\infty}}, \quad 	\mu=\frac{\mu^*}{\mu^*_{\infty}}, \quad \kappa=\frac{\kappa^*}{\kappa^*_{\infty}}, \quad \nu=\frac{\nu^*}{\nu^*_{\infty}}, \tag{\theequation$a$--$k$}
\label{eq:variables}
\end{gather}
\end{subequations}
where $(\cdot)^*$ denotes dimensional quantities, and $(\cdot)_\infty$ corresponds to free-stream flow conditions. The corresponding characteristic parameters include
\begin{subequations}
\begin{gather}
Re=\frac{\rho^*_{\infty}u^*_{\infty}\delta^*}{\mu^*_{\infty}}, \quad 	Ec_\mathit{\infty}=\frac{{u^{*2}_\infty}}{c^*_\mathit{p,\infty}T^*_\infty}, \quad 
Pr_\mathit{\infty}=\frac{c^*_\mathit{p,\infty} \mu^*_{\infty}}{\kappa^*_\infty}, \tag{\theequation$a$--$c$}
\label{eq:nondimnumbers}
\end{gather}
\end{subequations}
where $c^*_\mathit{p,\infty}$ is the isobaric heat capacity, and $Re$ is the Reynolds number based on the local Blasius length scale $\delta^*=\sqrt{\mu^*_\infty x^*/(\rho^*_\infty u^*_\infty)}$. In \eqref{eq:nondimnumbers}, $\Eck_\infty$ is the Eckert number, and $\Pran_\infty$ is the Prandtl number (all based on free-stream conditions). The Mach number $M_\infty=u^*_\infty/a^*_\infty$, with $a^*_\infty$ being the speed of sound, can be obtained from $\Eck_\infty$. 

\subsection{Equation of state and transport properties} \label{sec:equation_of_state} 
To close the conservation equations, thermal and caloric equations of state (EoS) must be defined by satisfying the compatibility condition defined as
\begin{equation}
e = e_{\mathit{ref}} + \int_{T_{\mathit{ref}}}^{T} c_\mathit{\upsilon,\infty}(\check{T}) \, \mathrm{d}\check{T} - \int_{\rho_{\mathit{ref}}}^{\rho} \left( T \frac{\partial p}{\partial T} \bigg|_{\rho} - \frac{p}{\check{\rho}^{2}} \right) \mathrm{d}\check{\rho},
\end{equation}
where $(\cdot)_{\mathit{ref}}$ denotes a reference state, $c_\mathit{\upsilon,\infty}$ is the ideal-gas specific heat capacity at constant volume, and $\check{(\cdot)}$ indicates an integration variable. In the present study, the Van der Waals (VdW) cubic EoS \citep{Vanderwaals1} is chosen to balance accuracy and computational efficiency \citep{Boldini2}, while still capturing sharp thermophysical properties near the vapour-liquid critical point and across the Widom line. The thermodynamic EoS are expressed in reduced form, denoted by the subscript $(\cdot)_\mathit{r}$, with dimensional quantities at the critical point indicated by the subscript $(\cdot)^*_\mathit{c}$. This formulation is independent of the specific fluid under consideration. The thermal and caloric EoS are given as
\begin{subequations}
\begin{gather}
p_\mathit{r}=\dfrac{8\rho_\mathit{r}T_\mathit{r}}{3-\rho_\mathit{r}} - 3\rho^2_\mathit{r}, \quad e_\mathit{r}=\dfrac{c_\mathit{\upsilon,r}T_\mathit{r}}{Z_\mathit{c}} -3\rho_\mathit{r}, \tag{\theequation$a{,}b$}
    \label{eq:eos}
\end{gather}
\end{subequations}
respectively, where $p_\mathit{r}=p^*/p^*_\mathit{c}$, $T_\mathit{r}=T^*/T^*_\mathit{c}$, $\rho_\mathit{r}=\rho^*/\rho^*_\mathit{c}$, and $e_\mathit{r}=e^*\rho^*_\mathit{c}/p^*_\mathit{c}$ are the reduced pressure, temperature, density, and internal energy, respectively. The reduced isochoric and isobaric heat capacity $c_\mathit{\upsilon,r}$ and $c_\mathit{p,r}$ are defined as 
\begin{subequations}
\begin{gather}
c_\mathit{\upsilon,r}=\dfrac{c^*_\upsilon}{R^*_\mathit{g}}=\dfrac{f}{2}, \quad  c_\mathit{p,r}=\dfrac{c^*_\mathit{p}}{R^*_\mathit{g}}=c_{\upsilon,r}+\left[1-\dfrac{\rho_\mathit{r}(3-\rho_\mathit{r})^2}{4T_\mathit{r}}\right]^{-1},   \tag{\theequation$a{,}b$} 
\end{gather}
\end{subequations}
where $f$ is the number of degrees of freedom. The VdW compressibility factor at the critical point, $Z_\mathit{c}=p^*_\mathit{c}/(R^*_\mathit{g}\rho^*_\mathit{c}T^*_\mathit{c})$, with $R^*_\mathit{g}$ as the specific gas constant, is equal to $3/8$.

The analytical expressions of \citet{Jossi1} and \citet{Stiel1} -- denoted as JST hereafter -- for the dynamic viscosity and thermal conductivity of non-polar supercritical fluids, respectively, are employed for the transport-property laws. These laws, which depend on reduced quantities, provide a general representation of thermodynamics near the critical point and are detailed in \citet{Boldini2}.

\begin{figure}
    \centering
    \includegraphics[angle=-0,trim=0 0 0 0, clip,width=1.0\textwidth]{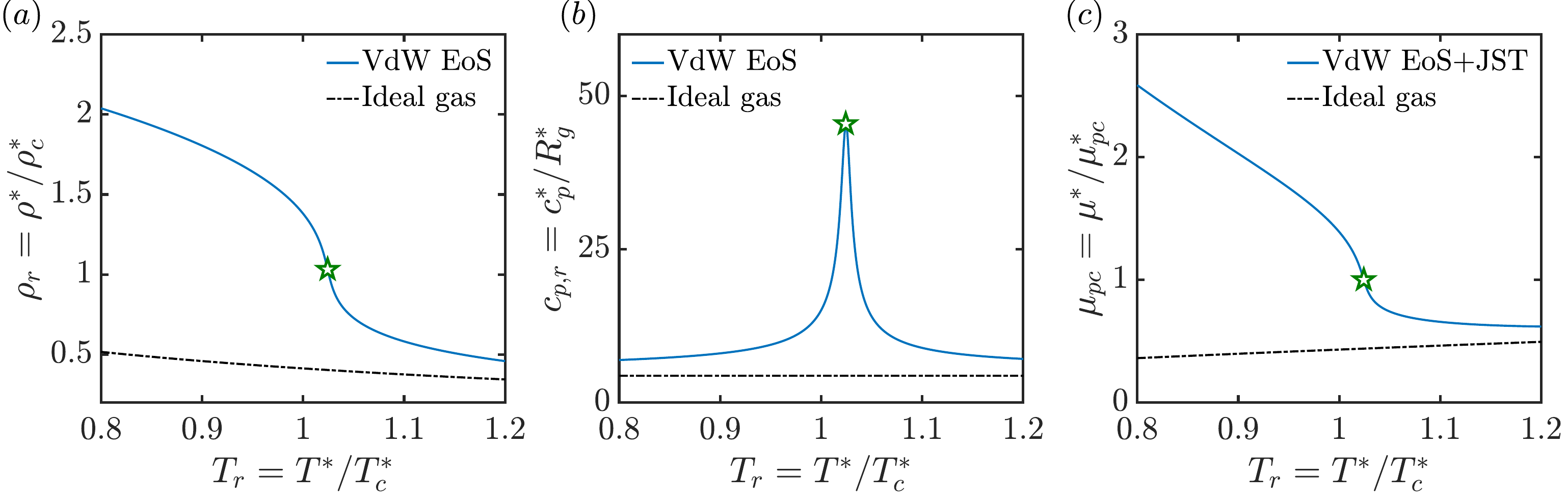}
    \captionsetup{justification=justified}
    \caption{Reduced thermodynamic and transport properties at $p_\mathit{r}=1.10$ for Van der Waals EoS and ideal-gas law (perfect gas:~$\rho=p/(R_\mathit{g}T)$ and $c_\mathit{p}=\gamma R_\mathit{g}/(\gamma-1)$, with heat capacity ratio $\gamma=1.4$; Sutherland's law: $\mu=T^{3/2}(1+T^*_{\mathit{ref}}/\SI{273.15}{K})/(T+T^*_\mathit{ref}/\SI{273.15}{K})$, with reference temperature $T^*_\mathit{ref}=\SI{110.4}{K}$): (a) density, (b) isobaric heat capacity, and (c) dynamic viscosity (reduced by the value at the pseudo-critical point $\mu^*_\mathit{pc}$). The location of $\max \{c_\mathit{p}(T)\}$, i.e. the pseudo-critical point, is marked by a star (\textcolor{matplotlibgreen}{$\medstar$}) symbol. Note that the number of degrees of freedom is $f=9$. }
    \label{fig_tp}
\end{figure}

The thermophysical properties for the VdW EoS with JST, and for the perfect-gas law with Sutherland's law, are presented in figure~\ref{fig_tp} for a reduced pressure of $p_\mathit{r}=p^*/p^*_c=1.10$, where $p^*_c$ is the critical pressure, in agreement with the flow cases in \S\,\ref{sec:flow}. The choice of this supercritical pressure is consistent with previous studies \citep{Ren1,Kawai1,Boldini1}, where reduced pressures ranged between $1.083$ and $1.56$.
At the pseudo-critical temperature $T_\mathit{r,pc}=T^*_\mathit{pc}/T^*_\mathit{c}$ -- or pseudo-critical point -- the isobaric heat capacity reaches a maximum, with the dense liquid-like ($T_\mathit{r}<T_\mathit{r,pc}$) and the low-density vapour-like ($T_\mathit{r}>T_\mathit{r,pc}$) region undergoing a sharp, yet continuous transition. The largest variations in fluid properties occur within a narrow temperature range around $T_\mathit{r,pc}$, where deviations from the ideal-gas behaviour are most pronounced.

\subsection{Linear stability analysis} \label{sec:linear_stability_analysis}

The linear stability analysis is performed within the framework of linear stability theory (LST) for non-ideal fluids \citep{Ren1,Boldini1}. The one-dimensional (1-D) base flow is based on the self-similar solution of the compressible boundary-layer equations with non-ideal thermophysical properties and either an adiabatic wall \citep{Ren1} or a diabatic wall \citep{Boldini1}; see \S\,\ref{sec:initial_conditions}. The perturbation vector, $\boldsymbol{q}^{\prime}=(p^{\prime},u^{\prime},v^{\prime},w^{\prime},T^{\prime})^\text{T}$ is expressed in the form of normal modes
\begin{equation}
    \boldsymbol{q}^{\prime}(x,y,z,t)=\hat{\boldsymbol{q}}(y)\exp[\mathrm{i}(\alpha x+\beta z-\omega t)] + \text{c.c.}, \label{eq:lst}
\end{equation}
where $\hat{\boldsymbol{q}}$ is the 1-D eigenfunction vector in the wall-normal direction, and c.c.~stands for complex conjugate. The linearised stability equations are recast into a compact matrix form, resulting in a nonlinear eigenvalue problem with Dirichlet boundary conditions, solved using a pseudo-spectral collocation method with Chebyshev collocation points and near-wall grid clustering \citep{Schmid1}. The spatial framework is adopted by prescribing spanwise wavenumber $\beta$ and angular frequency $\omega$. The streamwise wavenumber is complex ($\alpha=\alpha_\mathit{r}+\mathrm{i} \alpha_\mathit{i}$), where $\alpha_\mathit{i}$ represents the local spatial growth rate. Modal amplification occurs for $\alpha_\mathit{i}<0$.

\subsection{Flow solver} \label{sec:flow_solver}
The DNS are performed using the CUBENS solver \citep{Boldini2}, a parallel GPU-accelerated code that incorporates nonlinear thermodynamic and transport properties in the non-ideal thermodynamic region. The fully compressible NS equations in \eqref{eq:ns} are integrated on a Cartesian coordinate system. A sixth-order central finite-difference method is employed, combined with the non-dissipative kinetic-energy and entropy preserving (KEEP) scheme for the convective fluxes \citep{Kuya1}, while the diffusion fluxes are discretised using a fourth-order central finite difference scheme. Enhanced numerical stability is achieved through a split convective form, and the pressure-equilibrium discretisation \citep{Shima1} is applied to mitigate spurious grid-to-grid oscillations. Time integration is performed using a three-stage low-storage Runge--Kutta scheme. The time step $\Delta t$ is defined based on the frequency of the primary-wave disturbance, $\omega_{\text{2-D}}$, as $\Delta t=2\pi/(\omega_{\text{2-D}}LP)$, where $LP$ is a multiple of the number of samples saved during per forcing period \citep{Ren1}. The parameter $LP$ is chosen such that the maximum Courant--Friedrichs--Lewy number remains below $0.8$ in all directions. Non-reflecting boundary conditions for single-phase, non-ideal fluid flows \citep{Okongo1}, along with numerical sponge zones \citep{Mani1}, are applied at the domain boundaries. At the isothermal or adiabatic wall, no-slip and no-penetration conditions are imposed, while periodic boundary conditions are applied in the spanwise ($z$) direction. For post-processing, spectral analysis is performed in time and $z$-direction (for 3-D simulations), with Fourier components denoted as $( \omega / \omega_{\text{2-D}}, \beta / \beta_0)$, where $\omega_{\text{2-D}}$ and $\beta_0$ are the fundamental frequency and spanwise wavenumber of the disturbance-strip, respectively. In the streamwise direction, the wall-normal maximum amplitudes of the mass-flux perturbation $(\rho u)^{\prime}=\bar{\rho}u
^{\prime}+\bar{u}\rho^{\prime}+\rho^{\prime}u^{\prime}$, with $\bar{\rho}$ and $\bar{u}$ from the steady base-flow solution in \S\,\ref{sec:initial_conditions}, are used for quantification. For a quantitative analysis of the transitional (\S\,\ref{sec:quantities}) and turbulent (\S\,\ref{sec:turbulent}) boundary layer, time- and spanwise-averaged quantities are sampled every 50 time steps over approximately ten periods of the primary 2-D wave. 

\section{Flow cases and computational set-up} \label{sec:flow}
\begin{figure}
\centering
\includegraphics[angle=-0,trim=0 0 0 0, clip,width=0.9\textwidth]{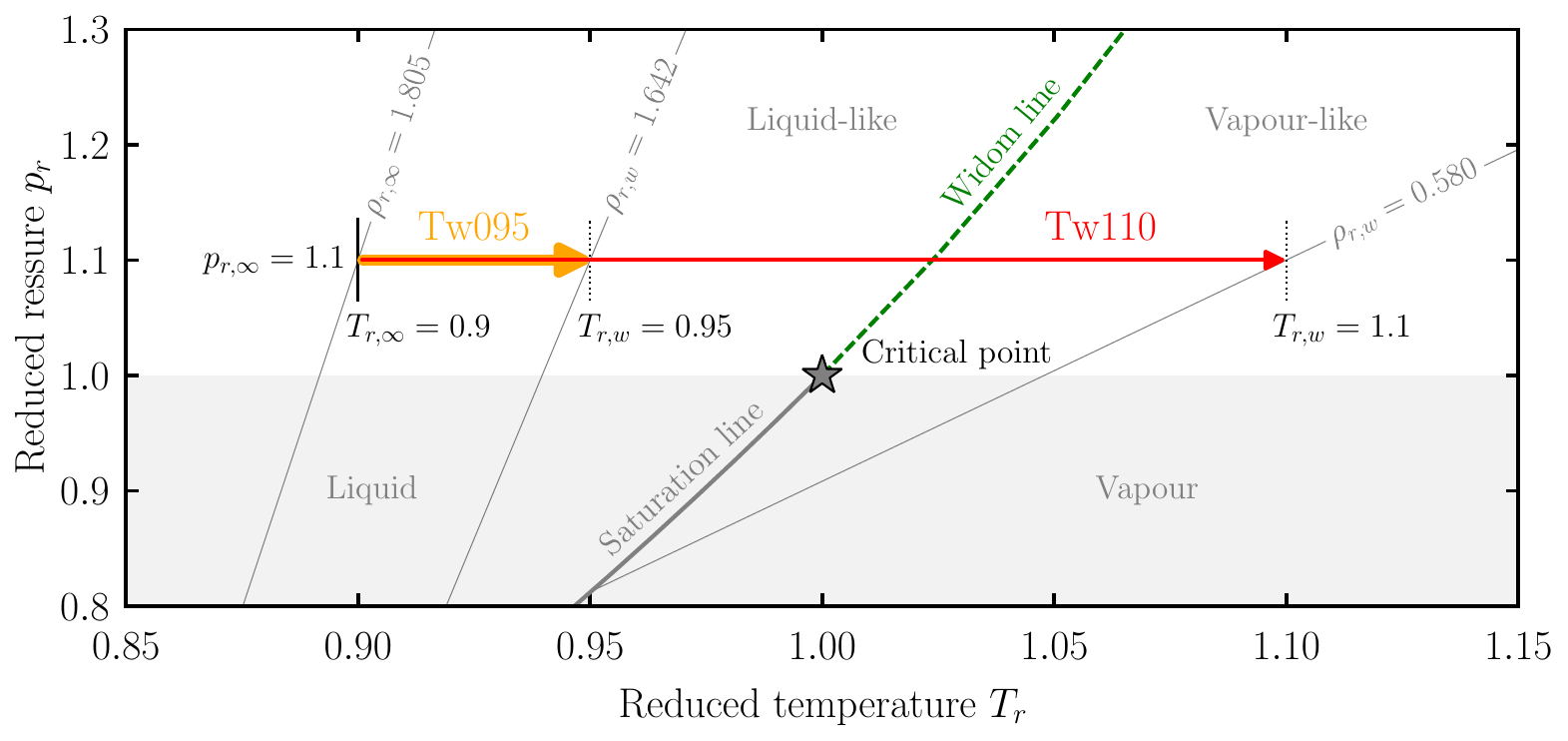}
\newcommand{\flowarrowA}{\tikz[baseline=-0.5ex]{
    \draw[line width=1.2pt, draw=orange, fill=orange, -latex] (0,0) -- (0.6,0);
}}
\newcommand{\flowarrowB}{\tikz[baseline=-0.5ex]{
    \draw[line width=1.2pt, draw=red, fill=red, -latex] (0,0) -- (0.6,0);
}}
\captionsetup{justification=justified}  
\caption{\label{fig_tp_diagram}Reduced temperature-pressure $(T_\mathit{r}$ -- $p_\mathit{r})$ diagram with isolines of reduced density $\rho_\mathit{r}$:~isobar at $p_\mathit{r,\infty}=1.10$ with cases at supercritical pressure of table~\ref{tab:tableBF}, i.e. Tw095 (\protect\flowarrowA) and Tw110 (\protect\flowarrowB). The saturation line and pseudo-critical (Widom) line, i.e. locus of the maxima of the isobaric specific heat capacity, follow the approximate generalised equation $p_\mathit{r}=\exp\{ (T_\mathit{r}-1) A_\mathit{VdW}/\min(T_\mathit{r},1)  \}$ with $A_\mathit{VdW}=4$ \citep{Banuti1}.}
\end{figure}

The flow and computational parameters for the ZPG transitional boundary-layer flows in this study are presented below. A Mach number of $0.2$, following \citet{Sayadi1}, and a free-stream reduced pressure of $p_\mathit{r,\infty}=1.10$ for the cases at supercritical pressure are selected. To illustrate the considered thermodynamic regimes, figure~\ref{fig_tp_diagram} shows the reduced temperature-pressure ($T_\mathit{r}$--$p_\mathit{r}$) diagram with selected isolines of reduced density $\rho_\mathit{r}=\rho^*/\rho^*_\mathit{c}$. Two thermodynamic regimes at supercritical pressure are considered, relative to the reduced pseudo-critical temperature of $T_\mathit{r,pc}=1.024$, both with a free-stream reduced temperature of $T_\mathit{r,\infty}=0.90$ and a weakly heated isothermal wall (subscript $(\cdot)_\mathit{w}$). In the subcritical (liquid-like) temperature case (denoted hereafter as Tw095), the wall temperature is $T_\mathit{r,w}=0.95$, such that the boundary-layer temperature remains below $T_\mathit{r,pc}$. It is important to note that the term `subcritical' here refers solely to the thermodynamic regime and should not be confused with subcritical growth below the critical Reynolds number in hydrodynamic stability theory \citep{Schmid1}. In the transcritical temperature case (denoted hereafter as Tw110), i.e. under pseudo-boiling conditions \citep{Banuti1}, the wall temperature reaches $T_{r,w}=1.10$, causing the boundary-layer temperature crossing the Widom line near the wall, which in turn triggers Mode-II instability. Note that, while the Widom line is formally defined in the temperature-pressure phase diagram (as shown in figure~\ref{fig_tp_diagram}), it is used hereafter as a convenient reference in spatial coordinates, where it corresponds to the local pseudo-critical point at a given supercritical pressure. In both cases, the free-stream compressibility factor $Z_\infty=Z_\mathit{c} p_\mathit{r,\infty}/(\rho_\mathit{r,\infty} T_\mathit{r,\infty})$ is equal to $0.254$. All relevant flow parameters are listed in table~\ref{tab:tableBF}.
\begin{table}
  \begin{center}
\def~{\hphantom{0}}
\begin{tabular}{w{c}{0.20\textwidth}w{c}{0.12\textwidth}w{c}{0.30\textwidth}w{c}{0.12\textwidth}cccccccc}
     & \multicolumn{2}{c}{\hspace{-1.3cm}Non-ideal fluid at supercritical pressure}  & Ideal gas  \\[0.6em]
    Case & Tw095 & Tw110 & TadIG  \\[0.2em]
    Regime & subcrit.~temp. &  \multicolumn{1}{c}{transcrit.~temp.} & - \\[0.2em]
    Wall & isotherm & isotherm & adiabatic  \\[0.2em]
    $T^*_\mathit{w}/T^*_\mathit{c}$ & 0.95 & 1.10 & -  \\[0.2em]
    $T^*_\mathit{w}/T^*_\infty$ & 1.056 & 1.222 & 1.007 \\[0.2em]
    Line style & \textcolor{orange}{\tikz[baseline=-0.5ex]{\draw[line width=1.2pt, fill=orange] (0,0) -- (1.2,0);}} & \textcolor{red}{\tikz[baseline=-0.5ex]{\draw[line width=1.2pt, fill=red] (0,0) -- (1.2,0);}} & \textcolor{blue}{\tikz[baseline=-0.5ex]{\draw[line width=1.2pt, fill=red] (0,0) -- (1.2,0);}}  \\
\end{tabular}
\captionsetup{justification=justified} 
\caption{Thermodynamic conditions for the three flow cases. For the supercritical pressure cases, the common flow parameters are the free-stream reduced pressure $p^*_\infty/p^*_\mathit{c}=1.10$ and reduced temperature $T^*_\infty/T^*_\mathit{c}=0.90$. For all cases, the Mach number is  $M_\infty=0.2$. The wall temperature is denoted by $T^*_\mathit{w}$. The non-ideal fluid flow cases at supercritical pressure are represented in the reduced temperature-pressure $(T_\mathit{r}$ --$p_\mathit{r})$ diagram in figure~\ref{fig_tp_diagram}.}
\label{tab:tableBF}
\end{center}
\end{table}
For cases Tw095 and Tw110, the free-stream parameters are set as follows: Eckert number $Ec_\infty=0.0159$, Prandtl number $Pr_\infty=1.0$, reduced speed of sound $a_\mathit{r,\infty}=\sqrt{a^{*2}_{\infty}/(p^*_\mathit{c} \upsilon^*_\mathit{c})}=2.766$ ($\upsilon^*=1/\rho^*$, specific volume), reduced specific heat at constant pressure $c_\mathit{p,r,\infty}=8.024$, and reduced specific heat at constant volume $c_\mathit{\upsilon,r,\infty}=9/2$. In contrast, for case TadIG, the free-stream Eckert number, Prandtl number, and heat capacity ratio are $Ec_\infty=0.016$, $Pr_\infty=0.75$, and $c^*_\mathit{p}/c^*_\mathit{\upsilon}=1.4$, respectively.

The computational domain is a rectangular box on top of the flat plate. The inlet boundary-layer thickness $\delta_{99,0}$ is used as the reference length scale and is set to unity at the inlet location, $x_0=x^*_0/\delta^*_{99,0}$. The inlet Reynolds number is defined based on the distance from the leading edge, $\Rey_{x,0}$. The domain extends to $x_\mathit{e}$ and is initialised with the self-similar boundary-layer laminar solution described in \S\,\ref{sec:initial_conditions}. The spanwise domain size corresponds to the disturbance spanwise wavelength $\lambda_\mathit{z}$, with $0<z/\delta_{99,0}<\lambda_\mathit{z}$. Further details on the DNS set-up, including a sensitivity analysis of the grid resolution, are provided in Appendix~\ref{sec:appA}.

\subsection{Initial conditions}
\label{sec:initial_conditions}
\begin{figure}
\centering
\includegraphics[angle=-0,trim=0 0 0 0, clip,width=1.0\textwidth]{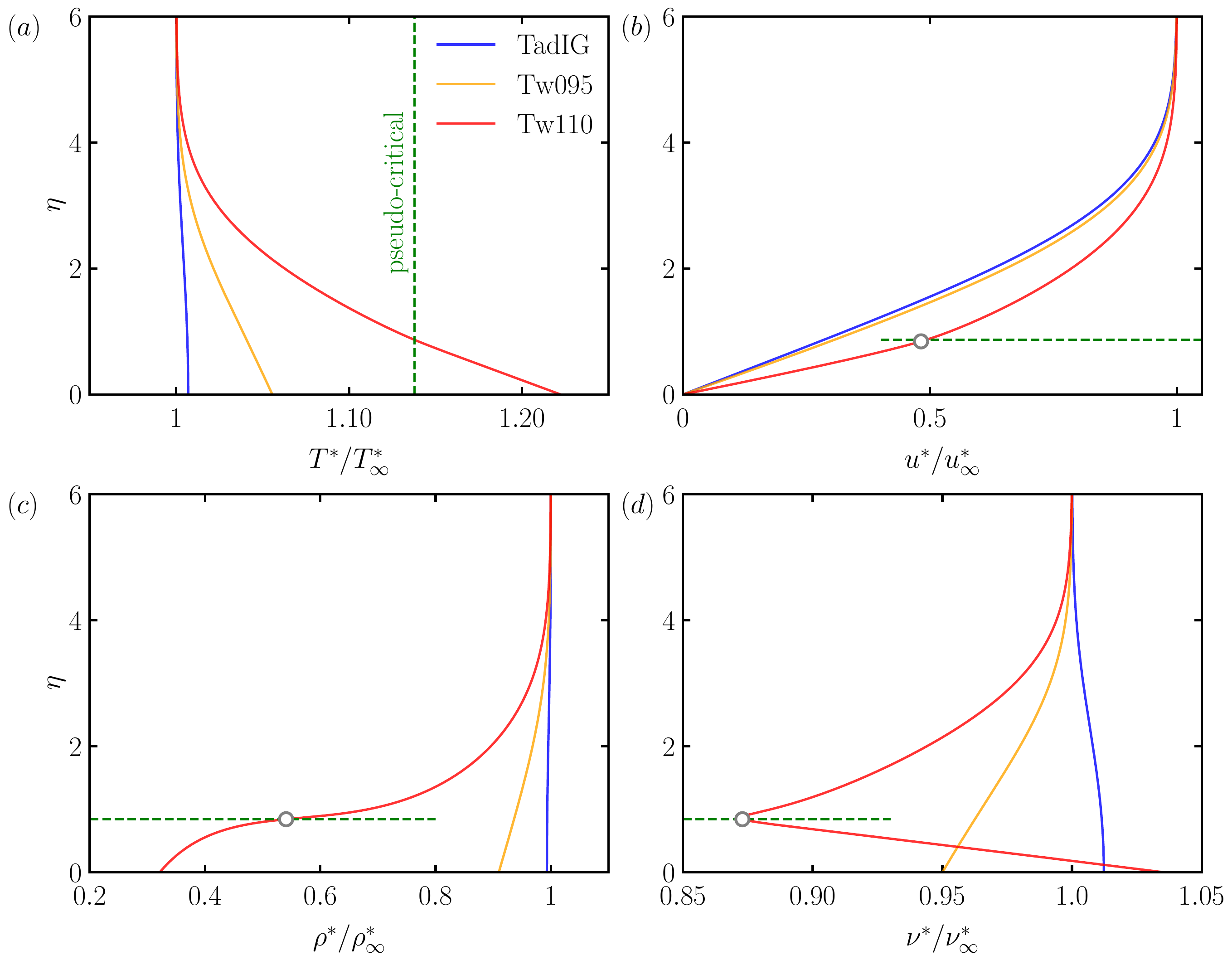}
\captionsetup{justification=justified}  
\caption{\label{fig_bl_laminar} Laminar profiles for the considered cases: (a) temperature $T^*/T^*_\infty$, (b) streamwise velocity $u^*/u^*_\infty$, (c) density $\rho^*/\rho^*_\infty$, and (d) kinematic viscosity $\nu^*/\nu^*_\infty$ as a function of the self-similar wall-normal coordinate $\eta$. The line legend is in agreement with table~\ref{tab:tableBF} for cases TadIG, Tw095, and Tw110. The dashed green line (\textcolor{matplotlibgreen}{\rule[0.5ex]{0.2cm}{1pt}} \textcolor{matplotlibgreen}{\rule[0.5ex]{0.2cm}{1pt}}) indicates the pseudo-critical point, i.e. at the pseudo-critical temperature $T^*=T^*_\mathit{pc}$. The location of the GIP for the transcritical case Tw110 is marked by the circle (\textcolor{gray}{$\circ$}) symbol in (b--d).}
\end{figure}

The computational domain is initialised using the self-similar boundary-layer profiles based on Lees-Dorodnitsyn variables \citep{Ren1,Boldini1}. The initial flow profiles for all cases listed in table~\ref{tab:tableBF} are plotted over the wall-normal coordinate $\mathrm{d}\eta=\rho^* u^*_\infty/\sqrt{2\xi} \,\mathrm{d}y^*$ in figure~\ref{fig_bl_laminar}. As temperature increases from liquid-like to vapour-like conditions (figure~\ref{fig_bl_laminar}a), the largest gradients in thermodynamic and transport properties -- particularly in density (figure~\ref{fig_bl_laminar}c) -- occur at the pseudo-critical point (green dashed line). This gives rise to an inflectional base-flow profile in case Tw110, as defined by the GIP-criterion, i.e. $\mathrm{d}(\bar{\rho} \ \mathrm{d}\bar{u}/\mathrm{d}y)/\mathrm{d}y=0$. As shown in figure~\ref{fig_bl_laminar}(d), the GIP coincides with the minimum of kinematic viscosity near the pseudo-critical point, and occurs for non-polar fluids at supercritical pressure and transcritical temperature \citep{Bugeat1,Bugeat2}. The streamwise velocity (figure~\ref{fig_bl_laminar}b) highlights the more pronounced profile of case Tw110, which resembles that of a transitional boundary-layer profile.
Appendix~\ref{sec:appB} presents a comparison between the unperturbed 2-D DNS solutions, which have reached a steady-state solution, and the initial self-similar solutions for both cases Tw095 and Tw110.

\subsection{Disturbance strip}
\label{sec:disturbance_strip}
Disturbances are introduced via a blowing and suction disturbance strip on the flat-plate surface. Once the laminar base flow reaches a steady-state solution, the localised three-dimensional disturbance \citep{Sayadi1} is activated as
\begin{equation}
    v(x,y=0,z,t)=f(x) \left[ A_{\text{2-D}} \sin (\omega_{\text{2-D}}t) + A_{\text{3-D}} \sin(\omega_{\text{3-D}} t) \cos( \beta_0 z) \right],
    \label{eq:diststrip}
\end{equation}
where $A_{\text{2-D}}=A^*_{\text{2-D}}/u^*_\infty$ and $A_{\text{3-D}}=A^*_{\text{3-D}}/u^*_\infty$ are the wave amplitudes for the primary 2-D and $z$-symmetric 3-D wave, respectively, and $\omega_{\text{2-D}}=\omega^*_{\text{2-D}}\delta^*_{99,0}/u^*_\infty$ and $\omega_{\text{3-D}}=\omega^*_{\text{3-D}}\delta^*_{99,0}/u^*_\infty$ are the corresponding angular frequencies. The spanwise wavenumber $\beta_0=\beta^*_0\delta^*_{99,0}=2\pi/\lambda_\mathit{z}$ equals to the spanwise size $z_\mathit{e}$ of the computational domain (see table~\ref{tab:numerical_setup}). The streamwise variation in \eqref{eq:diststrip} is governed by the function $f(x)=15.1875\xi^5-35.4375\xi^4+20.25\xi^3$, with $\xi=(x-x_1)/(x_\mathit{mid}-x_1)$ for $x_1<x<x_\mathit{mid}$ and $\xi=(x_2-x)/(x_2-x_\mathit{mid})$ for $x_\mathit{mid}<x<x_2$, where $x_\mathit{mid}=(x_1+x_2)/2$ corresponds to $Re_\mathit{x,mid}$. The disturbance strip is positioned upstream of branch I (see figure~\ref{fig_lst}). Since no experimental studies on controlled transition with supercritical fluids are currently available, the ideal-gas H-type breakdown scenario from \citet{Sayadi1}, validated by \citet{Boldini2}, is used as a reference, with $F_{\text{2-D}}=124 \times 10^{-6}$ and $\lambda_\mathit{z}=9.63$. 
In \S\,\ref{sec:2D} and \S\,\ref{sec:3D}, 2-D and 3-D simulations are performed, respectively. In the 2-D simulations, a 2-D wave ($\beta_\mathit{r}=0$) with frequency $F_{\text{2-D}}$ is forced, with amplitude $A_{\text{2-D}}=1.0 \times 10^{-8}$ in the linear regime, and increased to either $7.5 \times 10^{-4}$ or $7.5 \times 10^{-3}$ in the nonlinear regime. The latter corresponds to the amplitude used in the ideal-gas H-type breakdown reference simulation from \citet{Boldini2}. For the 3-D simulations, $A_{\text{2-D}}$ is set to $7.5 \times 10^{-3}$, while two amplitude levels for the $z$-symmetric 3-D wave with $F_{\text{3-D}}=62 \times 10^{-6}$ are considered: infinitesimally small (denoted as `IA') or finite (denoted as `LA'). The differences in forcing parameters for all simulations are summarised in table~\ref{tab:numerical_setup2}. The non-dimensional angular frequency $\omega$ and the frequency parameter $F$ are related by $\omega=F Re_{0}$, with $F=\omega^*\mu^*_\infty/(\rho^*_\infty u^{*2}_\infty)$ and local Reynolds number $Re_{0}=\sqrt{Re_{x,0}}$.
\begin{table}
  \begin{center}
\def~{\hphantom{0}}
  \begin{tabular}{l@{\hspace{6mm}}c@{\hspace{3mm}}c@{\hspace{3mm}}c@{\hspace{3mm}}c@{\hspace{3mm}}c@{\hspace{3mm}}c}
        Parameter & TadIG & Tw095-LA & Tw095-IA & Tw110-LA  & Tw110-IA \\[0.6em]
       $A_{\text{3-D}}$  &  $8.5 \times 10^{-5}$ &  $8.5 \times 10^{-5}$   & $1.0 \times 10^{-8}$ &  $8.5 \times 10^{-5}$  & $1.0 \times 10^{-8}$

  \end{tabular}
  \captionsetup{justification=justified}
  \caption{Forcing set-up: `LA' and `IA' denote finite 3-D amplitude forcing and infinitesimally small 3-D amplitude forcing, respectively. Others parameters are fixed: $A_{\text{2-D}}=7.5 \times 10^{-3}$ at $F_{\text{2-D}}=124 \times 10^{-6}$, $z$-symmetric 3-D wave at $F_{\text{3-D}}=62 \times 10^{-6}$, with $\Rey_\mathit{x,mid}=1.72 \times 10^5$ (cases TadIG and Tw095) or $\Rey_\mathit{x,mid}=9.61 \times 10^4$ (cases Tw110).}
  \label{tab:numerical_setup2}
  \end{center}
\end{table}

\section{Two-dimensional analysis: linear and nonlinear regime} \label{sec:2D}
This section focuses on the behaviour of Mode-II instability and its evolution from the linear (\S\,\ref{sec:linear_regime}) to the nonlinear regime (\S\,\ref{sec:nonlinear_regime}), highlighting the role of the Widom line (pseudo-boiling). This 2-D investigation serves as a precursor to the fully 3-D breakdown scenarios in \S\,\ref{sec:3D}.

\subsection{Linear evolution}
\label{sec:linear_regime}

\begin{figure}
\centering
\includegraphics[angle=-0,trim=0 0 0 0, clip,width=1.0\textwidth]{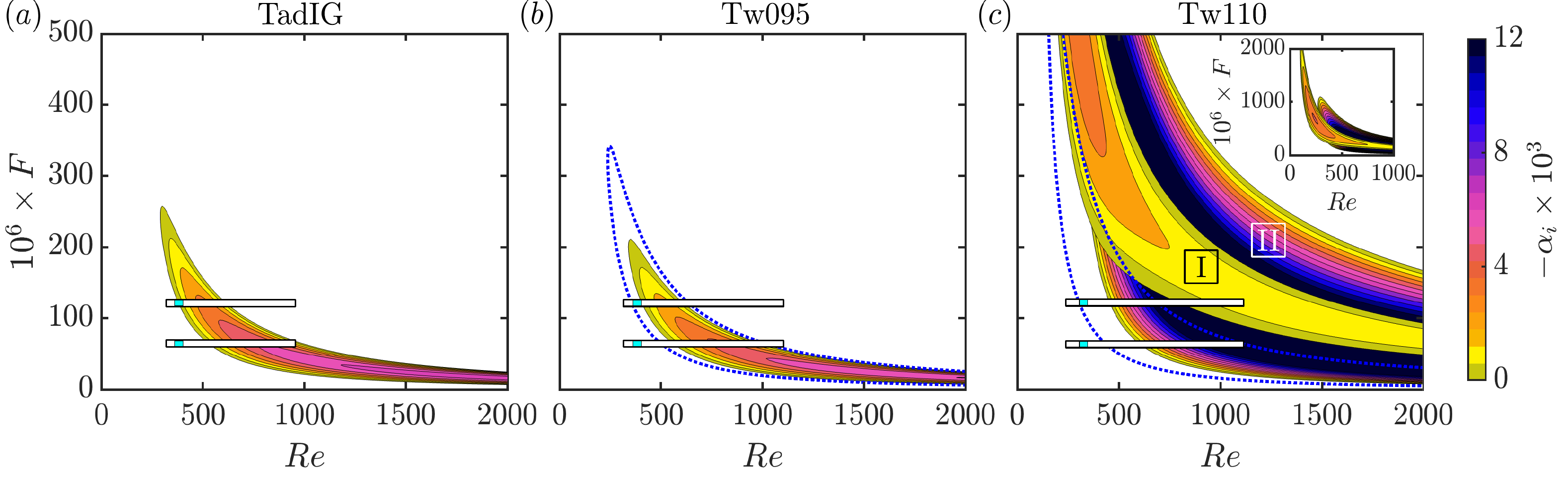}
\captionsetup{justification=justified}  
\caption{\label{fig_lst} Growth-rate ($-\alpha_\mathit{i}$) contours in the $Re$--$F$ stability diagram: (a) TadIG, (b) Tw095, and (c) Tw110 (Mode I and II). The dotted blue lines in (b,c) represent the ideal-gas neutral stability at equal $T^*_\mathit{w}/T^*_\infty$-ratios. In the inset of (c), the wide frequency band of Mode I and II is displayed. The location of the DNS domain and perturbation strip for subharmonic breakdown, i.e. $F_{\text{3-D}}=0.5F_{\text{2-D}}=62\times10^{-6}$, are marked by white and cyan bars, respectively, as described in \S\,\ref{sec:disturbance_strip}.}
\end{figure}

Starting from the base-flow profiles in \S\,\ref{sec:initial_conditions}, linear stability analysis is performed for the cases listed in table~\ref{tab:tableBF}. Figure~\ref{fig_lst} displays the growth rate $-\alpha_\mathit{i}$ in the $Re$--$F$ stability diagram, where $Re=\sqrt{Re_\mathit{x}}$.
In figure~\ref{fig_lst}(b), increasing the wall temperature toward $T_\mathit{r,pc}$ stabilises the flow. This trend is reversed under ideal-gas conditions, where a higher $T^*_\mathit{w}/T^*_\infty$-ratio leads to destabilisation (dotted blue line). In figure~\ref{fig_lst}(c), upon crossing the pseudo-critical point from the liquid-like free stream, Mode II appears \citep{Ren1}, becoming highly unstable even at low Reynolds numbers. Simultaneously, Mode I -- exhibiting growth rates an order of magnitude lower -- becomes unstable over a much broader frequency band, extending up to $F\approx 2 \times 10^{-3}$. This modal behaviour shifts the critical Reynolds number $Re_\mathit{cr}$, i.e. $\alpha_\mathit{i}=0$, toward lower values. The DNS domain, indicated by rectangular bars in figure~\ref{fig_lst}, is placed inside the linearly unstable region, with the disturbance strip (coloured in cyan) positioned somewhat upstream of, or close to, $Re_\mathit{cr}$ at the selected frequency $F$. As shown in figure~\ref{fig_lst}(c) for case Tw110, the DNS domain spans nearly the entire linearly unstable region of Mode II at $F_{\text{3-D}}=124 \times 10^{-6}$.

\begin{figure}
\centering
\includegraphics[angle=-0,trim=0 0 0 0, clip,width=1.0\textwidth]{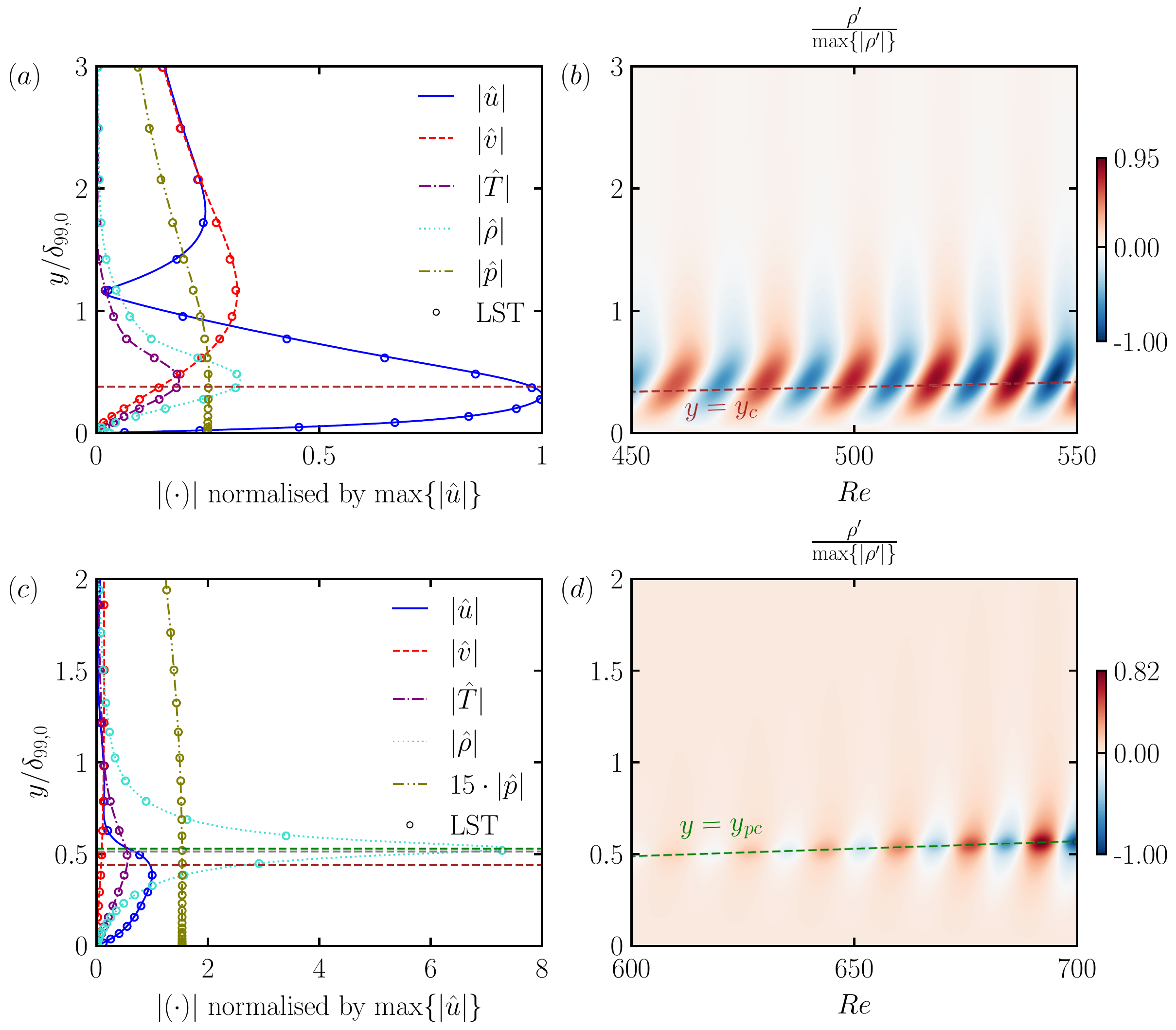}
\captionsetup{justification=justified}  
\caption{\label{fig_linear_eig}Case Tw095 (a,b) and Tw110 (c,d): (a,c) wall-normal eigenfunctions (lines DNS data, circles LST) of $u^{\prime}$, $v^{\prime}$, $
T^{\prime}$, $\rho^{\prime}$, and $p^{\prime}$ normalised by $\max\{|\hat{u}|\}$ at $\Rey_\mathit{\delta}=500$ (Tw095) and $\Rey_\mathit{\delta}=650$ (Tw110); contours $\rho^{\prime}$ (b,d) normalised by their respective maximum. The locations of the pseudo-critical point $y=y_\mathit{pc}$, i.e. where $\bar{T}^*=T^*_\mathit{pc}$, the GIP $y=y_\mathit{GIP}$, and the critical layer $y=y_\mathit{c}$ are indicated in green (\textcolor{matplotlibgreen}{\rule[0.5ex]{0.1cm}{1pt}} \textcolor{matplotlibgreen}{\rule[0.5ex]{0.1cm}{1pt}}), grey (\textcolor{gray}{\rule[0.5ex]{0.1cm}{1pt}} \textcolor{gray}{\rule[0.5ex]{0.1cm}{1pt}}), and brown (\textcolor{matplotlibbrown}{\rule[0.5ex]{0.1cm}{1pt}} \textcolor{matplotlibbrown}{\rule[0.5ex]{0.1cm}{1pt}}), respectively.}
\end{figure}

In the context of the 2-D DNS, a steady laminar solution is first obtained, see \S\,\ref{sec:initial_conditions}, and LST-DNS comparisons of growth rate and phase speed are provided in Appendix~\ref{sec:appC}. The eigenfunctions, normalised by $\max\{|\hat{u}|\}$, are successfully compared in figures~\ref{fig_linear_eig}(a,c) at $Re=500$ for case Tw095 and at $Re=650$ for case Tw110. In the subcritical regime, $u^{\prime}$ shows a phase jump near $y/\delta_{99,0} \approx 1$, and $\rho^{\prime}$ is confined near the wall around the critical layer $y=y_\mathit{c}$, defined by $\bar{u}(y_\mathit{c})=c_\mathit{r}$, resembling the ideal-gas case TadIG (not shown here). For Tw110, Mode II is affected by the pseudo-critical point (solid green line at $y=y_\mathit{pc}$), with $\max\{|\hat{u}|\}$ located in the vapour-like regime. Figures~\ref{fig_linear_eig}(b,d) provide an overview of the modal instability in Tw095 and Tw110 via contours of $\rho^{\prime}$. In the latter case, the density fluctuations form `rope-shaped' patterns around the pseudo-critical point -- where gradients in transport and thermodynamic properties are largest (see figure~\ref{fig_bl_laminar}) --  near the GIP (horizontal grey line), in agreement with \citet{Ren1}. While similar density patterns occur for the second-mode instability in a hypersonic boundary layer \citep{Unnikrishnan1}, two key distinctions apply here: (i) transcritical Mode II is not linked to Mack's second mode \citep{Ren1}, and (ii) at $M_\infty=0.2$, pressure perturbations are predominantly of hydrodynamic nature. Their amplitude is higher at the wall and does not peak at the pseudo-critical point (see figure~\ref{fig_linear_eig}c), in contrast to the findings of \citet{Ren1}.

A key question concerning Mode-II instability, as it manifests under transcritical heating conditions, is its physical linear mechanism. \citet{Bugeat2} demonstrated in a plane Couette flow that shear and baroclinic effects interact to generate two vorticity waves around the central layer, coinciding with the location of the minimum of kinematic viscosity and the critical layer. Similarly, the present 2-D boundary-layer flow is analysed using the linearised vorticity equation, where $\bar{\Omega}=\partial \bar{v}/\partial x - \partial \bar{u}/\partial y$ is the base-flow vorticity, for the disturbance vorticity $\xi=\partial v^{\prime}/\partial x-\partial u^{\prime}/\partial y$ as:
\begin{equation}
	\begin{split}
		\underbrace{\dfrac{D \xi}{D t }}_{\mathit{LHS}} \approx \underbrace{v^{\prime} \dfrac{\partial^2 \bar{u}}{\partial y^2}}_{S_\mathit{\xi}}  +  \underbrace{\dfrac{\partial \bar{u}}{\partial y} \left( \dfrac{\partial u^{\prime}}{\partial x} + \dfrac{\partial v^{\prime}}{\partial y} \right)}_{C_\mathit{\xi}}  \,  \underbrace{-\dfrac{1}{\bar{\rho}^2}  \dfrac{\partial \bar{\rho}}{\partial y}\dfrac{\partial p^{\prime}}{\partial x}}_{B_\mathit{\xi}} + O(\mu).
	\end{split}
 \label{eq:span_budget}
\end{equation}
Here, $S_\mathit{\xi}$, $C_\mathit{\xi}$, and $B_\mathit{\xi}$ denote the shear, compressible stretching, and baroclinic terms, respectively. The viscous term is represented by $O(\mu)$. The term $1/\bar{\rho}^2 \, \partial \rho^{\prime}/\partial x \, \partial \bar{p}/\partial y$, which belongs to $B_\mathit{\xi}$, is negligible (see Appendix~\ref{sec:appB}). Figure~\ref{fig_linear_budget} evaluates \eqref{eq:span_budget} for case Tw110. For case Tw095, the only significant term of \eqref{eq:span_budget} far from the wall is the shear contribution $S_\mathit{\xi}$, where $|\partial^2 \bar{u}/\partial y^2|$ is maximal.
\begin{figure}
\centering
\includegraphics[angle=-0,trim=0 0 0 0, clip,width=1.0\textwidth]{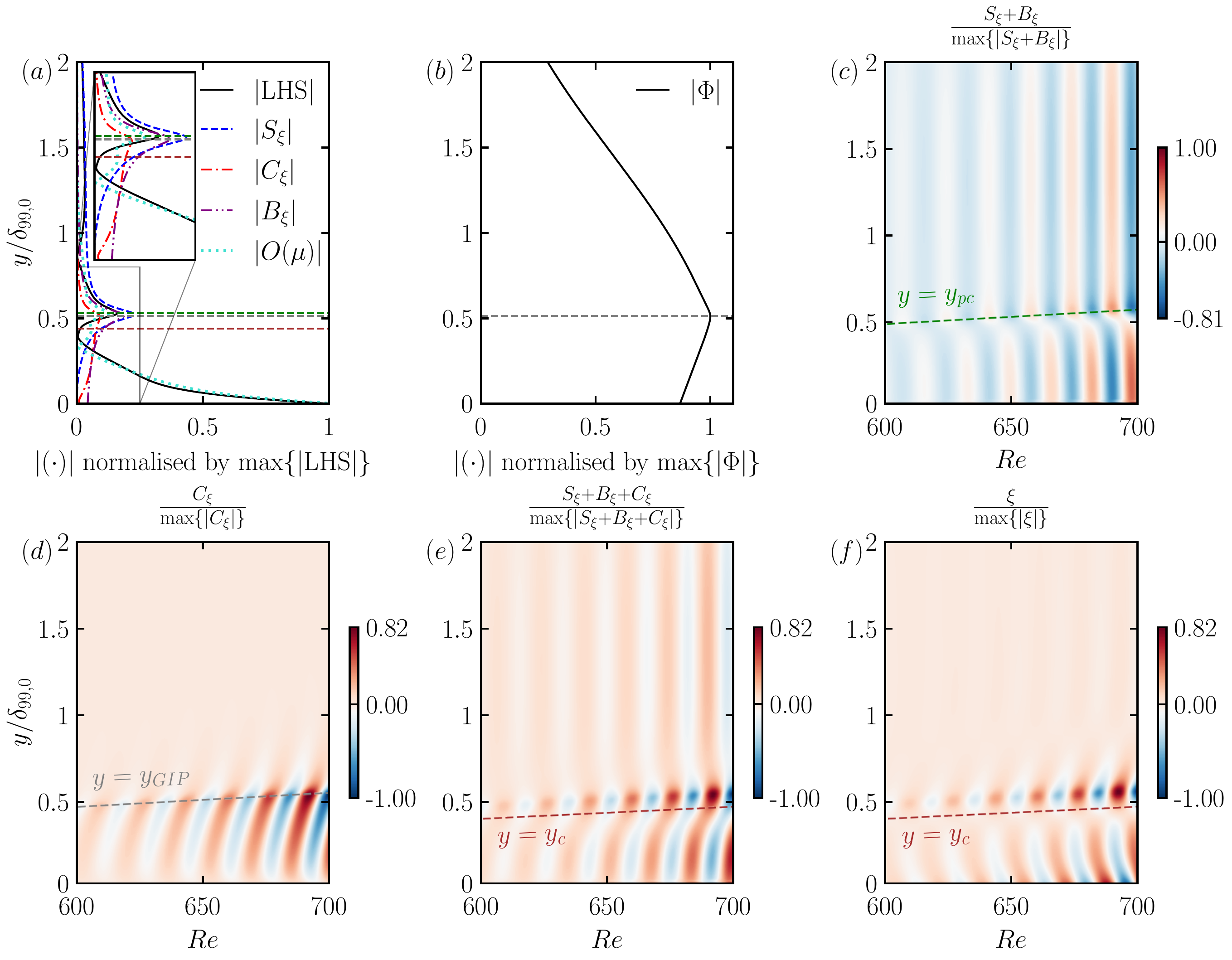}
\captionsetup{justification=justified}  
\caption{\label{fig_linear_budget}Case Tw110. Terms of the vorticity perturbation equation \eqref{eq:span_budget}: (a) spectral domain at $Re=650$ (all terms are normalised by $\max\{|D \xi/D t|\}$), (b) normalised density-weighted vorticity $|\Phi|=|\bar{\rho}\bar{\Omega}|$, (c) $S_\mathit{\xi}+B_\mathit{\xi}$, (d) $C_{\xi}$, (e) $S_\mathit{\xi}+B_\mathit{\xi}+C_\mathit{\xi}$, and (f) $\xi$. The locations of the pseudo-critical point $y=y_\mathit{pc}$, i.e. where $\bar{T}^*=T^*_\mathit{pc}$, the GIP $y=y_\mathit{GIP}$, and the critical layer $y=y_\mathit{c}$ are indicated in green (\textcolor{matplotlibgreen}{\rule[0.5ex]{0.1cm}{1pt}} \textcolor{matplotlibgreen}{\rule[0.5ex]{0.1cm}{1pt}}), grey (\textcolor{gray}{\rule[0.5ex]{0.1cm}{1pt}} \textcolor{gray}{\rule[0.5ex]{0.1cm}{1pt}}), and brown (\textcolor{matplotlibbrown}{\rule[0.5ex]{0.1cm}{1pt}} \textcolor{matplotlibbrown}{\rule[0.5ex]{0.1cm}{1pt}}), respectively.}
\end{figure} 
In spectral domain at $Re=650$, figure \ref{fig_linear_budget}(a) confirms that both $S_\mathit{\xi}$ and $B_\mathit{\xi}$ reach a maximum near the pseudo-critical point, where the minimum of $\bar{\nu}$ is located. At the GIP, where the density-weighted vorticity $\Phi=\bar{\rho}\bar{\Omega}$ is maximal (figure~\ref{fig_linear_budget}b), the compressible stretching term $C_\mathit{\xi}$ also peaks and remains large in the vapour-like region. The viscous term $O(\mu)$ exhibits a local maximum at the pseudo-critical point due to its direct dependence on $|\partial^2 \bar{u}/\partial y^2|$, which is largest at $y_\mathit{pc}$. In figure~\ref{fig_linear_budget}(c), the sum of $S_\mathit{\xi}$ and $B_\mathit{\xi}$ exhibits two out-of-phase waves with a phase difference of $\pi$ (not shown), located around $y_\mathit{pc}$ and exhibiting asymmetry due to the structure of $B_\mathit{\xi}$. When the out-of-phase $C_\mathit{\xi}$-term is included, as shown in figure~\ref{fig_linear_budget}(e), the two vorticity waves shift around the critical layer at $y=y_\mathit{c}$. This behaviour is consistent with \citet{Bugeat2}, where $C_\mathit{\xi}$ was absent and the critical layer coincided with the GIP, i.e., the centre line of the plane Couette flow under the parallel flow assumption. In other words, the misalignment between the critical layer and the GIP in the boundary layer results in a corresponding shift of the two vorticity waves. With the addition of the viscous $O(\mu)$-term, the vorticity perturbation $\xi$ in figure~\ref{fig_linear_budget}(f) is further amplified near the wall, tilting the near-wall vorticity wave in the upstream direction.

\subsection{Nonlinear evolution}
\label{sec:nonlinear_regime}

We now focus on the nonlinear response of the boundary layer to a finite amplitude perturbation, with the blowing-suction amplitude increased to $A_{\text{2-D}}=7.5 \times 10^{-3}$, compared to the infinitesimal amplitude used in \S\,\ref{sec:linear_regime}. Fourier modes are denoted using the double-spectral notation $(\omega /\omega_{\text{2-D}},0)$, where $\omega_{\text{2-D}}$ is the fundamental frequency.

\begin{figure}
\centering
\includegraphics[angle=-0,trim=0 0 0 0, clip,width=1.0\textwidth]{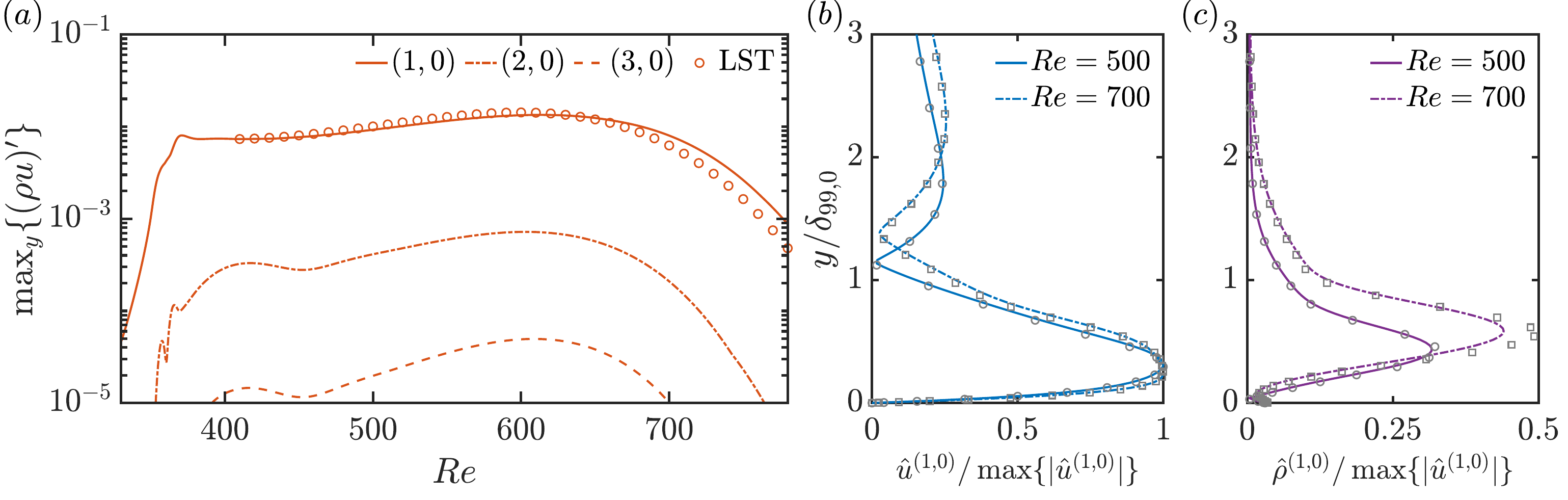}
\captionsetup{justification=justified}  
\caption{\label{fig_amp2D_T09w095}Case Tw095 with $A^{(1,0)}_{\text{2-D}}=7.5 \times 10^{-3}$: (a) maximum wall-normal mass-flux amplitude for mode (1,0) (solid line), (2,0) (dash-dotted line), and (3,0) (dashed line); streamwise velocity (b) and density (c) perturbations as a function of the wall-normal coordinate $y/\delta_{99,0}$ at $Re=500$ (solid line) and $Re=700$ (dash-dotted line), normalised by their respective $\max\{|\hat{u}^{(1,0)}|\}$. In (b,c), the scaled LST solution is represented with circle (\textcolor{gray}{$\circ$}) symbols at $Re=500$ and with square (\textcolor{gray}{$\smallsquare$}) symbols at $Re=700$.}
\end{figure}

Figure~\ref{fig_amp2D_T09w095}(a) shows the downstream modal evolution for case Tw095.
Once the $1\%$-threshold is crossed at $Re \approx 600$, the primary (1,0) mode decays. Higher harmonics (2,0), (3,0), and beyond are slaved to the primary wave through weakly nonlinear effects in the receptivity region. Despite being linearly stable, they follow the streamwise growth -- and subsequent stabilisation -- of the forced (1,0), with amplitudes about two order of magnitude lower. Wall-normal profiles of streamwise velocity and density at $Re=500$ and $700$ (figures~\ref{fig_amp2D_T09w095}b,c) reveal only minor nonlinear effects, with the boundary-layer receptivity resembling the incompressible TadIG case.

\begin{figure}
\centering
\includegraphics[angle=-0,trim=0 0 0 0, clip,width=1.0\textwidth]{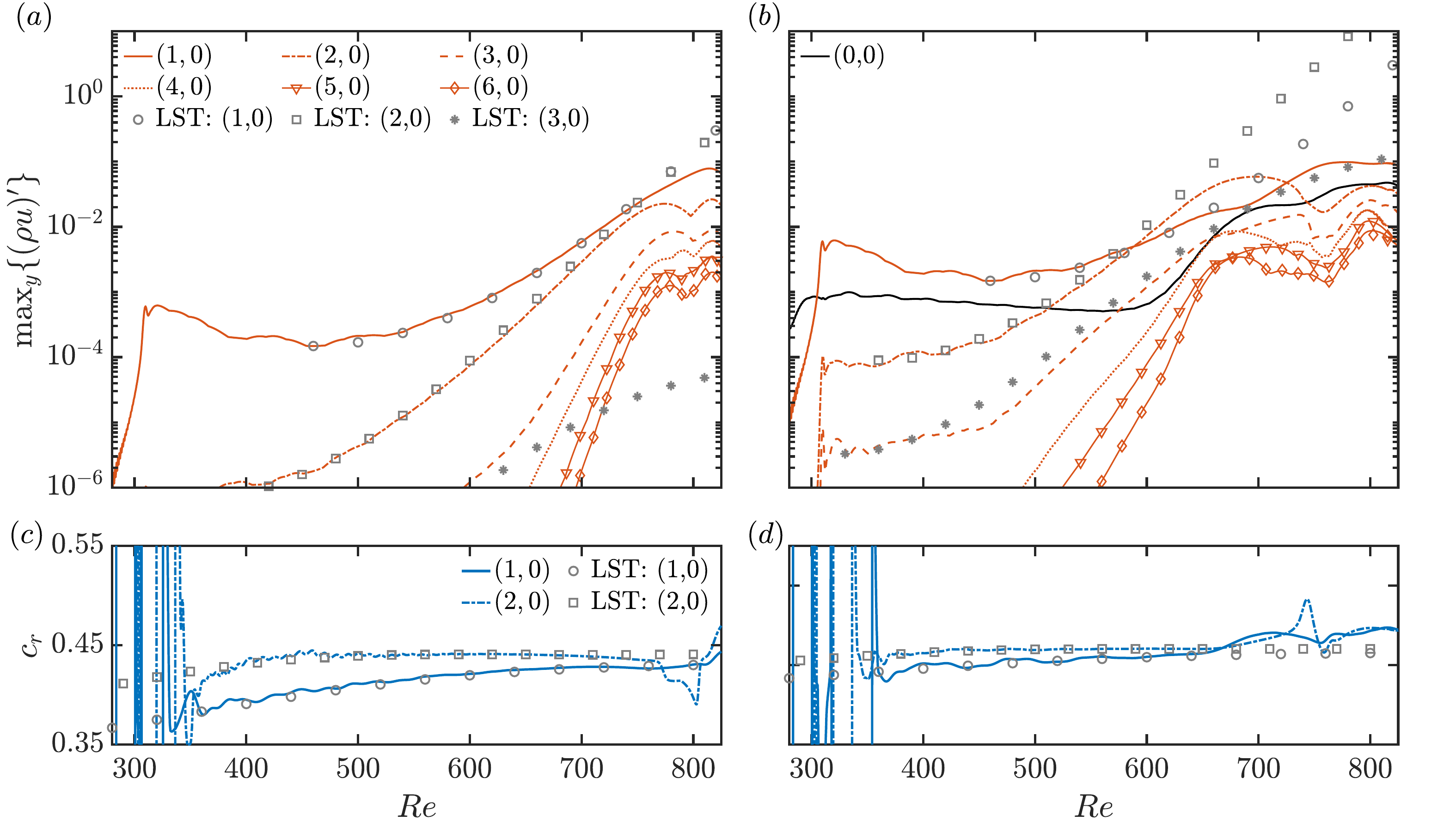}
\captionsetup{justification=justified}  
\caption{\label{fig_amp2D_T09w110}Case Tw110: (a,c) $A^{(1,0)}_{\text{2-D}}=7.5 \times 10^{-4}$ and (b,d) $A^{(1,0)}_{\text{2-D}}=7.5 \times 10^{-3}$; (a,b) maximum wall-normal mass-flux amplitude for mode $(1,0)$ (solid line), $(2,0)$ (dash-dotted line), $(3,0)$ (dashed line), $(4,0)$ (dotted line), $(5,0)$ (solid line with \textcolor{mycolor2}{$\smalltriangledown$}), and $(6,0)$ (solid line with \textcolor{mycolor2}{$\smalldiamond$}); (c,d) phase speed $c_\mathit{r}$ for mode $(1,0)$ (solid line) and $(2,0)$ (dash-dotted line). In (b), the mean-flow distortion $(0,0)$ is indicated with a black solid line. The LST solution is represented with circle (\textcolor{gray}{$\circ$}) symbols for mode $(1,0)$, square (\textcolor{gray}{$\smallsquare$}) symbols for mode $(2,0)$, and asterisk (\textcolor{gray}{$\ast$}) symbols for mode $(3,0)$.}
\end{figure}

In contrast to the subcritical regime, nonlinearity significantly influences the 2-D modal evolution in case Tw110. To highlight this, $A_{\text{2-D}}$ is increased from $7.5 \times 10^{-4}$ in figure~\ref{fig_amp2D_T09w110}(a) to $7.5 \times 10^{-3}$ in figure~\ref{fig_amp2D_T09w110}(b). In the low-amplitude case, modes $(1,0)$ and $(2,0)$ follow the LST prediction up to $Re \approx 700$, and, as shown in figure~\ref{fig_lst}(c), linear instability also arises at $2 F_{\text{2-D}}$, with a larger growth rate than that of $(1,0)$. Higher harmonics ($\omega/\omega_{\text{2-D}}\geq 3$) deviate from their linear evolution, growing rapidly to high amplitudes -- unlike in the subcritical regime (figure~\ref{fig_amp2D_T09w095}a). This behaviour resembles the Kelvin-Helmholtz (KH) instability in a mixing layer \citep{Babucke1}. In the high-amplitude case in figure~\ref{fig_amp2D_T09w110}(b), higher harmonics exceed the $0.1\%$ amplitude threshold at lower values of $Re$. Mode $(2,0)$, which is more unstable than mode $(1,0)$ in the linear regime, surpasses $(1,0)$ at $Re \approx 600$, before nonlinearly saturating after reaching the $2\%$-amplitude threshold. As it peaks at $Re \approx 700$, a subharmonic resonance mechanism with its subharmonic mode $(1,0)$ -- typically observed as vortex pairing in mixing layers \citep{Monkewitz1} -- emerges, destabilising the latter. Thus, mode $(1,0)$ undergoes strong destabilisation and becomes the most dominant mode again further downstream. At this streamwise location, the mean-flow distortion, i.e. mode $(0,0)$, reaches up to $5\%$, and all higher harmonics are fully nonlinear. It is worth noting that, contrary to the conclusions of \citet{Kachanov2}, who reported that a threshold amplitude of the 2-D fundamental TS wave on the order of $29\%$ of the mean flow was required to amplify 2-D subharmonics, the amplitude of mode $(2,0)$ reaches only $5\%$ in the present case under transcritical conditions.
The resonant interaction between modes $(1,0)$ and $(2,0)$ is further evidenced by their phase speeds $c_\mathit{r}$ in figures~\ref{fig_amp2D_T09w110}(c,d). For the low-amplitude case ($A^{(1,0)}_{\text{2-D}}=7.5 \times 10^{-4}$), the phase speeds are sufficiently close near $Re \approx 800$. In contrast, for the high-amplitude case, $c^{(1,0)}_\mathit{r}$ and $c^{(2,0)}_\mathit{r}$ remain nearly identical across the entire computational domain triggering subharmonic resonance at $Re\approx 700$. Notably, around $Re \approx 730$, $c^{(2,0)}_\mathit{r}$ increases sharply, deviating from its linear evolution (in grey) and indicating strong nonlinear stabilisation (figure~\ref{fig_amp2D_T09w110}b).

\begin{figure}
\centering
\includegraphics[angle=-0,trim=0 0 0 0, clip,width=1\textwidth]{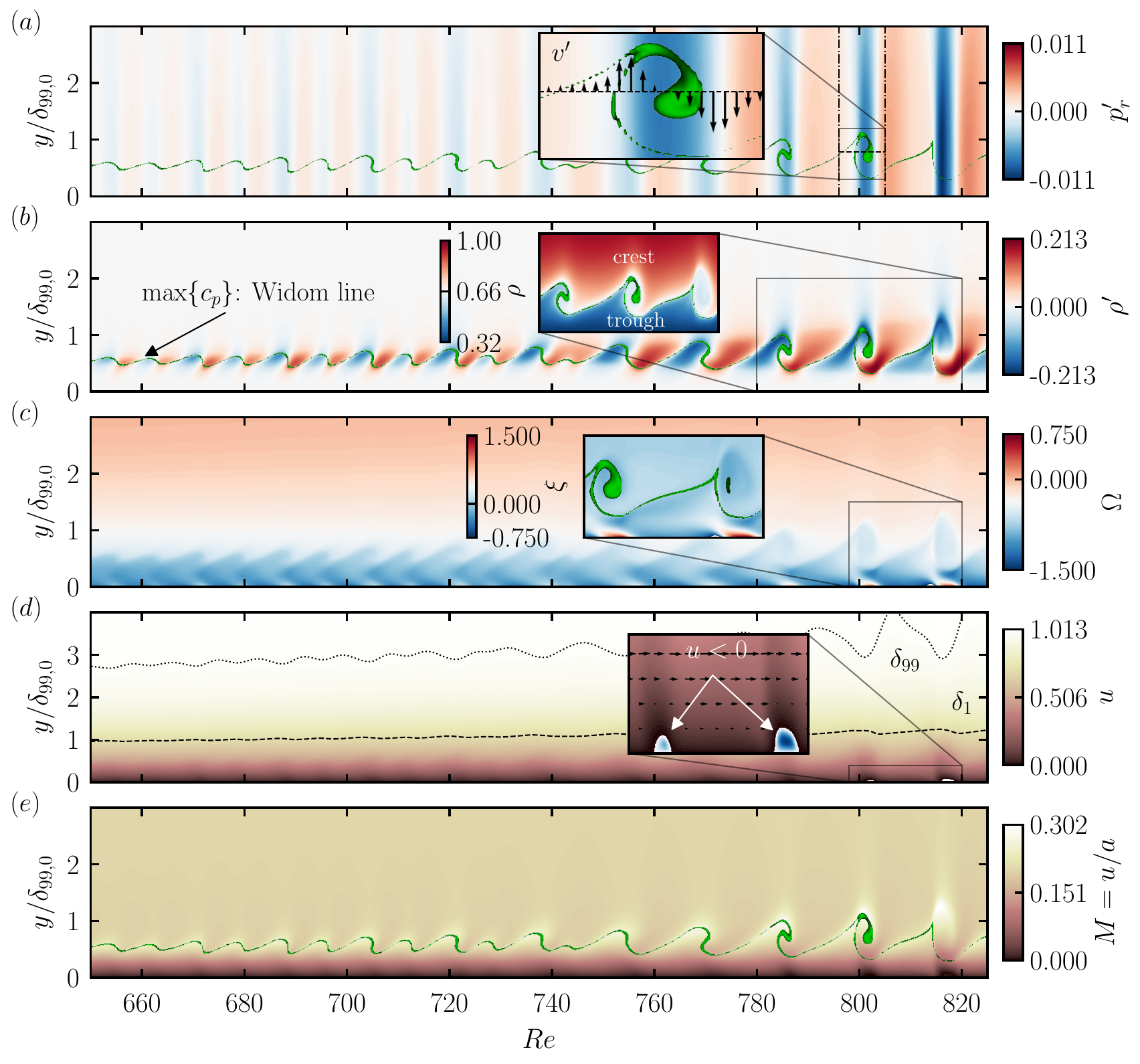}
\captionsetup{justification=justified}  
\caption{\label{fig_amp2D_T09w110_total}Case Tw110. Instantaneous contours at $T/T_0=0$, where $T_0=2\pi/\omega_0$ (fundamental frequency $\omega_0$), with $A^{(1,0)}_{\text{2-D}}=7.5 \times 10^{-3}$: (a) reduced pressure fluctuation $p^\prime_r=p^{*\prime}/p^*_\mathit{c}$, (b) density fluctuation $\rho^\prime$, (c) vorticity $\Omega$, (d) streamwise velocity $u$, with boundary-layer thickness $\delta_{99}$ and displacement thickness $\delta_1$ indicated by dotted and dashed lines, respectively, and (e) Mach number $M=u/a$. The Widom line $y=y_\mathit{WL}$ lies within the green region, i.e. between $98\% \max\{c_\mathit{p}\}$ and $\max\{c_\mathit{p}\}$. Note that the Widom line is used here as a spatial reference for the local pseudo-critical point at supercritical pressure. Insets in panels (a--c) show the wall-normal velocity fluctuation $v^\prime$, density $\rho$, and vorticity fluctuation $\xi$, respectively. The inset in (d) highlights separation zones ($u < 0$) in blue and includes velocity vectors $|\vec{V}|=\sqrt{u^2+v^2}$. A supplementary movie of the billow roll-ups is available at \url{https://github.com/pcboldini/DNSvisualisation}.}
\end{figure}

To qualitatively assess the nonlinear evolution of Mode II, selected perturbation contours for $A^{(1,0)}_{\text{2-D}}=7.5 \times 10^{-3}$ are shown in figures~\ref{fig_amp2D_T09w110_total}(a,b). The region corresponding to highest specific heat at constant pressure -- between $98\% \max\{c_\mathit{p}\}$ and $\max\{c_\mathit{p}\}$ -- is shaded in green. At this high disturbance level, the constant-pressure assumption for a laminar boundary layer breaks down, and $\max\{c_\mathit{p}\}$ depends on both reduced pressure and temperature. It is estimated using the analytical Widom-line relation $T_\mathit{r,pc}=1/A_{\text{VdW}} \ln(p_\mathit{r}) + 1$, where the Widom line -- defined as the locus of pseudo-critical points at supercritical pressure (figure~\ref{fig_tp_diagram}) -- is characterised by the critical slope $A_{\text{VdW}}=T_\mathit{c}/p_\mathit{c} (\mathrm{d}p/\mathrm{d}T)_\mathit{c}=4$ \citep{Banuti1}. Unlike in the linear regime (figures~\ref{fig_linear_eig}c,d), where the pseudo-critical point height grows with $\sqrt{x}$, the nonlinear disturbance wave propagation leads to increasing distortion of the Widom line, proportional to the perturbation magnitude (figures~\ref{fig_amp2D_T09w110_total}a,b). The large harmonic mode $(2,0)$, prominent between $Re\approx 600$ and $720$, halves the streamwise wavelength of the perturbation wave ($(1,0)$ as the fundamental), thereby doubling the oscillation frequency of the Widom line. As shown in the inset of figure~\ref{fig_amp2D_T09w110_total}(a), the crests and troughs of the Widom line align with regions of positive and negative wall-normal velocity perturbation $v^{\prime}$. Since $v^{\prime}$ and the pressure perturbation $p^{\prime}$ are out of phase \citep{Luhar1,Bugeat2}, the upward and downward displacements of the Widom line correspond to local pressure decreases and increases, respectively. Density perturbations in figure~\ref{fig_amp2D_T09w110_total}(b) -- initially confined near the Widom line in the linear regime (figure~\ref{fig_linear_eig}d) -- now reach up to $20\%$ of the free-stream value in regions where $p^{\prime}_r$ is negative, i.e. closer to the critical point. These perturbations mirror the billowing behaviour of the Widom line, with $\rho^{\prime}<0$ around the Widom-line crests and $\rho^{\prime}>0$ around the troughs. 

The term `billowing' is used by analogy with classical shear-layer flows, where a periodic train of large, rolling wave-like structures forms as the KH instability evolves nonlinearly \citep{Klaassen1,Liu1}. Here, a train of streamwise-growing density billows arises  (see inset of figure~\ref{fig_amp2D_T09w110_total}b). In both shear layers and the present case, billowing arises from an excess of vorticity concentrated in a localised flow region (see figure~\ref{fig_linear_budget}b). Although the train of billowing flow patterns is caused by the transcritical Mode-II instability rather than by the KH instability, \citet{Bugeat2} proved that the resulting vorticity fields are identical in the linear regime. Thus, it is not surprising that, in the nonlinear regime, the billowing patterns observed here resemble those seen in classical KH-type roll-ups, despite the absence of a true shear-layer mechanism.

Figures~\ref{fig_amp2D_T09w110_total}(b,c) highlight the nonlinear evolution of these billows via contours of $\rho^{\prime}$ and vorticity $\Omega=\partial v/\partial x- \partial u/\partial y$. Their deformation and streamwise growth progressively narrow the near-wall vapour-like region, which shifts significantly closer to the wall at $Re \approx 800$ compared to the linear regime. Simultaneously, the upper vapour-like fluid is lifted up by the rolling billows. Peaks in $\Omega$, which were previously aligned with the Widom line in the laminar regime, now appear below the Widom-line troughs in the near-wall region and at the wall. Starting from $Re \approx 780$, and recurring periodically downstream, the $\Omega$-peaks strengthen (high-shear regions), whose magnitude scales with their fluctuation $\xi$ (inset of figure~\ref{fig_amp2D_T09w110_total}c). The contours of streamwise velocity in figure~\ref{fig_amp2D_T09w110_total}(d) reveal the emergence of near-wall flow-reversal regions ($u<0$), located just beneath the near-wall $\Omega$-peaks at the Widom-line troughs, indicating the formation of a developing shear layer near the wall. In the vicinity of these flow reversal zones, a region of low streamwise velocity forms, leading to the reduction of the boundary-layer thickness $\delta_{99}$ by approximately $10\%$ relative to the unperturbed profile. At $Re \approx 800$, the near-wall separation zone reaches a height of about $2\%$ of $\delta_{99}$, while the billow crest extends beyond $y/\delta_{99,0}=1$. Here, the local Mach number reaches its maximum, as shown in figure~\ref{fig_amp2D_T09w110_total}(e). This behaviour results from both the boundary-layer displacement effect and the reduction in the local speed of sound at the Widom line. Downstream of the crest, the Mach number rapidly returns to its laminar value. 

To further investigate the near-wall high-shear region during billow roll-up, wall-normal profiles are extracted at $Re=802$, corresponding to the first flow reversal found in figure~\ref{fig_amp2D_T09w110_total}(d). 
\begin{figure}
\centering
\includegraphics[angle=-0,trim=0 0 0 0, clip,width=1.0\textwidth]{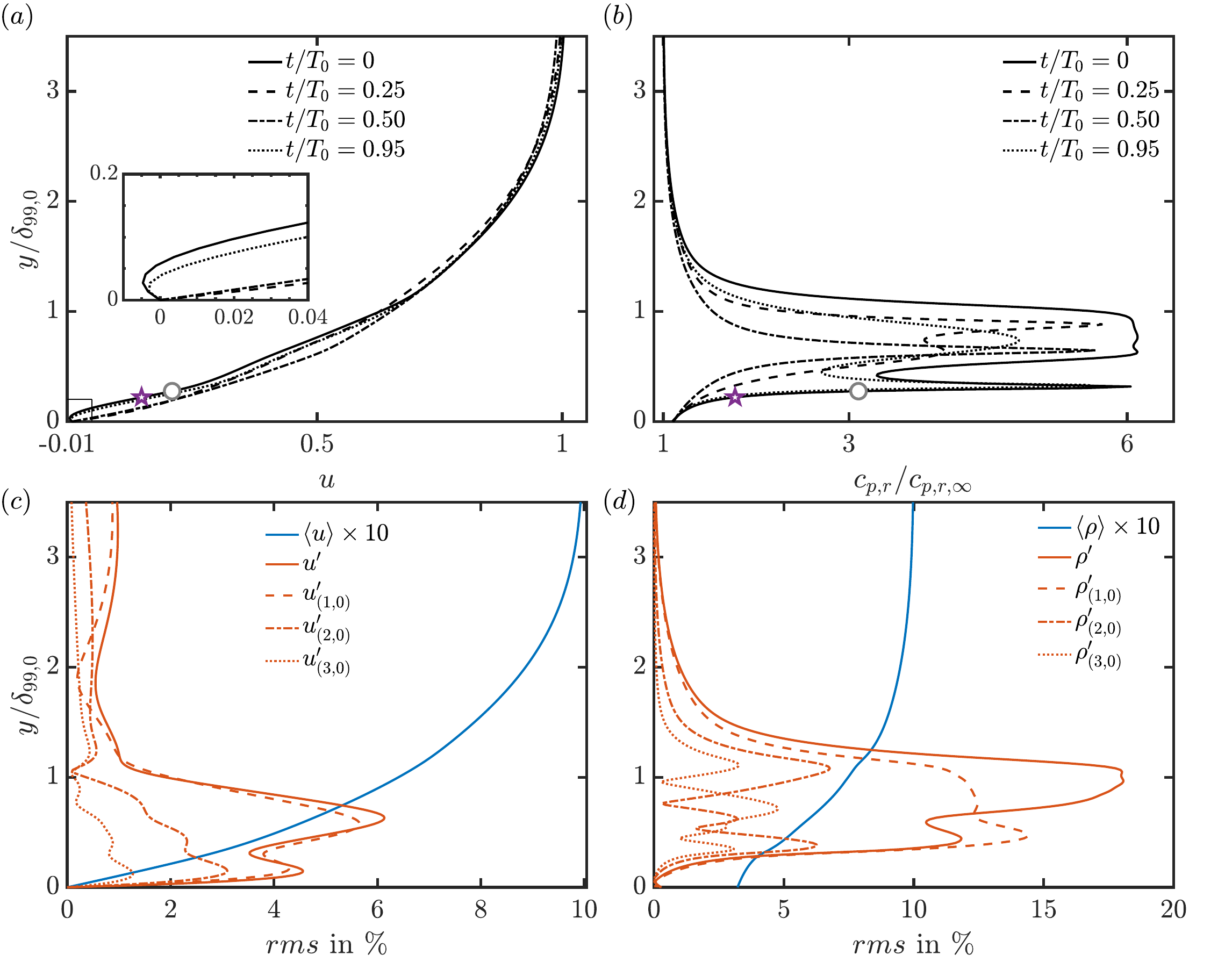}
\captionsetup{justification=justified}  
\caption{\label{fig_amp2D_T09w110_EF}Case Tw110 for $A^{(1,0)}_{\text{2-D}}=7.5 \times 10^{-3}$. Wall-normal slice at $Re=802$ showing: (a,b) instantaneous streamwise velocity $u$ and reduced specific heat at constant pressure $c_\mathit{p,r}/c_\mathit{p,r,\infty}$ at time periods $t/T_0=[0,0.25,0.5,0.95]$, where $T_0=2\pi/\omega_0$ is the fundamental forcing period, (c,d) time-averaged streamwise velocity $\langle u \rangle$ and density $\langle \rho \rangle$ profiles. In (c,d), the $rms$ (root-mean-square) of $u^{\prime}$ and $\rho^{\prime}$, respectively, and higher harmonics (modes $(1,0)$, $(2,0)$, and $(3,0)$) are shown. The locations of GIP and IP at $t/T_0=0$ are marked in insets (a,b) by circle (\textcolor{gray}{$\circ$}) and star (\textcolor{mycolor4}{$\smallstar$}) symbols, respectively. }
\end{figure}
Figures~\ref{fig_amp2D_T09w110_EF}(a,b) show the streamwise velocity $u$ and the reduced specific heat at constant pressure $c_\mathit{p,r}/c_\mathit{p,r,\infty}$, respectively, at four time instants within one forcing period $T_0=2\pi/\omega_0$. Between $t/T_0=0.95$ and $t/T_0=0$, the billow roll-up induces two distinct kinks in the $u$-profile: one at the billow crest around $y/\delta_{99,0}\approx 1.0$, and another at the trough near $y/\delta_{99,0}\approx 0.3$. These locations coincide with sharp peaks in $c_\mathit{p}$ (figure~\ref{fig_amp2D_T09w110_EF}b), with the highest values occurring at $t/T_0=0$, linked to a local pressure drop $p^{\prime}_r<0$ (see figure~\ref{fig_amp2D_T09w110_total}a). Beneath the near-wall kink in the vapour-like region, flow reversal at the wall is observed (inset of figure~\ref{fig_amp2D_T09w110_EF}a), along with a GIP, where $\Phi=\rho\Omega$ is maximal (grey circle). Here, the vorticity experiences a maximum, with an inviscid instability found according to the generalised Fjørtoft's criterion \citep{Bugeat2}. By $t/T_0=0.25$, no further GIP or flow reversal is observed, confirming the periodic nature of the billows generation. At $t/T_0=0.50$, the profiles resemble those of the unperturbed boundary layer (figure~\ref{fig_bl_laminar}). The large near-wall shift in the instantaneous $u$ observed over one $T_0$ is caused by strong near-wall velocity fluctuations -- reaching up to $5\%$ of the $rms$ value, (figure~\ref{fig_amp2D_T09w110_EF}c) -- sustained by both the primary mode $(1,0)$ and its first harmonic $(2,0)$. A secondary peak in $u^{\prime}_\mathit{rms}$ at $y/\delta_{99,0} \approx 0.7$ is primarily attributed to mode $(1,0)$. Density fluctuations (figure~\ref{fig_amp2D_T09w110_EF}d) are dominant in the high-$c_\mathit{p}$ region, with the $\rho^{\prime}_\mathit{rms}$-peak located at the height of the far-wall $u$-kink.

In conclusion, under transcritical conditions, we observe localised flow reversal beneath the billow roll-up at $Re=802$, along with the formation of a shear layer just above it. This behaviour is linked to the sharp near-wall increase in $c_\mathit{p}$, which amplifies near-wall $u$-disturbances and destabilises the local mean flow. In other words, this effect is tied to the upward and downward displacements of the Widom line, which correspond to decreases and increases in reduced pressure $p_\mathit{r}$, respectively. As $p_\mathit{r}$ decreases, steep gradients in thermodynamic properties intensify. Ultimately, the initial assumption of a ZPG laminar boundary layer breaks down. When the billow rolls up, a net $\mathrm{d}p/\mathrm{d}x<0$ (APG) forms. Simultaneously, the mean-flow deformation is insufficient to counteract the large near-wall (Mode-II) $u$-fluctuations, leading to a travelling region of flow reversal.
While \citet{Taylor1} attributed transition in incompressible flows to unsteady pressure gradients caused by `external' disturbances leading to separation (see note by \citet{Kloker3}), similar localised separation zones -- such as those in figure~\ref{fig_amp2D_T09w110_total}(d) -- were found under strong APG in the K-type breakdown of an ideal-gas boundary layer \citep{Kloker1,Kloker2} and in the laminar-to-turbulent transition experiments of \citet{Kosorygin1}. In the current study, however, the disturbance is `internal', originating from the displacement of the Widom line, and is absent under weakly non-ideal gas conditions at the same forcing amplitude.

\section{Three-dimensional breakdown to turbulence}\label{sec:3D}
Building on the 2-D nonlinear analysis of the Mode-II instability, we perform 3-D DNS to assess the complete breakdown to turbulence. A $z$-symmetric oblique wave with spanwise wavenumber $\beta_0$ and half the frequency of the primary wave is introduced alongside to the 2-D wave. The oblique wave is imposed with either infinitesimal (`IA') or finite amplitude (`LA'). The 2-D wave amplitude, $A_{\text{2-D}}=7.5 \times 10^{-3}$, is consistent with \S\,\ref{sec:nonlinear_regime}. In the subcritical regime, case Tw095-IA is reported in \S\,\ref{sec:modal_analysis} for comparison but ultimately excluded from further analysis, as it does not trigger transition.

\subsection{Modal analysis}
\label{sec:modal_analysis}
\begin{figure}
\centering
\includegraphics[angle=-0,trim=0 0 0 0, clip,width=1.0\textwidth]{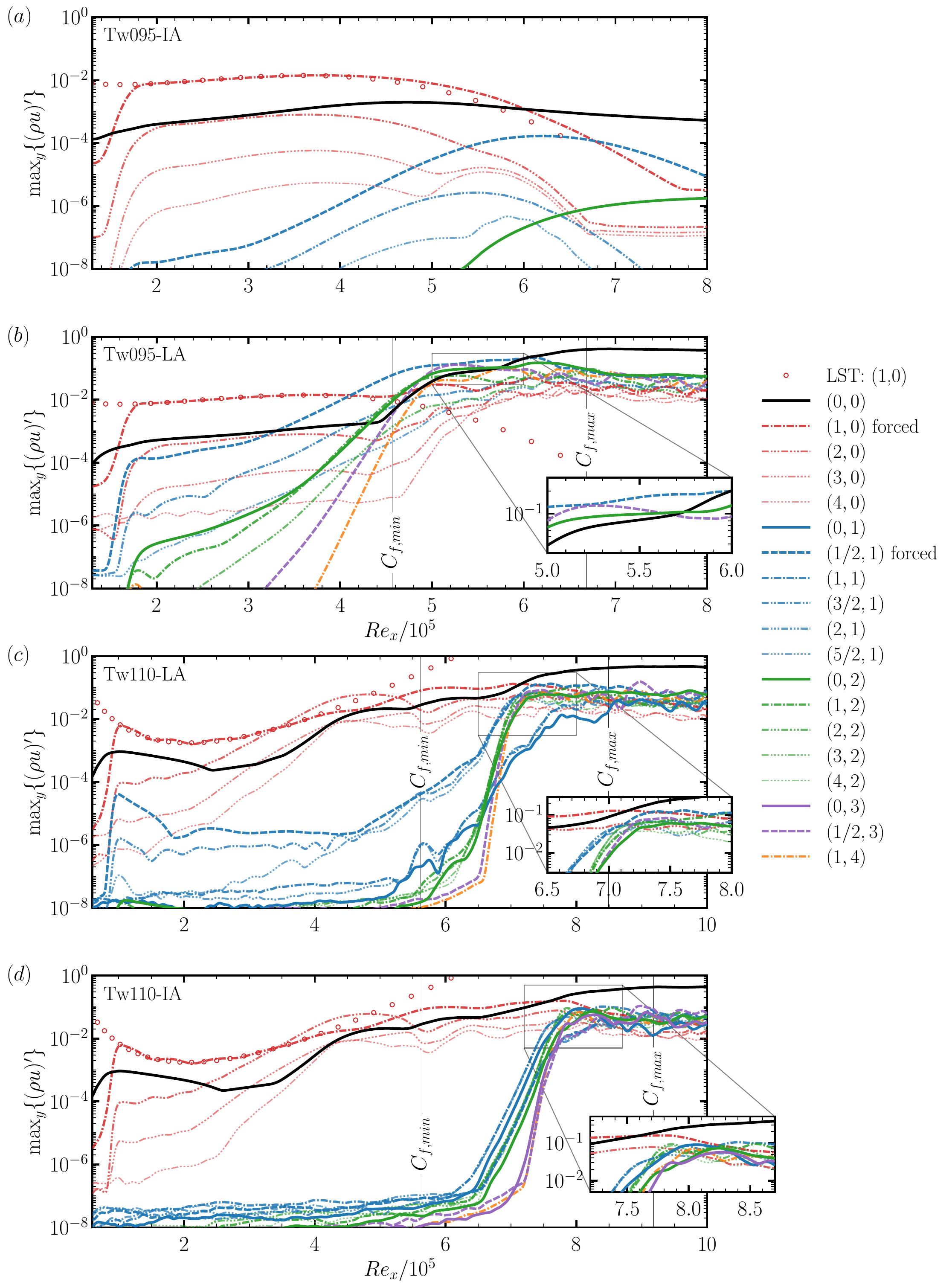}
\captionsetup{justification=justified}  
\caption{\label{fig_Tw_fft}Streamwise evolution of the $y$-maximum of $(\rho u)^\prime$ for the most relevant modes $( \omega / \omega_{\text{2-D}}, \beta / \beta_0)$ for case: (a) Tw095-IA, (b) Tw095-LA, (c) Tw110-LA, and (d) Tw110-IA. The minimum and maximum values of the time- and spanwise-averaged skin-friction coefficient are indicated as $C_\mathit{f,min}=\min\{C_\mathit{f}\}$ and $C_\mathit{f,max}=\max\{C_\mathit{f}\}$, respectively. Insets in (c) and (d) highlight the relevant modes in the breakdown region. Note the different $Re_\mathit{x}$-axis limits between the subcritical (a,b) and (c,d) transcritical cases.}
\end{figure}

Figure~\ref{fig_Tw_fft} presents the relevant modes for the subcritical cases with infinitesimal and finite amplitude, Tw095-IA and Tw095-LA, and the transcritical cases with infinitesimal and finite amplitude, Tw110-LA and Tw110-IA. All share the same fundamental frequency and spanwise wavenumber; the `LA'-cases in figures~\ref{fig_Tw_fft}(b,c) use the same 2-D and 3-D forcing amplitudes, while the `IA'-cases in figures~\ref{fig_Tw_fft}(a,d) have an infinitesimal 3-D forcing amplitude of $A_{\text{3-D}}=1.33 \times 10^{-6} A_{\text{2-D}}$ (see table~\ref{tab:numerical_setup2}). 

In case Tw095-IA (figure~\ref{fig_Tw_fft}a), no transition occurs. Although the oblique mode $(1/2,1)$ undergoes secondary subharmonic growth starting at $Re_\mathit{x}/10^5 \approx 3$, its amplitude remains insufficient to trigger nonlinear interactions. The primary 2-D mode $(1,0)$ decays in line with figure~\ref{fig_amp2D_T09w095}(a), while mode $(0,2)$ continues to grow due to transient growth \citep{Boldini1}.

In contrast, case Tw095-LA (figure~\ref{fig_Tw_fft}b) transitions to turbulence via the subharmonic instability, which rapidly induces three-dimensionality. Its modal evolution closely resembles that of case TadIG (not shown). At the disturbance strip, higher harmonics such as $(2,0)$ and $(3/2,1)$ are also triggered but remain only moderately amplified. When the primary wave $(1,0)$ reaches $\sim\!1\%$ at $\Rey_\mathit{x}/10^5 \approx 2.9$, subharmonic resonance sets in and the forced subharmonic oblique mode $(1/2,1)$ -- absent in the 2-D DNS set-up in figure~\ref{fig_amp2D_T09w095}(a) -- grows rapidly, surpassing $(1,0)$ near $\Rey_\mathit{x}/10^5 \approx 4$. Compared to TadIG, this process is delayed due to the slower destabilisation of $(1,0)$ as the wall temperature is increased toward the Widom line. During the secondary-instability growth, higher modes experience strong nonlinear amplification, contributing to the increase of the mean-flow distortion $(0,0)$ near $C_\mathit{f,min}=\min\{C_\mathit{f}\}$. Later, all modes saturate except for $(1/2,3)$, which matches the amplitude of $(1/2,1)$ at $Re_\mathit{x}/10^5 \approx 5.3$. Hereafter, 
a wave-vortex triad typical of the oblique breakdown \citep{Chang1} forms between the strongly nonlinear $(1/2,3)$ and steady vortex mode $(0,2)$, further destabilising $(1/2,1)$ (see inset of figure~\ref{fig_Tw_fft}b). Further downstream where nonlinear saturation sets in, mode $(0, 0)$ reaches its maximum, marking the onset of turbulence at $C_\mathit{f,max}=\max\{C_\mathit{f}\}$.

For case Tw110-LA in figure~\ref{fig_Tw_fft}(c), although the evolution of the fundamental wave and its higher harmonics closely follows the 2-D development in figure~\ref{fig_amp2D_T09w110}(b) up to $Re_\mathit{x}/10^5 \approx 6.5$, the subharmonic resonance is significantly delayed due to the strong 2-D nonlinearity of the higher harmonics. The oblique mode $(1/2,1)$ is initially strongly damped between $Re_\mathit{x}/10^5 \approx 1.0$ and $2.0$, and maintains a low amplitude level, along with $(3/2,1)$, up to $Re_\mathit{x}/10^5 \approx 4.7$. Before mode $(1,0)$ undergoes subharmonic resonance with $(2,0)$ (see \S\,\ref{sec:nonlinear_regime}), it saturates and a phase-speed locking process, cf.~\citet{Hader1}, occurs between mode $(1,0)$ and its secondary wave $(1/2,1)$, triggering the growth of $(1/2,1)$ via the secondary instability mechanism \citep{Herbert1}. Mode $(1/2,1)$ and its higher harmonics grow to nonlinear amplitude levels and saturate at $Re_\mathit{x}/10^5 \approx 6.0$. In contrary, modes with spanwise wavenumber parameter $ \beta / \beta_0=1$ briefly grow near $C_\mathit{f,min}$ due to nonlinear interaction but do not contribute to the later breakdown stage. When the amplitudes of subharmonic modes ($\omega / \omega_0=1/2$) approach those of the primary wave and higher harmonics, additional higher modes with $ \beta / \beta_0 \geq 2$ are nonlinearly amplified, rapidly growing downstream and enhancing both the destabilisation of $(1,0)$ and the mean-flow distortion. In the early breakdown stage, following the saturation of mode $(1,0)$ at $Re_\mathit{x}/10^5 \approx 7.1$, mode $(1/2,1)$ grows nonlinearly, surpassing mode at $(1,0)$ at $Re_\mathit{x}/10^5 \approx 7.35$, while the mean-flow deformation $(0,0)$ exceeds the $20\%$-amplitude threshold. In the late transitional stage, before $C_\mathit{f,max}$ at $Re_\mathit{x}/10^5 \approx 8.5$, mode $(1/2,1)$ remains dominant, eventually followed by the abrupt growth of $(1/2,3)$ at $Re_\mathit{x}/10^5 \approx 9.0$.

Figure~\ref{fig_Tw_fft}(d) illustrates the modal evolution of case Tw110 with infinitesimal 3-D forcing (`IA'). As expected, up to $\min\{C_\mathit{f}\}$, the evolution of modes $(h,0)$, with $h \geq 0$, matches that of case Tw110-LA in figure~\ref{fig_Tw_fft}(c). No subharmonic resonance of $(1/2,1)$ occurs further downstream in this case, as no phase-speed locking mechanism between the primary wave and its subharmonic is present. In contrast, fundamental resonance with $(1,1)$ sets in at $Re_\mathit{x}/10^5 \approx 6.2$, along with the amplification of 3-D harmonic modes with spanwise wavenumber parameter $\beta/\beta_0 = 2$, before mode $(1/2,1)$ grows nonlinearly only from $Re_\mathit{x}/10^5 \approx 6.5$ onward. Around $Re_\mathit{x}/10^5 \approx 7.2$, the spectrum rapidly fills up with all other nonlinearly amplified 3-D modes, contributing to a significant increase in the mean-flow deformation $(0,0)$. Note that mode $(1,1)$ arises initially due to numerical background noise, and the forcing of $(1/2,1)$ remains likewise at the noise level; therefore, both are insignificant and not representative of classical boundary-layer modes in the early nonlinear stage. Further downstream around $Re_\mathit{x}/10^5 \approx 6.0$, both mode $(1,1)$ and $(0,1)$ (forming a wave-vortex triad with $(1,0)$) begin to behave as discrete boundary-layer wave modes, whereas $(1/2,1)$ continues to be a continuous, non-resonant mode. Under such numerical background-noise conditions and in the absence of primary instability of the investigated 3-D modes, the fundamental resonance mechanism proves to be more robust than the subharmonic one. Eventually, the steady mode $(0,1)$, which dominates the transitional stage in the K-type breakdown \citep{Boldini3}, reaches the amplitude level of $(1,0)$. In the final breakdown stage, $(1,1)$ and $(1/2,3)$ alternate as the dominant modes before $C_\mathit{f,max}$ is reached. Compared to Tw110-LA, the transition is more gradual, but follows a K-type breakdown scenario. Notably, despite the infinitesimal 3-D forcing, breakdown to turbulence is triggered solely by the 2-D fundamental wave and numerical background noise -- a behaviour not observed in case Tw095-IA under weakly non-ideal gas conditions.

\subsection{Flow structures} \label{sec:flow_structures}
\begin{figure}
\centering
\includegraphics[angle=-0,trim=0 0 0 0, clip,width=1.0\textwidth]{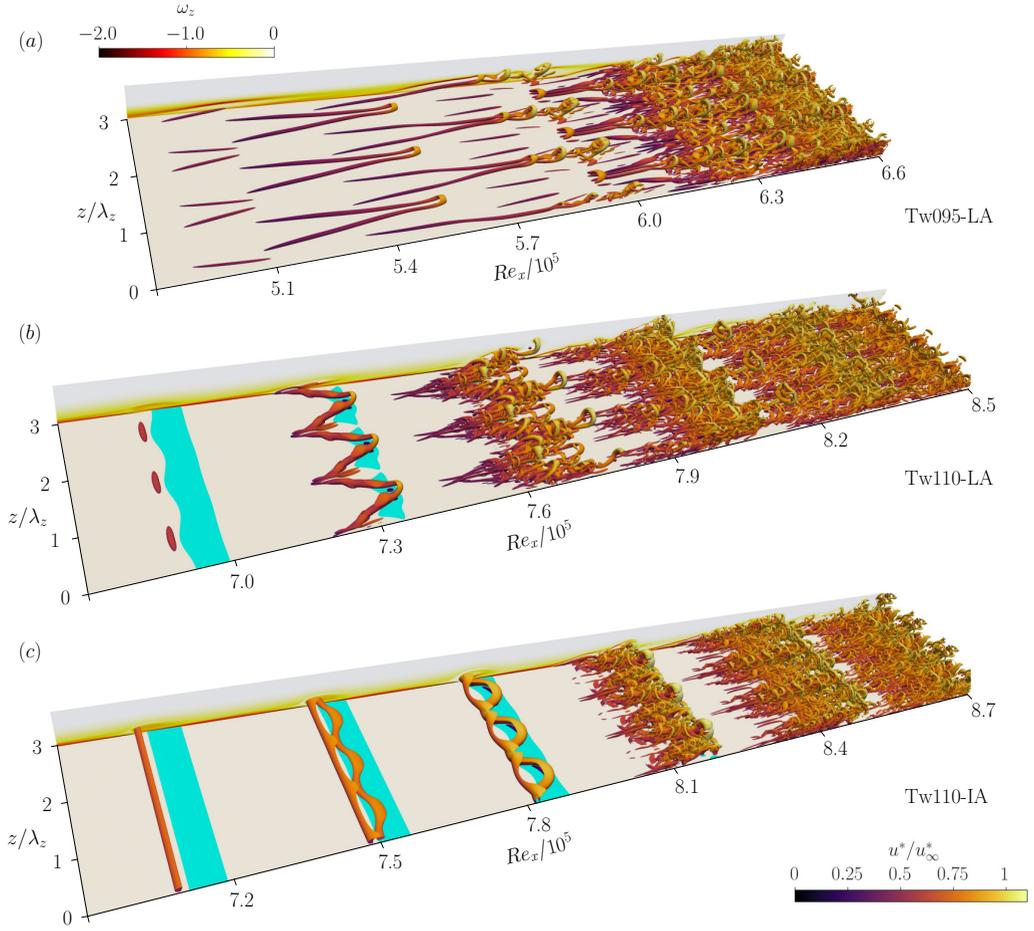}
\captionsetup{justification=justified}  
\caption{\label{fig_Tw_qvort}Instantaneous isosurfaces of the $Q$-criterion, coloured by the streamwise velocity magnitude: (a) case Tw095-LA ($Q = 0.015$) at $t/T_0=0$, (b) case Tw110-LA ($Q = 0.020$) at $t/T_0=0.5$, and (c) case Tw110-IA ($Q = 0.020$) at $t/T_0=0.5$. $T_0$ is the period of the fundamental wave. The side $xy$-plane shows the instantaneous spanwise vorticity $\omega_\mathit{z}$. Isosurfaces of the separation zones, i.e. regions with $u< 0$, are coloured in cyan. For better visualisation, the domain is copied three times in the spanwise direction. Supplementary movies are available at \url{https://github.com/pcboldini/DNSvisualisation}.}
\end{figure}

To investigate the mechanisms driving the breakdown in both thermodynamic regimes, figure~\ref{fig_Tw_qvort} displays the streamwise evolution of the relevant transitional flow structures using isocontours of the $Q$-criterion.

\begin{figure}
\centering
\includegraphics[angle=-0,trim=0 0 0 0, clip,width=1.0\textwidth]{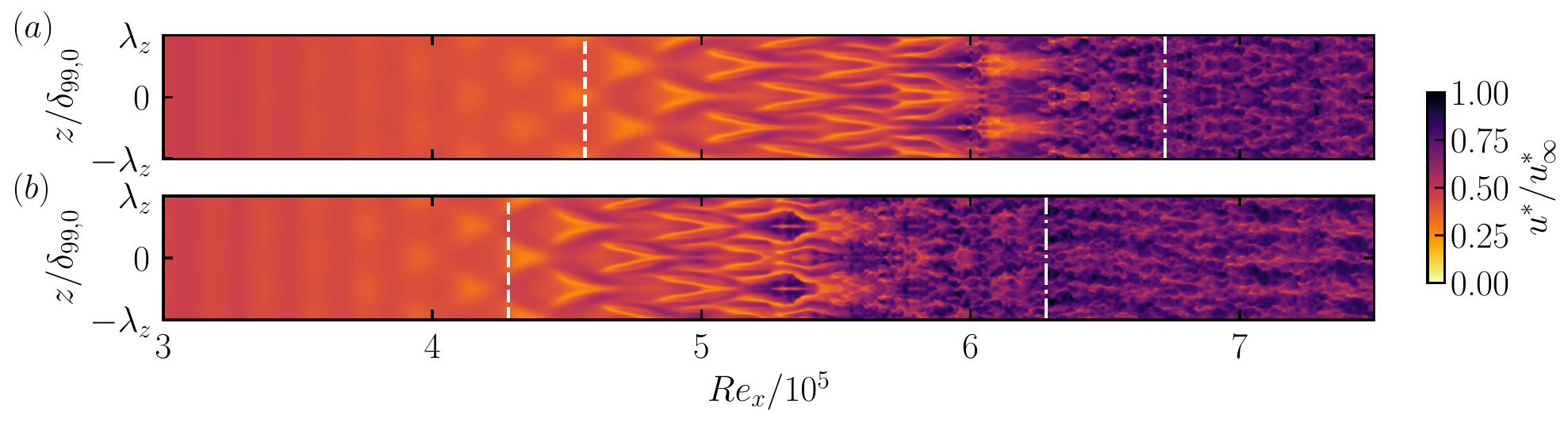}
\captionsetup{justification=justified}  
\caption{\label{fig_Tw095_ucont}Contours of instantaneous streamwise velocity ($xz$-plane at $y/\delta_{99,0}=0.49$):~(a) Tw095-LA and (b) TadIG. The dashed and dash-dotted white vertical lines represent the location of $C_\mathit{f,min}=\min\{C_\mathit{f}\}$ and $C_\mathit{f,max}=\max\{C_\mathit{f}\}$, respectively. For better visualisation, the domain is copied once in the spanwise direction.}
\end{figure}

The subcritical case Tw095-LA (figure~\ref{fig_Tw_qvort}a) exhibits the classical H-type breakdown, similar to the ideal-gas case TadIG, with staggered $\Lambda$-vortices and high-shear layers \citep{Bake1} above the vortices' tips, characterised by peaks in $\omega_\mathit{z}\!\propto\!\partial u/\partial y$. Conversely, no vortices are present at $\lambda_\mathit{z}/2$, i.e. half a spanwise wavelength apart. While the early breakdown stage closely resembles that of the ideal-gas case, the $\Lambda$-vortices are elongated in the streamwise direction from $\Rey_\mathit{x}/10^5 \approx 5.2$ onward, coinciding with the large amplitude of 3-D modes $(3/2,1)$ and $(0,2)$ in figure~\ref{fig_Tw_fft}(b). This elongation of the $\Lambda$-vortices, along with longitudinal structures on their sides, is particularly visible between $5.2 \leq \Rey_\mathit{x}/10^5 \leq 6$ in figure~\ref{fig_Tw095_ucont}(a), which shows a horizontal flow snapshot ($xz$-plane) of the streamwise velocity at $y/\delta_{99,0}=0.49$. In contrast, the staggered $\Lambda$-vortex arrangement in case TadIG (figure~\ref{fig_Tw095_ucont}b) breaks down more rapidly during the transitional stage.

Compared to the subcritical regime, figures \ref{fig_Tw_qvort}(b,c) reveal distinct vortical structures in the transcritical-heating regime. During the early stage of breakdown -- dominated by mode $(1,0)$ -- two-dimensional billow structures with near-wall separation zones (highlighted in cyan) are convected downstream. At later stages, cases Tw110-LA and Tw110-IA exhibit fundamentally different breakdown dynamics. In Tw110-LA, the forced 3-D mode $(1/2,1)$ rapidly grows from $0.01\%$ to $13\%$ between $Re_\mathit{x}/10^5 \approx 6$ and $7.3$, leading to the sudden appearance of staggered $\Lambda$-vortices lifting off from the billow roll-ups. Unlike the elongated staggered alignment observed in Tw095-LA, $\Lambda$-vortices here emerge more abruptly and locally, with secondary structures forming at half a spanwise wavelength apart (see \S\,\ref{sec_Tw110-LA}). By $Re_\mathit{x}/10^5 \approx 7.5$, where mode $(0,0)$ reaches around $30\%$, the first row of hairpin vortices becomes visible, with small-scale structures at their legs. Farther downstream, the complete breakdown of the preceding 2-D billow roll-up becomes evident, characterised by near-wall longitudinal structures between $Re_\mathit{x}/10^5 \approx 7.5$ and $7.7$. 

In contrast, case Tw110-IA exhibits a delayed breakdown of each 2-D billow, as indicated by the extended near-wall separation zones. The onset of three-dimensionality is significantly delayed, consistent with the more gradual growth of 3-D modes in figure~\ref{fig_Tw_fft}(d). $\Lambda$-like vortices emerge within the 2-D billow roll-ups, initially displaying an aligned peak-valley splitting -- characteristic of the fundamental K-type breakdown (see \S\,\ref{sec_Tw110-IA}). The $\Lambda$-vortices exhibit a shorter streamwise than spanwise wavelength, consistent with the large amplitude of mode $(2,1)$ in the early breakdown stage in figure~\ref{fig_Tw_fft}(d). No strong lift-up of hairpin-like vortices is observed; however, near-wall longitudinal structures -- resembling those in Tw110-LA -- gradually emerge. Ultimately, each billow independently breaks into small-scale turbulent structures, eventually merging with the turbulent structures originating from the previous billow. From $Re_\mathit{x}/10^5 \approx 8.0$ onward, as mode $(0,0)$ reaches $\sim\!30\%$, the resulting turbulent flow pattern resembles that of Tw110-LA between $Re_\mathit{x}/10^5 \approx 7.5$ and $7.7$, as shown in figure~\ref{fig_Tw_qvort}(b). 

\subsubsection{Case Tw110-LA: detailed breakdown analysis}
\label{sec_Tw110-LA}
\begin{figure}
\centering
\includegraphics[angle=-0,trim=0 0 0 0, clip,width=1.0\textwidth]{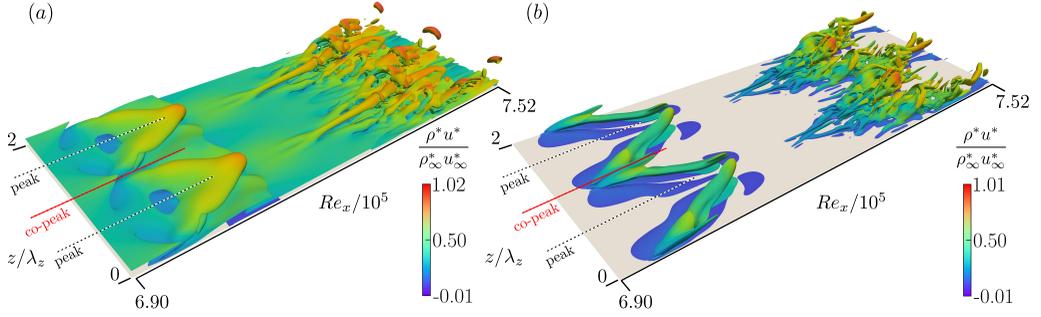}
\captionsetup{justification=justified}  
\caption{\label{fig_Tw110LA_vorticity}Case Tw110-LA. Instantaneous isosurfaces of (a) spanwise vorticity $|\omega_\mathit{z}|=0.45$ and (b) streamwise vorticity $|\omega_\mathit{x}|=0.45$, coloured by the streamwise momentum $\rho u$ magnitude at $t/T_0=0.5$. $T_0$ is the period of the fundamental wave. For better visualisation, the domain is copied once in the spanwise direction.}
\end{figure}

Figure~\ref{fig_Tw110LA_vorticity} shows close-up views of the initial breakdown region in case Tw110-LA ($6.90 \leq Re_\mathit{x}/10^5 \leq 7.52$), with isosurfaces of spanwise $\omega_\mathit{z}$ and streamwise $\omega_\mathit{x}$ vorticity in figures~\ref{fig_Tw110LA_vorticity}(a,b), respectively, at the same time instant as in figure~\ref{fig_Tw_qvort}(b). The visualisation spans two spanwise wavelengths to capture the full structure.
The $\omega_\mathit{z}$-distribution reveals a delta-wing-like shear layer at $z=0.5\lambda_\mathit{z},1.5\lambda_\mathit{z},...$ (dashed black lines), with peak spanwise vorticity at its tip. These structures, featuring round, slightly inclined heads, are aligned above the oppositely-signed $\omega_\mathit{x}$-legs of the $\Lambda$-shaped vortex (figure~\ref{fig_Tw110LA_vorticity}b). At these locations -- referred to hereafter as `peak' locations -- the strongest $u^{\prime}$ and $\rho^{\prime}$ perturbations occur with $u^{\prime}_\mathit{rms}\approx 0.2$ and $\rho^{\prime}_\mathit{rms}\approx 0.18$. This layer is hereafter referred to as the `upper' high-shear layer and is denoted as `UL'. Half a spanwise length apart, at $z=0,\lambda_\mathit{z},...$ (dashed red line), $u^{\prime}$ and $\rho^{\prime}$ -- unlike in case Tw095-LA -- do not decay but exhibit the second-largest $rms$ amplitudes in spanwise direction. This location is thus referred to hereafter as `co-peak', cf.~the ideal-gas APG case \citep{Kloker1,Kloker2,Kosorygin1}. Here, the adjacent $\Lambda$-vortices are closely spaced, and a longitudinal vortical structure appears in between. 

\begin{figure}
\centering
\includegraphics[angle=-0,trim=0 0 0 0, clip,width=1.0\textwidth]{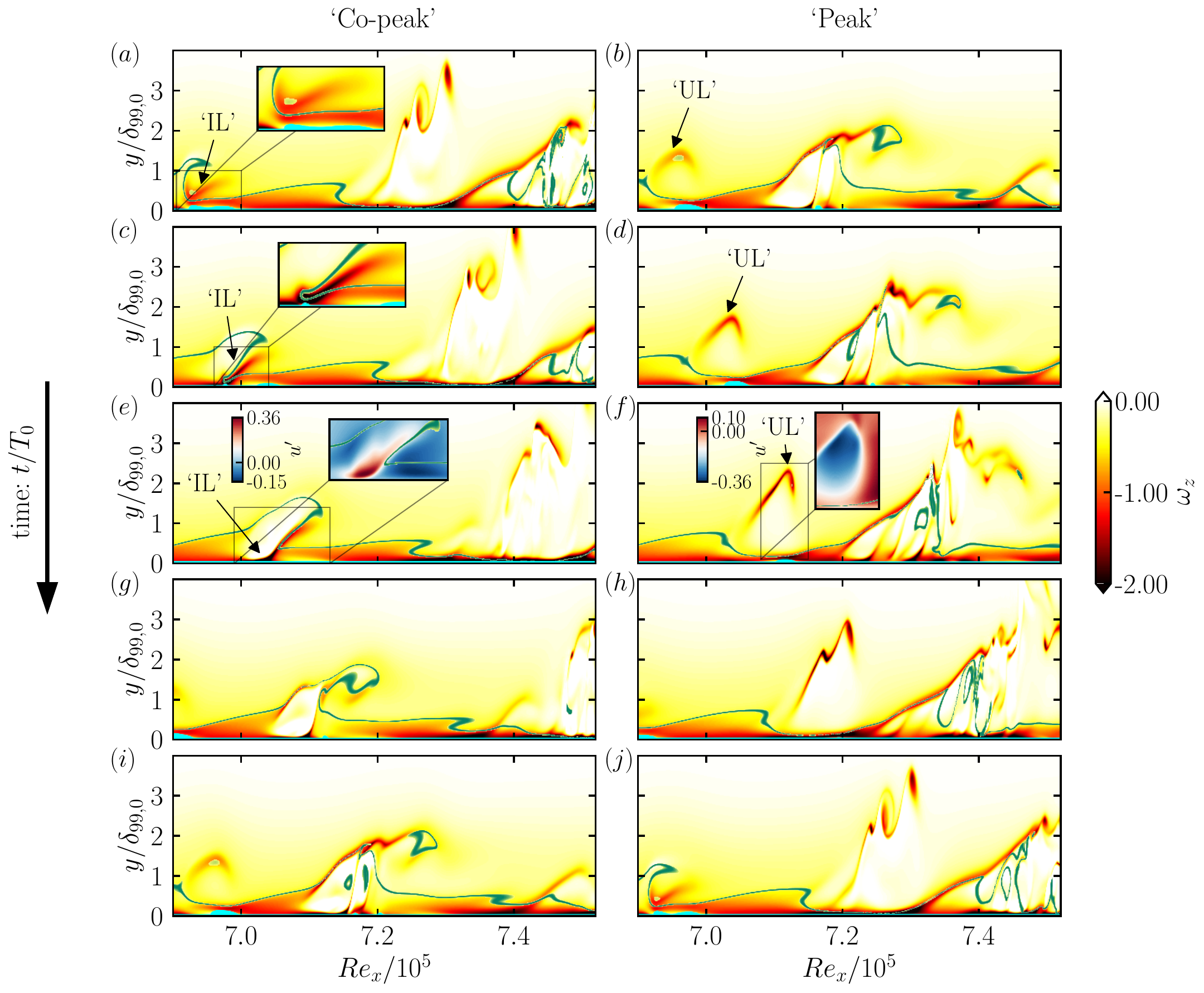}
\captionsetup{justification=justified}  
\caption{\label{fig_Tw110LA_yplanes_vorty}Case Tw110-LA. Instantaneous contours of spanwise vorticity $\omega_\mathit{z}$ in the $Re_\mathit{x}$--$y/\delta_{99,0}$ plane at: (a,b) $t/T_0=0$, (c,d) $t/T_0=0.25$, (e,f) $t/T_0=0.5$, (g,h) $t/T_0=0.75$, and (i,j) $t/T_0=1.0$. $T_0$ is the period of the fundamental wave. The first column (a,c,e,g,i) corresponds to the spanwise `co-peak' location at $z/\lambda_z=0$ (see figure~\ref{fig_Tw110LA_vorticity}), while the second column in (b,d,f,h,j) corresponds to the spanwise `peak' location at $z/\lambda_\mathit{z}=0.5$ (see figure~\ref{fig_Tw110LA_vorticity}). The `upper' and `inverted lower' high-shear layers are labelled as `UL' and `IL', respectively. The near-wall region for which $u<0$ is coloured in cyan. The Widom line $y=y_\mathit{WL}$ lies within the green region, i.e. between $95\% \max\{c_\mathit{p}\}$ and $\max\{c_\mathit{p}\}$. }
\end{figure}

The origin of the secondary vortex system is illustrated in figure~\ref{fig_Tw110LA_yplanes_vorty}, which shows instantaneous contours of $\omega_\mathit{z}$ in a longitudinal planes (between $Re_\mathit{x}/10^5 = 6.90$ and $7.52$) at five time steps ($t/T_0=[0,0.25,0.5,0.75,1.0]$).
At the `co-peak' plane ($z/\lambda_\mathit{z}=0$, figures~\ref{fig_Tw110LA_yplanes_vorty}a,c,e,g,i), a local maximum of $\omega_\mathit{z}$ forms near the wall -- hereafter referred to as the `inverted lower' high-shear layer and labelled as `IL' -- resembling that of the ideal-gas APG case. Its origin is linked to the downstream-convected near-wall separation zone (highlighted in cyan) below the 2-D billow trough, which induces a local high-shear velocity profile (see \S\,\ref{sec:nonlinear_regime} in figure~\ref{fig_amp2D_T09w110_EF}a). This layer is unstable to 3-D perturbations, cf.~stationary laminar separation bubble (LSB) \citet{Maucher1}, due to strong near-wall $u^{\prime}$-disturbances (see inset in figure~\ref{fig_Tw110LA_yplanes_vorty}c), which are transported downstream with the shear layer. Similar to the 2-D investigation (figure~\ref{fig_amp2D_T09w110_EF}c), they correspond to a strong near-wall amplitude of mode $(1,0)$ and its higher harmonics. Mode $(1/2,1)$, which was absent in \S\,\ref{sec:nonlinear_regime}, reaches its peak at $y/\delta_{99,0}\approx 1$ (not shown). Over time, the `inverted lower' layer evolves:~its head is pushed toward the wall, while its leg is lifted upward (figure~\ref{fig_Tw110LA_yplanes_vorty}e). The secondary vortex system shown in figure~\ref{fig_Tw110LA_vorticity}(b) then develops above it -- again reminiscent of the ideal-gas APG case. The two-oppositely signed vortical structures transport low-velocity, low-density fluid upward, as seen in the $\rho u$-contours in figure~\ref{fig_Tw110LA_vorticity}(b). At later times (figures~\ref{fig_Tw110LA_yplanes_vorty}g and \ref{fig_Tw110LA_yplanes_vorty}i), the `inverted lower' layer merges with a second high-shear layer, crossing the entire boundary layer. 

At $t/T_0=1$, the flow structures at the `co-peak' location match those at the `peak' location at $t/T_0=0$ in figure~\ref{fig_Tw110LA_yplanes_vorty}(b), indicating the staggered, periodic breakdown pattern of case Tw110-LA. The newly formed shear layer initially consists of two adjacent, streamwise-aligned, counter-rotating vortex filaments. As it becomes increasingly unstable in figures~\ref{fig_Tw110LA_yplanes_vorty}(b,d,f), it rolls up and breaks into small-scale structures (see the secondary vortex system at `peak' locations near $Re_\mathit{x}/10^5 \approx 7.35$ in figure~\ref{fig_Tw110LA_vorticity}). Simultaneously, the `upper' high-shear layer (`UL' in figures~\ref{fig_Tw110LA_yplanes_vorty}b,d,f), also destabilises and breaks up into numerous small vortical structures (figures~\ref{fig_Tw110LA_yplanes_vorty}h and  \ref{fig_Tw110LA_yplanes_vorty}j), generating an isolated hairpin-like vortex, captured by the $\omega_\mathit{z}$-isocontours at $Re_\mathit{x}/10^5 \approx 7.5$ in figure~\ref{fig_Tw110LA_vorticity}(a). Note that the breakdown to turbulence is initiated by the earlier breakup of the `inverted lower' layer at the `co-peak' positions, which precedes the breakup of the `upper' layer at the `peak' positions. This sequence is also evident in figure~\ref{fig_Tw_qvort}(b), where near-wall turbulence triggered by the `inverted lower' layer emerges significantly farther upstream than the small-scale structures in the boundary-layer core, which are caused by the break-up of the `upper' layer.

\subsubsection{Case Tw110-IA: detailed breakdown analysis}
\label{sec_Tw110-IA}
\begin{figure}
\centering
\includegraphics[angle=-0,trim=0 0 0 0, clip,width=1.0\textwidth]{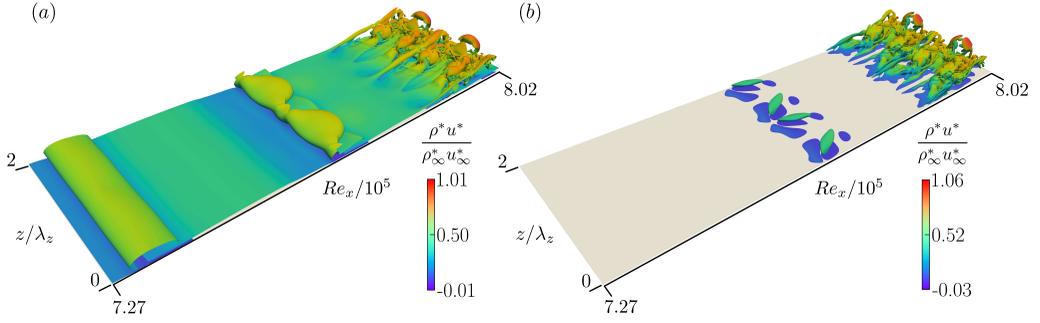}
\captionsetup{justification=justified}  
\caption{\label{fig_Tw110IA_vorticity}Case Tw110-IA. Instantaneous isosurfaces of (a) spanwise vorticity $|\omega_\mathit{z}|=0.45$ and (b) streamwise vorticity $|\omega_\mathit{x}|=0.45$, coloured by the streamwise momentum $\rho u$ magnitude at $t/T_0=0.5$. $T_0$ is the period of the fundamental wave. For better visualisation, the domain is copied once in the spanwise direction.}
\end{figure}

\begin{figure}
\centering
\includegraphics[angle=-0,trim=0 0 0 0, clip,width=1.0\textwidth]{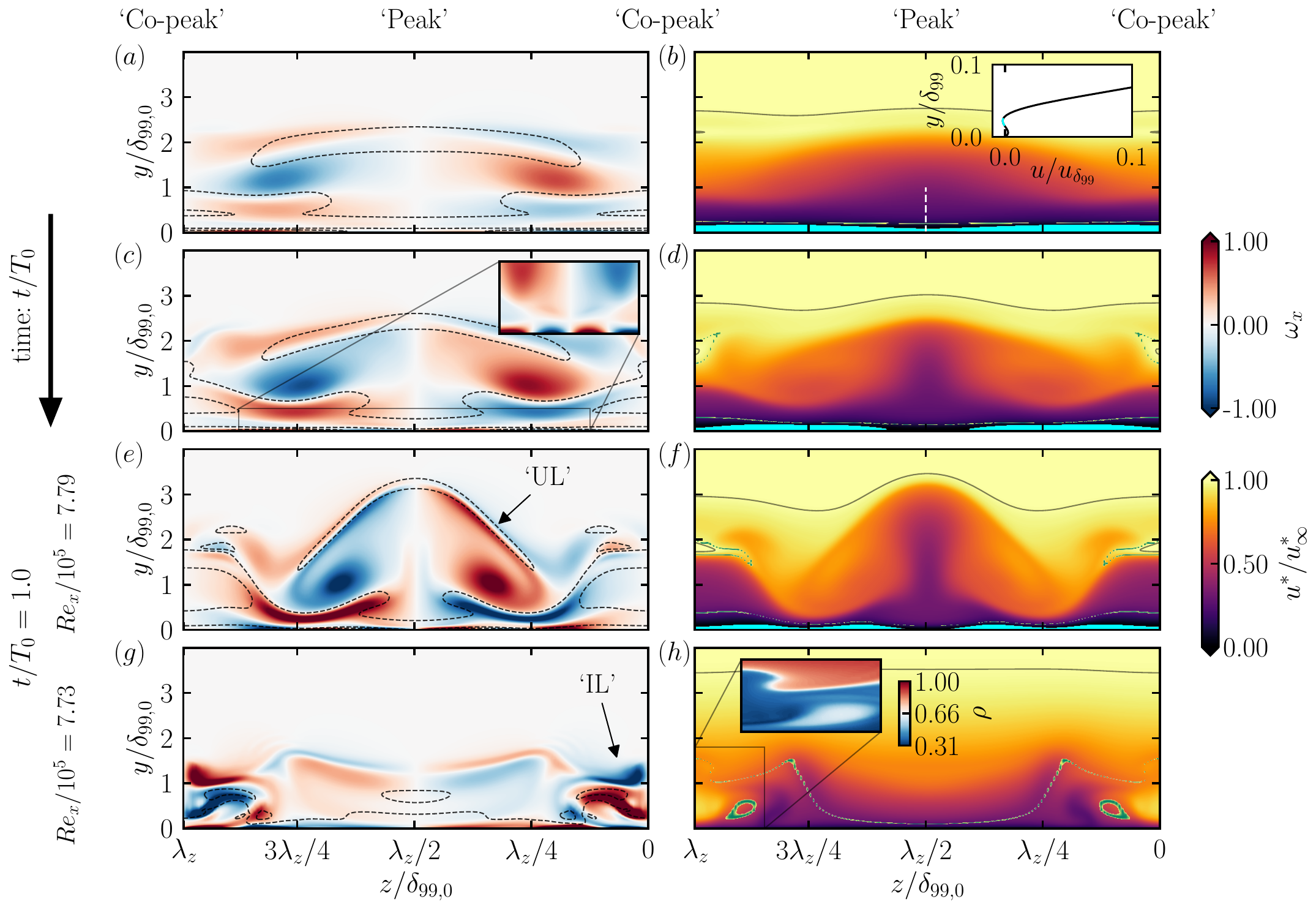}
\captionsetup{justification=justified}  
\caption{\label{fig_Tw110IA_zplanes_vortx}Case Tw110-IA. Instantaneous contours of streamwise vorticity $\omega_\mathit{x}$ and velocity $u^*/u^*_\mathit{\infty}$ in the $z/\delta_{99,0}$-$y/\delta_{99,0}$ plane at (a,b) $t/T_0=0.5$ ($Re_\mathit{x}/10^5 = 7.62$), (c,d) $t/T_0=0.75$ ($Re_\mathit{x}/10^5 = 7.70$), and (e--h) $t/T_0=1.0$. $T_0$ is the period of the fundamental wave. The dashed black line in (a,c,e,g) indicates contours of $|\omega_\mathit{z}|=0.45$, while the black line in (b,d,f,h) corresponds to $\delta_{99}$. The `upper' and `inverted lower' high-shear layers are labelled as `UL' and `IL', respectively. The near-wall region for which $u<0$ is coloured in cyan. The Widom line $y=y_\mathit{WL}$ lies within the green region, i.e. between $95\% \max\{c_\mathit{p}\}$ and $\max\{c_\mathit{p}\}$.}
\end{figure}

Figures~\ref{fig_Tw110IA_vorticity}(a,b) show isocontours of $\omega_\mathit{z}$ and $\omega_\mathit{x}$ for case Tw110-IA, respectively, at the same time step as in figure~\ref{fig_Tw_qvort}(c). The visualisation spans two spanwise wavelengths to capture the full structure.
Unlike case Tw110-LA, mode $(1,0)$ retains a high amplitude over a longer streamwise distance, allowing the 2-D billow roll-ups to grow farther downstream, with intensified billow crests and larger near-wall separation zones. A key difference between Tw110-IA and Tw110-LA lies in the 3-D disturbance evolution during the breakdown. In the former, following the nonlinear amplification of 3-D modes, $\Lambda$-like vortices (see figure~\ref{fig_Tw110IA_vorticity}) form gradually inside each streamwise-convected 2-D billow roll-up. Unlike the abrupt, staggered onset in case Tw110-LA, these vortices emerge naturally in an aligned formation without the external forcing of mode $(0,1)$ or $(1,\pm1)$, cf.~\citet{Rist1}. This development inside the single billow roll-up is highlighted in figure~\ref{fig_Tw110IA_zplanes_vortx} between $t/T_0=0.5$ and $t/T_0=1.0$ ($7.42 \leq Re_\mathit{x}/10^5 \leq 7.79$).
As the billow roll-up travels downstream, the separation zone at the spanwise `peak' location weakens (see inset of figure~\ref{fig_Tw110IA_zplanes_vortx}b) due to the recirculation of high-velocity fluid toward the wall by the $\Lambda$-vortex legs. The rapid growth of the 3-D modes around $Re_\mathit{x}/10^5 \approx 7.6$ intensifies the `upper' layers at the `peak' locations ($z=0.5\lambda_\mathit{z},1.5\lambda_\mathit{z},...$), which emerge from the billow crest. Meanwhile, the near-wall separation zones generate small, oppositely-signed longitudinal vortices (inset of figure~\ref{fig_Tw110IA_zplanes_vortx}c). At $t/T_0=1$, the spanwise `peak' features an `upper' high-shear $\Lambda$-shaped layer (see $|\omega_\mathit{z}|=\mathrm{const.}$-contours in figure~\ref{fig_Tw110IA_zplanes_vortx}e), which enhances the spanwise flow modulation, while its legs are pushed closer to the wall, giving rise to regions of high $\omega_\mathit{z}$ and additional secondary vortical structures, cf.~\citet{Rist1}. This `upper' layer is analogous to that of case Tw110-LA in figure~\ref{fig_Tw110LA_yplanes_vorty}. Upstream, the `inverted lower' layer develops inside the boundary layer in a manner analogous to that observed in case Tw110-LA, transforming the initial `peak'-`valley' splitting into a `peak'-`co-peak' breakdown scenario. This high-shear layer originates at the trough of the billow roll-up and generates strong secondary low-density co-rotating vortices at the intermediate spanwise `co-peak' locations (figures~\ref{fig_Tw110IA_zplanes_vortx}g,h), which travel faster than the flow in the near-wall region of the spanwise `peak'. These longitudinal structures, visible in figure~\ref{fig_Tw110IA_vorticity} around $Re_\mathit{x}/10^5 \approx 7.8$, break up earlier than the far-wall hairpin-like vortices and contribute to the formation of turbulent structures also at the spanwise `co-peak' locations.

\setcounter{equation}{1}

\subsection{Integral quantities} \label{sec:quantities}
\begin{figure}
\centering
\includegraphics[angle=-0,trim=0 0 0 0, clip,width=1.0\textwidth]{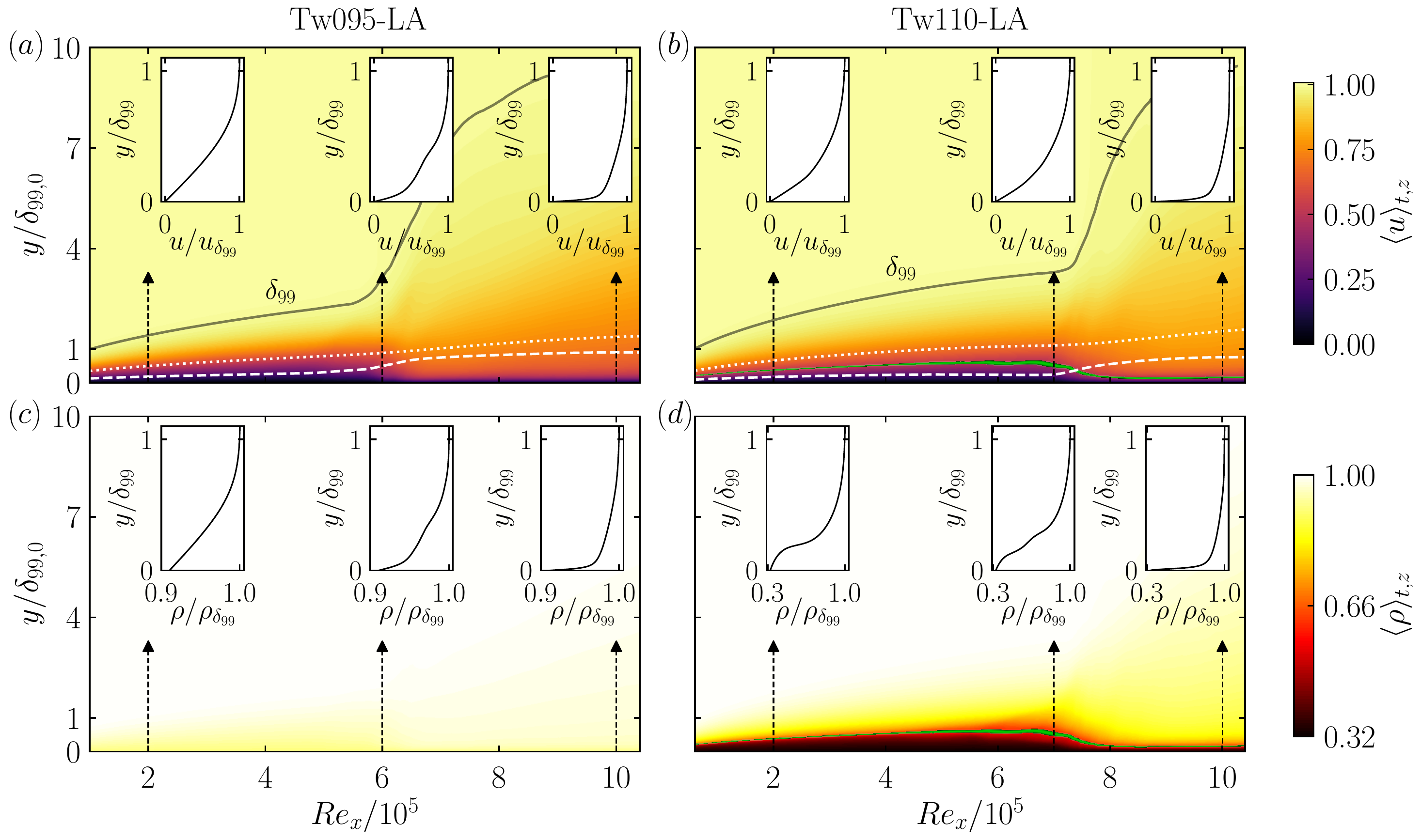}
\captionsetup{justification=justified}  
\caption{\label{fig_qavg}Contours of the time- and spanwise-averaged (a,b) streamwise velocity $\langle u\rangle_{t,z}$ and (c,d) density $\langle \rho\rangle_{t,z}$ for (a,c) case Tw095-LA and (b,d) case Tw110-LA. In (a,b), the displacement thickness $\delta_{1}$ and momentum thickness $\theta$ are indicated by a white dotted and dashed line, respectively. Insets in (a,b) show selected streamwise velocity profiles normalised by the corresponding $u(\delta_{99})=0.99$. The insets in (c,d) depict selected density profiles normalised by the corresponding $\rho(\delta_{99})$. In (b) and (d), the Widom line $y=y_\mathit{WL}$ lies within the green-shaded region, i.e. between $95\% \max\{c_\mathit{p}\}$ and $\max\{c_\mathit{p}\}$.}
\end{figure}

To illustrate the transition process, figure~\ref{fig_qavg} presents isocontours of time- and spanwise-averaged streamwise velocity and density. The velocity profiles in figures~\ref{fig_qavg}(a,b) illustrate the transition to turbulence for cases Tw095-LA and Tw110-LA, respectively. In the latter, the near-wall separation zones are absent, as they are convected downstream and disappear on average. In the turbulent regime, the velocity and density profiles of both cases become qualitatively similar, regardless of whether the fluid is weakly or strongly non-ideal. In the strongly non-ideal case Tw110-LA, after transition to turbulence, the Widom line (coloured in green in figure~\ref{fig_qavg}d) shifts significantly closer to the wall, considerably reducing the height of the vapour-like region compared with the laminar regime. The implications of this shift on the turbulent boundary layer are discussed in \S\,\ref{sec:turbulent}.

\begin{figure}
\centering
\includegraphics[angle=-0,trim=0 0 0 0, clip,width=1.0\textwidth]{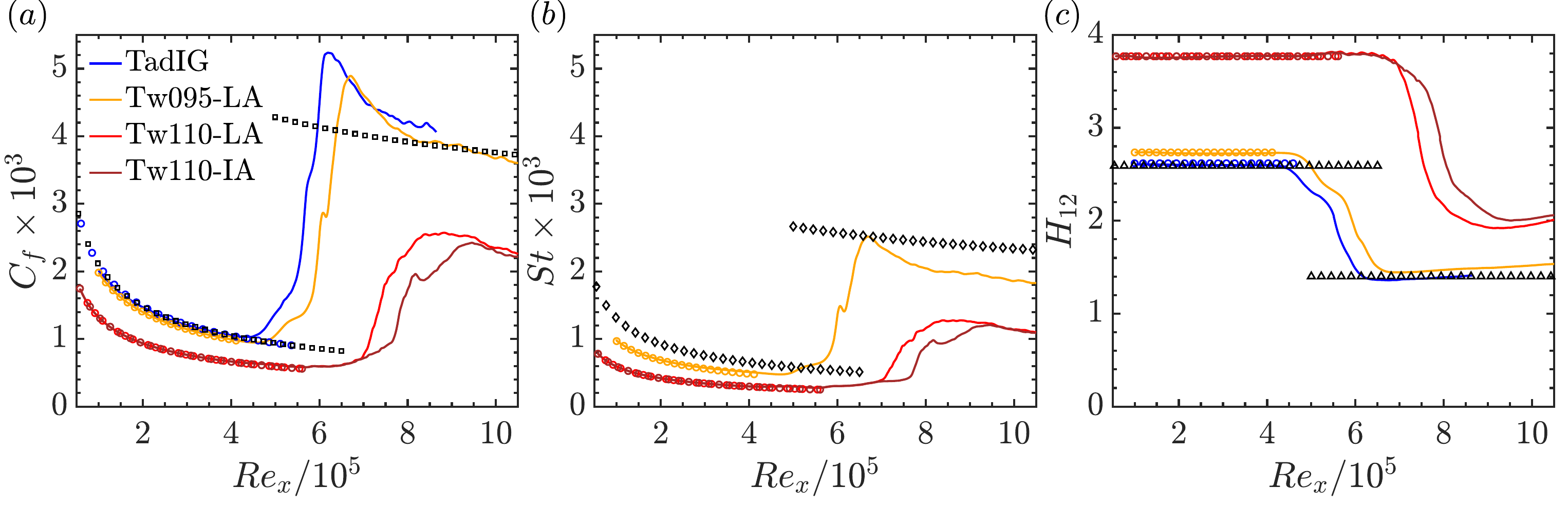}
\captionsetup{justification=justified}  
\caption{\label{fig_Cf_St}Time- and spanwise-averaged: (a) skin-friction coefficient $C_\mathit{f}$, (b) Stanton number $St$, and (c) shape factor $H_{12}$ as functions of $Re_\mathit{x}$. Solid lines denote DNS results, while circle (\textcolor{black}{$\circ$}) symbols represent the self-similar laminar correlations from \eqref{eq:cfSt_lam} with initial conditions as in \S\,\ref{sec:initial_conditions}. In (a,b), the theoretical incompressible skin-friction coefficient and the theoretical incompressible Stanton number using the Reynolds analogy $St=0.5C_fPr_\infty^{-2/3}$ are represented  by square (\textcolor{mycolor10}{$\smallsquare$}) and diamond (\textcolor{mycolor10}{$\smalldiamond$}) symbols, respectively \citep{White1}. Note that for TadIG, $St=0$ due to adiabatic wall conditions. In (c), the theoretical incompressible shape factor is indicated with triangle (\textcolor{mycolor10}{$\smalltriangleup$}) symbols \citep{White1}. }
\end{figure}

To quantitatively characterise the transitional boundary layers following the qualitative analysis in \S\,\ref{sec:flow_structures}, the streamwise evolution of the time- and spanwise-averaged skin-friction coefficient $C_\mathit{f}$ and the Stanton number $St$ is examined. These quantities are defined in dimensionless form as
\begin{equation}
C_\mathit{f}=\dfrac{\overline{\tau}^*_\mathit{w}}{0.5\rho^*_\infty u^{*2}_\infty}=\dfrac{2\overline{\tau}_\mathit{w}}{Re}, \quad     St=\dfrac{\overline{q}^*_\mathit{w}}{\rho^*_\infty u^*_\infty (h^*_\mathit{aw}-h^*_\mathit{w})}=\dfrac{\overline{q}_\mathit{w}}{Re Pr_\infty Ec_\infty (h_\mathit{aw}-h_\mathit{w})},  \tag{\theequation$a{,}b$}
    \label{eq:cfSt}
\end{equation}
where the non-dimensional characteristic parameters are defined in \eqref{eq:nondimnumbers}. The adiabatic wall enthalpy, $h^*_\mathit{aw}=h^*_\infty+ru^{*2}_\infty/2$, where the recovery factor $r=(h_\mathit{aw}-h_\infty)/(h_0-h_\infty)$ ($h_0$ is the total enthalpy) is given as $Pr^{1/3}_\infty=1$ \citep{White1}, can be expressed in reduced form as $h_\mathit{r,aw}=h_\mathit{r,\infty}+rM^2_\mathit{\infty} a^2_\mathit{r,\infty}/2$, where the free-stream reduced enthalpy is defined as $h_\mathit{r,\infty}=e_\mathit{r,\infty}+p_\mathit{r,\infty}/\rho_\mathit{r,\infty}$. It is important to note that the assumption $r \approx 1$ has been verified in the laminar boundary layer under transcritical conditions (not shown). Figures~\ref{fig_Cf_St}(a,b) show the streamwise evolution of $C_\mathit{f}$ and $St$, respectively, for all cases in table~\ref{tab:numerical_setup2}. The theoretical laminar curves, $C_\mathit{f,lam.}$ and $St_\mathit{lam.}$, for a non-ideal fluid are derived from \eqref{eq:cfSt} using the self-similar Lees-Dorodnitsyn variables as
\begin{subequations}
\begin{gather}
C_\mathit{f,lam.}=\dfrac{\sqrt{2}C_\mathit{w}}{Re}\dfrac{\partial u}{\partial \eta}\bigg|_\mathit{w}, \quad St_\mathit{lam.}=\dfrac{1}{\sqrt{2} Re Ec_\infty (h_\mathit{aw}-h_\mathit{w})} \dfrac{C_\mathit{w} \, c_\mathit{p,r,w}}{Pr_\mathit{w} \, c_\mathit{p,r,\infty}}\dfrac{\partial T}{\partial \eta}\bigg|_\mathit{w}
    \label{eq:cfSt_lam}, \tag{\theequation$a{,}b$}
\end{gather}
\end{subequations}
where $\partial u/\partial \eta|_\mathit{w}$ and $\partial T/\partial \eta|_\mathit{w}$ are the wall-normal gradients of the laminar streamwise velocity and temperature in figure~\ref{fig_bl_laminar}, respectively, and $C_\mathit{w}=\rho_\mathit{w}\mu_\mathit{w}$ is the Chapman-Rubesin parameter at the wall. In contrast, turbulent correlations for non-ideal boundary-layer flows are not yet available. An accurate estimation is presented in \S\,\ref{sec:turbulent}.

The skin-friction and Stanton-number curves initially follow the laminar trend in all cases. In Tw095-LA, both $\partial u/\partial \eta|_\mathit{w}$ and $C_\mathit{w}$ in \eqref{eq:cfSt_lam} resemble the ideal-gas case, resulting in a $C_\mathit{f,lam.}$ distribution that closely matches the ideal-gas case. In contrast, for both Tw110-LA and Tw110-IA, despite a fuller laminar velocity profile, $C_\mathit{w}$ is significantly below unity due to density and viscosity stratification. Thus, $C^{\text{Tw110}}_\mathit{f,lam.}<C^{\text{Tw095}}_\mathit{f,lam.}$. A similar trend holds for $St$ in figure~\ref{fig_Cf_St}(b). In Tw095-LA, $St$ closely matches the incompressible limit, i.e. $St=0.332Pr^{-2/3}/Re$. In Tw110-LA and Tw110-IA, the enthalpy difference $(h_\mathit{aw}-h_\mathit{w})$ is significantly larger, yielding $St^{\text{Tw110}}_\mathit{lam.}<St^{\text{Tw095}}_\mathit{lam.}$. 

In Tw095-LA, transition begins as mode $(1/2,1)$ overtakes $(1,0)$ and its amplitude exceeds $1\%$. Between $\Rey_\mathit{x}/10^5 \approx 5.3$ and $6.0$ coinciding with the saturation of $(1/2,1)$ and other 3-D modes, both $C_\mathit{f}$ and $St$ level off as elongated $\Lambda$-vortices become the predominant flow structures. At $\Rey_\mathit{x}/10^5 \approx 6.0$, a sharp rise in $C_\mathit{f}$ and $St$ is observed, as the amplification of modes $(1/2,1)$, $(0,2)$, and higher 3-D modes leads to strong mean-flow deformation and the roll-up of the $\Lambda$-vortices into hairpin vortices that evolve into ring-like structures. Consequently, both $C_\mathit{f}$ and $St$ overshoot, consistent with the ideal-gas case TadIG. 

In Tw110-LA and Tw110-IA, transition is delayed. The $C_\mathit{f}$ and $St$ curves begin to deviate from the laminar predictions around $Re_\mathit{x}/10^5\approx 5.7$, following the subharmonic resonance of modes $(1,0)$ and $(2,0)$ and the appearance of near-wall separation zones. However, these separation zones are convected downstream such that the average $C_\mathit{f}$ remains positive - in contrast to classic LSB, where a negative $C_\mathit{f}$ is observed due to a larger, steady separation zone \citep{Alam1}. The $C_\mathit{f}$ and $St$ curves for cases Tw110-LA and Tw110-IA remain identical up to $Re_\mathit{x}/10^5\approx 6.9$, beyond which Tw110-LA exhibits a rapid rise in $C_\mathit{f}$ due to the abrupt amplification of 3-D disturbances. In contrast, Tw110-IA shows a sharp increase in both $C_\mathit{f}$ and $St$ only at $Re_\mathit{x}/10^5\approx 7.65$, followed by a kink in the evolutions around $Re_\mathit{x}/10^5\approx 8.2$, associated with the strong saturation of the steady modes $(0,1)$ and $(0,2)$. Qualitatively, the minor decrease in $C_\mathit{f}$ and $St$ corresponds to the region between the break-up of two spanwise rollers, where no prominent vortical structures are observed (see figure~\ref{fig_Tw_qvort}c between $Re_\mathit{x}/10^5\approx 8.0$ and $8.1$). Further downstream, the distributions of both transcritical cases gradually level off without a pronounced overshoot. Although mode $(0,0)$, representing the wall-normal amplitude maximum, is about as large as in the subcritical case, the longitudinal vortex mode $(0,2)$ is significantly smaller, and no distinct overshoot appears. Accordingly, the reduction in $C_\mathit{f}$ and $St$ is more pronounced in the turbulent regime than in the laminar region, if compared to the subcritical case. The contribution of $\overline{\mu^{\prime}\partial u^{\prime}/\partial y|_\mathit{w}}$ to the wall-shear stress $\overline{\tau}_\mathit{w}$ is found to be negligible. Despite different initial forcings, the $C_\mathit{f}$ and $St$ curves of Tw110-LA and Tw110-IA converge as expected toward the same turbulent values.
 
The streamwise development of the shape factor $H_{12}=\delta_{1}/\theta$, where $\delta_{1}=\int_0^{\infty} (1 - \rho u) \mathrm{d}y$ is the displacement thickness and $\theta=\int_0^{\infty} (\rho u)(1-u) \mathrm{d}y$ is the momentum thickness, is shown in figure \ref{fig_Cf_St}(c). In the transcritical cases, the increase of $H_{12}$ ($H_{12} \approx 3.76$) in the laminar boundary layer is driven by a significant rise in $\delta_1$, caused by the reduction in streamwise momentum $\rho u$ in the near-wall vapour-like regime, and a decrease in $\theta$ resulting from the fuller velocity profile compared to the subcritical case. Before $H_{12}$ drops to approximately $2.0$ -- primarily due to the sharp rise in the turbulent $\theta$-value (see figure \ref{fig_qavg}b) -- it exhibits minor oscillations between $\Rey_\mathit{x}/10^5 \approx 5.0$ and $6.0$, attributed to the localised increase and decrease in $\delta_1$ over the near-wall separation zones.

\section{Turbulent boundary layer} \label{sec:turbulent}

After breakdown, we apply mean-flow scaling theories to the turbulent boundary layers of the subcritical case Tw095-LA and transcritical case Tw110-LA (see table~\ref{tab:tableBF}). Specifically, we examine whether the velocity profiles, after transformation, collapse onto the velocity profile of a constant-property boundary layer. If such a collapse is observed, the transformed profiles will be used to estimate the turbulent skin-friction coefficient 
$C_\mathit{f}$ and Stanton number $St$, in order to assess whether these parameters can be predicted for fluids at supercritical pressure.

\subsection{Mean velocity and enthalpy-velocity transformations}
We consider the transformation proposed by \citet{Patel1} and \citet{Trettel1}, which extends the \citet{VanDriest1}
velocity transformation (subscript $(\cdot)_\mathit{vD}$) defined as $\overline{u}^+_\mathit{vD} = \int^{\overline{u}^+}_0 \sqrt{\overline{\rho}/\overline{\rho}_\mathit{w}} \, \mathrm{d}\overline{u}^+$, by accounting for spatial variations in the viscous length scale $\delta_\mathit{v}^\star$ in the near-wall region. It is formulated as 
\begin{equation}
\overline{u}^\star = \int^{\overline{u}^+}_0 \left( 1 - \dfrac{y}{\delta_\mathit{v}^\star} \dfrac{\mathrm{d}\delta_\mathit{v}^\star}{\mathrm{d}y} \right) 
\underbrace{\frac{u_\tau}{u^\star_\tau}\, \mathrm{d}\overline{u}^+}_{\mathrm{d}\overline{u}^+_\mathit{vD}}, 
\label{eq:vDPatel} 
\end{equation}
where the superscripts $(\cdot)^{+}$ and $(\cdot)^{\star}$ indicate non-dimensionalisation by the viscous length scale $\delta_\mathit{v}$ and friction velocity $u_\tau$ (as defined in Appendix~\ref{sec:appA}), and by the semi-local viscous length scale $\delta^{\star}_\mathit{v}=\overline{\mu}/(\overline{\rho} {u^{\star}_\mathit{\tau}})$ and semi-local friction velocity ${u^{\star}_\mathit{\tau}}=\sqrt{{\overline\tau_\mathit{w}}/\overline{\rho}}$, respectively. The Reynolds (time and spanwise) average of a given variable $\chi$ is expressed as $\overline{\chi}=\chi-\chi^{\prime}$, with $\chi^{\prime}$ denoting the corresponding fluctuation. It is worth noting that the semi-local Mach number $M_\tau=u_\tau/\overline a_\mathit{w}$ remains much smaller than unity across the entire domain (with a maximum value of $0.025$). Therefore, the transformation recently proposed by \citet{Hasan1}, which extends \eqref{eq:vDPatel} to account for intrinsic compressibility effects, would yield results equivalent to $\overline{u}^{\star}$.

\begin{figure}
\centering
\includegraphics[angle=-0,trim=0 0 0 0, clip,width=1.0\textwidth]{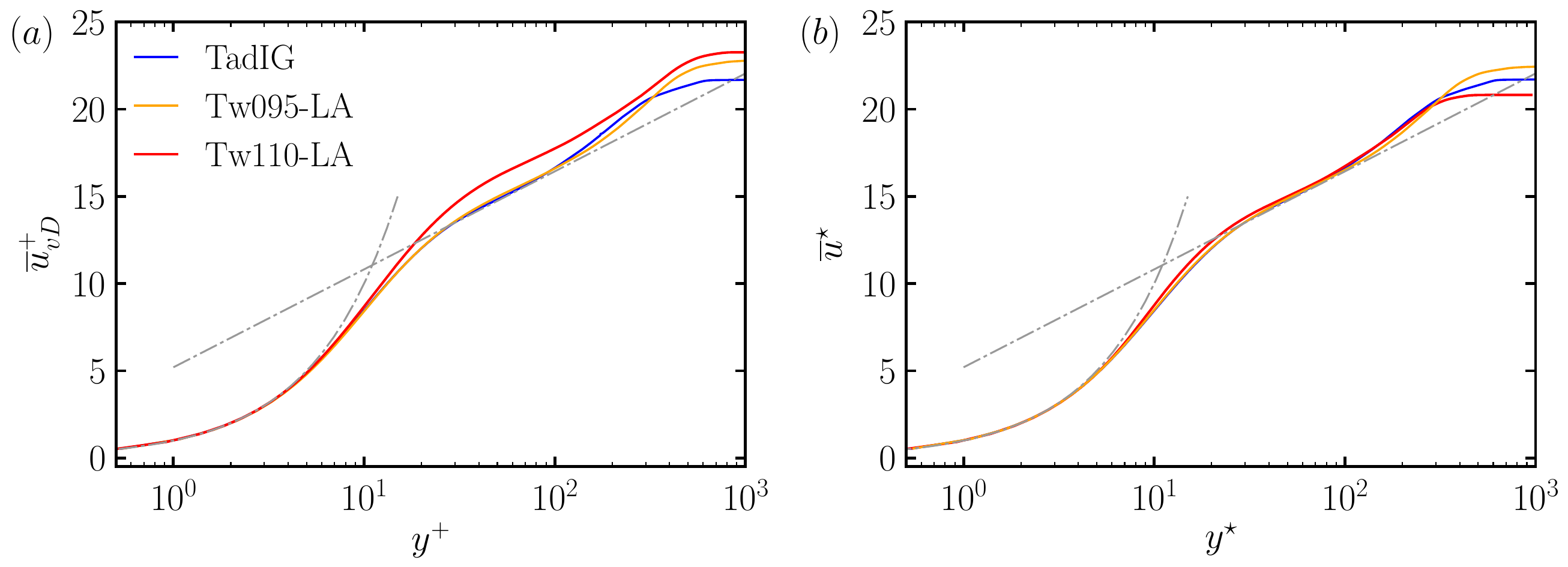}
\captionsetup{justification=justified}  \caption{\label{fig_Tw110LA_ustar} Wall-normal profiles of the transformed streamwise velocity using (a) \citet{VanDriest1} and (b) \citet{Patel1}. Case Tw095-LA and Tw110-LA are shown in orange at $Re_\mathit{\theta}=1387$ ($Re_\mathit{x}=1.0\cdot 10^6$) and in red at $Re_\mathit{\theta}=881$ ($Re_\mathit{x}=1.06\cdot 10^6$), respectively. Grey dash-dotted lines denote the linear and logarithmic laws ((a):~$(1/\kappa)\log y^+ + C$, (b):~$(1/\kappa)\log y^{\star} + C$) with $\kappa=0.41$ and $C=5.2$. In blue, ideal-gas case TadIG at $Re_\mathit{\theta}=1190$ ($Re_\mathit{x}=0.8\cdot 10^6$).}
\end{figure}
Figures~\ref{fig_Tw110LA_ustar}(a,b) show the transformed Reynolds-averaged mean velocity profiles for $\overline{u}^+_\mathit{vD}$ and $\overline u^\star$ on a logarithmic scale of $y^+$ and $y^\star$, respectively. The turbulent profiles are extracted near the end of the computational domains to ensure well-developed turbulent boundary layers: for Tw095-LA at $Re_\mathit{\theta}=1387$ ($Re_\mathit{x}=1.0\cdot 10^6$), for Tw110-LA at $Re_\mathit{\theta}=881$ ($Re_\mathit{x}=1.06\cdot 10^6$), and for TadIG at $Re_\mathit{\theta}=1190$ ($Re_\mathit{x}=0.8\cdot 10^6$). 
As expected, both velocity transformations of the subcritical case Tw095-LA show a good collapse with the velocity of the constant-property turbulent boundary layer (case TadIG), due to the low variation of thermophysical properties across the boundary layer. In contrast, the \citet{VanDriest1} scaling shows a large offset of the logarithmic region for the transcritical case Tw110-LA. For this case, the large near-wall variation of the viscous length scale must be taken into account. With this correction, \eqref{eq:vDPatel} clearly improves the collapse with the constant-property boundary layer (see figure~\ref{fig_Tw110LA_ustar}b). 

Contrary to the findings of \citet{Kawai1}, this collapse is achieved for the following reasons:~in case Tw110-LA, the 
wall-normal velocity in the boundary layer remains relatively small, with $\overline{v}/u_\infty \approx 0.2 \%$. Furthermore, the density and viscosity fluctuations, which peak in the buffer layer, are moderate on the order of $\sqrt{\overline{\rho^{\prime}\rho^{\prime}}}/\overline{\rho}$ and $\sqrt{\overline{\mu^{\prime}\mu^{\prime}}}/\overline{\mu} \approx 0.3$, respectively. By contrast, in the transcritical case investigated by \citet{Kawai1} at a higher reduced pressure ($p_\mathit{r} \approx 1.28$) but significantly larger wall-to-free-stream temperature ratios ($T^*_w/T^*_\infty \approx 4$--$8$), the wall-normal velocity reached approximately $1.5\%$, while the density and viscosity fluctuations reached $\sqrt{\overline{\rho^{\prime}\rho^{\prime}}}/\overline{\rho} \approx 1.0$, and $\sqrt{\overline{\mu^{\prime}\mu^{\prime}}}/\overline{\mu} \approx 0.4$, particularly in the logarithmic layer. These strong fluctuations led to large near-wall convective flux and an overshoot of the Reynolds shear stress surpassing unity in the logarithmic region, ultimately causing the failure of the velocity transformation. In our case, however, the total stress balance -- on which the \citet{Patel1} transformation relies -- is still mainly governed by the viscous and turbulent stresses under transcritical conditions. Overall, the transformed velocity profiles indicate that the flow has transitioned into the fully turbulent regime by the end of the computational domain. 

Next, we focus on estimating the mean thermodynamic properties to develop a predictive model for $C_\mathit{f}$ and $St$, which remains unknown for turbulent boundary layers with highly non-ideal fluids. One possible approach is to directly integrate the total heat flux from the wall to the free stream to reconstruct the temperature field. However, this method is cumbersome, as it requires an iterative procedure on the heat flux to match the prescribed wall-to-free-stream temperature ratio, combined with a turbulence model that itself depends on the velocity field as well as the density- and temperature-dependent thermodynamic properties. As a more practical alternative, we examine the mean temperature–velocity correlations, which offer a simpler route. When non-ideal gas effects become significant, enthalpy-based relations must be employed, as enthalpy is no longer a linear function of $T$ but instead depends on another thermodynamic property. Furthermore, classical enthalpy-based relations such as Crocco-Busemann \citep{Smits1}, or the relations of \citet{Walz1} and \citet{Duan1}, assume a constant mixed Prandtl number $Pr_\mathit{m}$ ($Pr_\mathit{m}=1$ in Crocco-Busemann), defined as 
\begin{equation}
Pr_\mathit{m}=\dfrac{(\overline{\mu}+\mu_t)\overline{c}_\mathit{p}}{\overline{\kappa}+\kappa_t}, \quad
\text{with:}~\mu_\mathit{t}=\dfrac{-\overline{\rho}\widetilde{u^{\prime \prime}v^{\prime \prime}}}{\partial \widetilde{u}/\partial y}~ \text{and}~\kappa_\mathit{t} = \dfrac{-\overline{c}_\mathit{p} \overline{\rho} \widetilde{v^{\prime \prime}T^{\prime \prime}} }{\partial \widetilde{T}/\partial y},
\label{eq:Prandtl}
\end{equation}
where $\mu_\mathit{t}$ and $\kappa_\mathit{t}$ are the eddy viscosity and eddy conductivity, respectively. Note that $\widetilde{\chi}=\overline{\rho \chi}/\overline{\rho}$ denotes the Favre average, with $\chi^{\prime \prime}$ as the Favre fluctuation.
 
\begin{figure}
\centering
\includegraphics[angle=-0,trim=0 0 0 0, clip,width=1.0\textwidth]{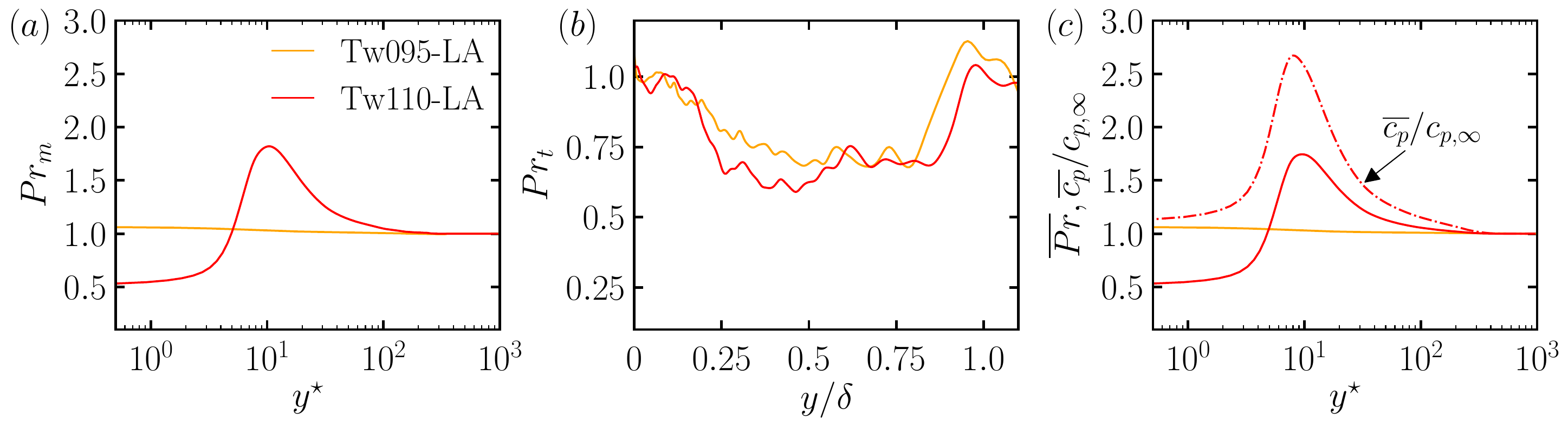}
\captionsetup{justification=justified} 
\caption{\label{fig_Tw_Pr}Case Tw095-LA ($Re_\mathit{\theta}=1387$) in orange and Tw110-LA ($Re_\mathit{\theta}=881$) in red: (a) mixed Prandtl number $Pr_\mathit{m}$, (b) turbulent Prandtl number $Pr_\mathit{t}$, and (c) mean molecular Prandtl number $\overline{Pr}$ and $\overline{c}_\mathit{p}/c_\mathit{p,\infty}$ (red dash-dotted line) for case Tw110-LA.}
\end{figure}

The mixed Prandtl number $Pr_\mathit{m}$, turbulent Prandtl number $Pr_\mathit{t}=\mu_\mathit{t}\overline{c}_\mathit{p}/\kappa_\mathit{t}$, and molecular Prandtl number $\overline{Pr}$ are plotted in figures~\ref{fig_Tw_Pr}(a--c), respectively, for the subcritical and transcritical cases.
In case Tw095-LA, the mixed Prandtl number remains approximately constant and close to unity across the boundary layer ($Pr_\mathit{m,w} \approx 1.06$), despite the turbulent Prandtl number decreasing to approximately $0.7$ in the outer region (see figure~\ref{fig_Tw_Pr}b). The evolution of $Pr_\mathit{m}$ confirms the assumption underlying the aforementioned enthalpy-based relations with constant Prandtl number. Under transcritical conditions, however, $Pr_\mathit{m}$ deviates from unity in the logarithmic region, reaching $1.84$ in the buffer layer before dropping below unity in the viscous sub-layer ($Pr_\mathit{m,w} \approx 0.51$). When observing the mean molecular Prandtl number $\overline{Pr}$ in figure \ref{fig_Tw_Pr}(c), its behaviour underlines that the evolution of the $Pr_\mathit{m}$ is dominated by molecular diffusion. The strong increase and subsequent decrease of $Pr_\mathit{m}$ towards the wall are proportional to the variation of $\overline{c}_\mathit{p}$, indicating that the Widom line is predominantly located in the buffer layer in the turbulent regime. In contrast, the turbulent Prandtl number moderately decreases from unity in the near-wall region to approximately $0.6$.

To address the variable $Pr_\mathit{m}$ in the highly non-ideal case Tw110-LA, the theory of \citet{VanDriest2} is considered. For a turbulent boundary layer in mean steady state over a ZPG flat plate, it holds that
\begin{equation}
    \left(\frac{\overline{i}^{\prime}}{Pr_{m}}\right)^{\prime}+\left(1-Pr_{m}\right) \frac{\tau^{\prime}}{\tau}\left(\frac{\overline{i}^{\prime}}{Pr_{m}}\right)=-\frac{u_{\infty}^{2}}{h_{\infty}},
    \label{eq:VanDriest1}
\end{equation}
where $\overline{i}=\overline{h}/h_\infty$ is the mean enthalpy ratio, $\tau$ is the total shear stress, and the superscript $(\cdot)^{\prime}$ indicates differentiation with respect to $\overline{u}_\mathit{r}=\overline{u}/u_\infty$. Integrating \eqref{eq:VanDriest1} yields
\begin{equation}
\dfrac{\overline{h}}{h_\infty} = \dfrac{h_w}{h_\infty} 
- \dfrac{\mathcal{S}}{\mathcal{S}_\infty} \left( \dfrac{h_{w}}{h_\infty} - 1 \right) 
+ \dfrac{u_\infty^2}{h_\infty} \left[ \dfrac{\mathcal{S}}{\mathcal{S}_\infty} \mathcal{R}_\infty - \mathcal{R} \right],
\label{eq:VanDriest_enth}
\end{equation}
where the functions $\mathcal{S}$ and $\mathcal{R}$ are defined as:
\begin{subequations}
\begin{gather}
\mathcal{S}=\int_{0}^{\overline{u}_r} Pr_m \cdot \exp \left[-\int_{\tau_{w}}^{\tau} \frac{\left(1-Pr_{m}\right)}{\tau} \mathrm{d} \tau\right] \mathrm{d} \overline{u}_{r}, \notag \\
\mathcal{R}=\int_{0}^{\overline{u}_r} Pr_{m} \cdot \exp \left[-\int_{\tau_{w}}^{\tau} \frac{\left(1-Pr_{m}\right)}{\tau} \mathrm{d} \tau\right]\left\{\int_{0}^{\overline{u}_{r}} \exp \left(\int_{\tau_{w}}^{\tau} \frac{\left(1-Pr_{m}\right)}{\tau} \mathrm{d} \tau\right) \mathrm{d} \overline{u}_{r}\right\} \mathrm{d} \overline{u}_{r}. \tag{\theequation$a{,}b$}
\end{gather}
\end{subequations}
Note that for $Pr_\mathit{m}=1$ and $Pr_\mathit{m}=\mathrm{const.}$, the classical relations of Crocco-Busemann \citep{Smits1} and \citet{Walz1} are recovered, respectively.

\begin{figure}
\centering
\includegraphics[angle=-0,trim=0 0 0 0, clip,width=1.0\textwidth]{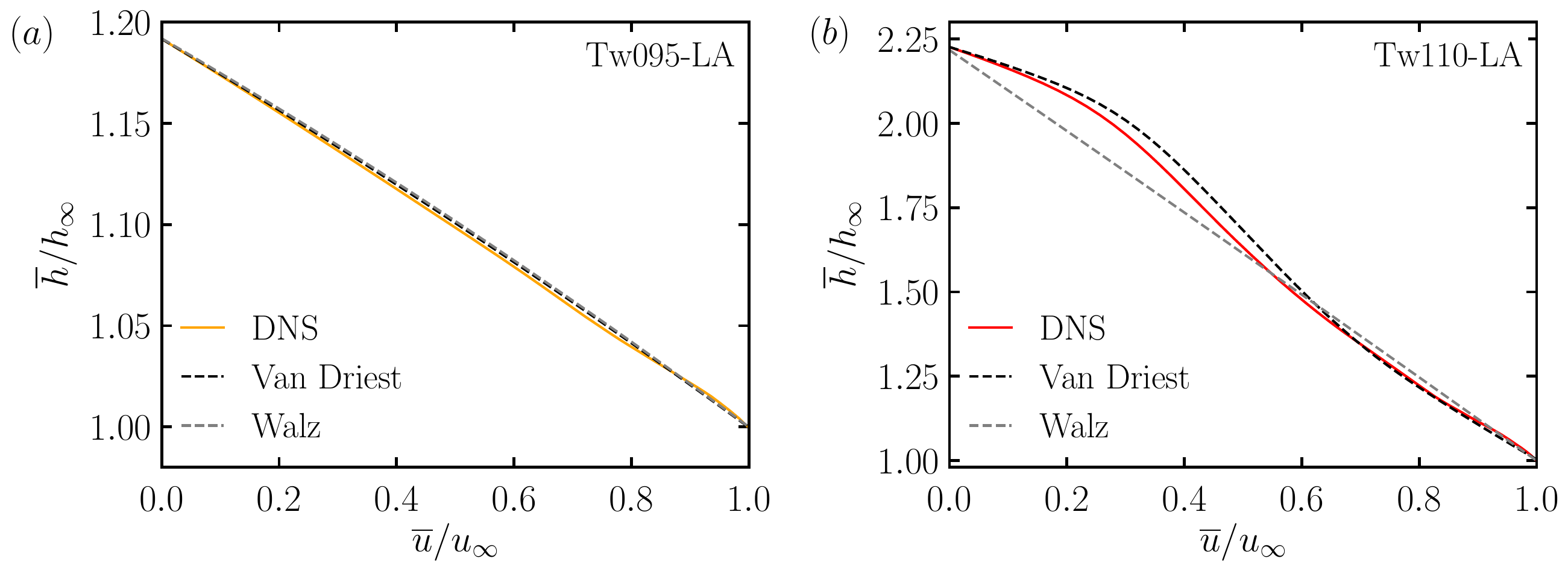}
\captionsetup{justification=justified} 
\caption{\label{fig_Tw_hhe}Reynolds-averaged mean enthalpy from \citet{VanDriest2} in black for (a) Tw095-LA at $Re_\mathit{\theta}=1387$ (DNS profile in orange) and (b) Tw110-LA at $Re_\mathit{\theta}=881$ (DNS profile in red). In grey, the relation of \citet{Walz1} as $\overline{h}/h_\infty=h_\mathit{w}/h_\infty+(h_\mathit{aw}-h_\mathit{w})(\overline{u}/u_\infty)/h_\infty -ru^2_\infty(\overline{u}/u_\infty)^2/(2h_\infty)$. }
\end{figure}

Figures~\ref{fig_Tw_hhe}(a,b) compare \eqref{eq:VanDriest_enth} against DNS data for the heated case Tw095-LA ($T^*_\mathit{w}/T^*_\infty=1.056$) and Tw110-LA ($T^*_\mathit{w}/T^*_\infty=1.222$), respectively.
For the subcritical case, Van Driest's relation accurately predicts $\overline{h}/h_\infty$ and aligns with Walz's relation as $Pr_\mathit{m} \approx \mathrm{const.}$ However, for the highly non-ideal fluid with strongly varying mixed Prandtl number, as seen in figure~\ref{fig_Tw_Pr}(a), Walz's relation become particularly inaccurate in the inner layer. From $\overline{u}/u_\infty \approx 0.6$, or $y^\star \approx 20$, the enthalpy undergoes a significant increase, which only Van Driest's relation is able to capture in good agreement with the DNS data.

\begin{figure}
\centering
\includegraphics[angle=-0,trim=0 0 0 0, clip,width=0.8\textwidth]{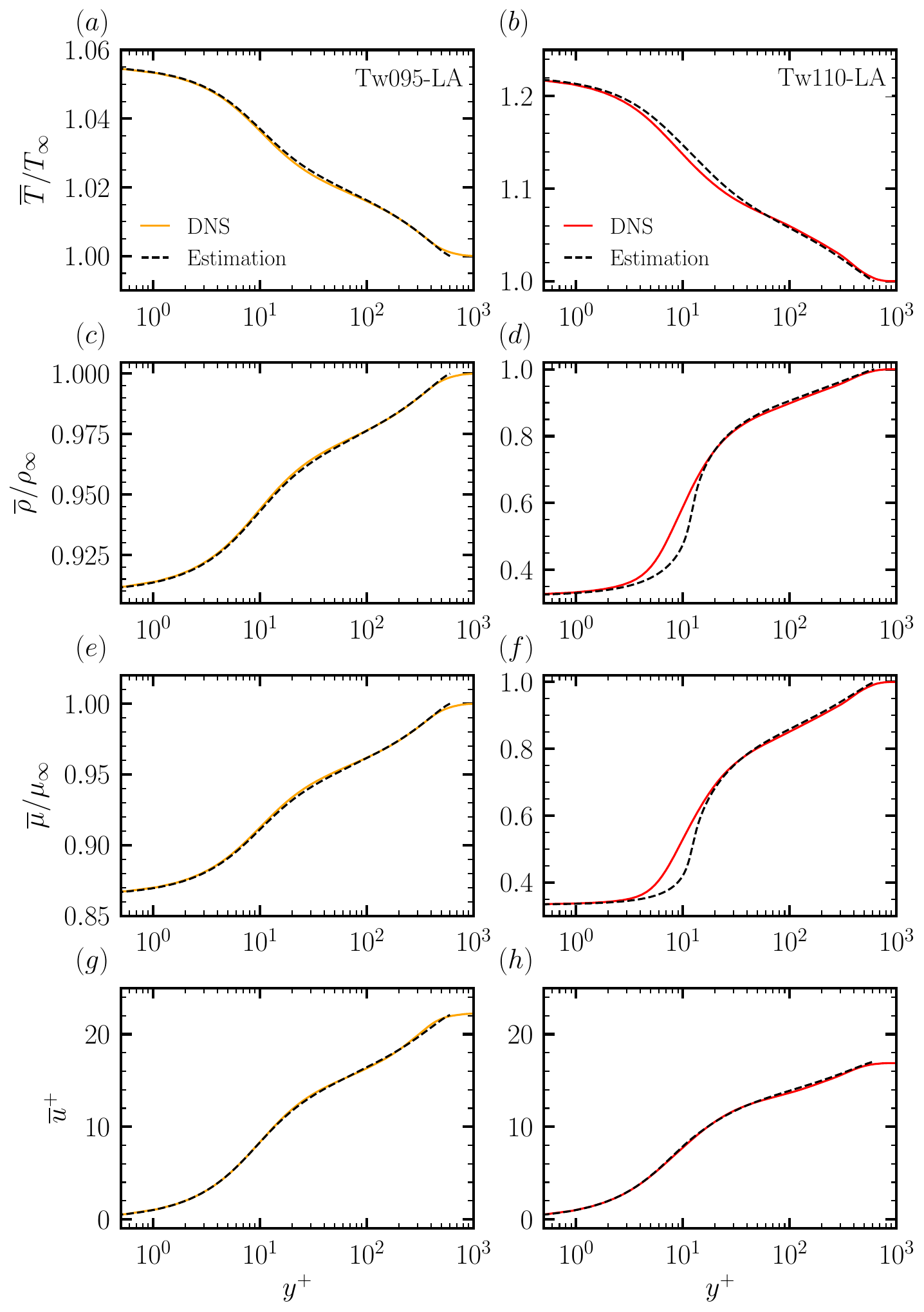}
\captionsetup{justification=justified} 
\caption{\label{fig_profiles_DNS_esti}Estimated mean (a,b) temperature $\overline{T}/T_\infty$, (c,d) density $\overline{\rho}/\rho_\infty$, (e,f) viscosity $\overline{\mu}/\mu_\infty$, and (g,h) streamwise velocity $\overline{u}^{+}$ profiles (dashed grey lines) compared to DNS results (solid lines, case Tw095-LA ($Re_\mathit{\theta}=1387$) in orange and case Tw110-LA ($Re_\mathit{\theta}=881$) in red).}
\end{figure}

\subsection{Estimating mean profiles and fluxes}

We now follow the approach of \citet{Hasan2} to predict the drag and heat transfer and compare the results to the available DNS data (see figure~\ref{fig_Cf_St}). In this approach, the mean shear is integrated from the wall to the free stream to obtain the streamwise velocity, using a combination of inner and outer layer modelling approximations. The inner-layer eddy viscosity is modelled using the Johnson–King mixing length formulation \citep{Johnson1}, accounting for variations in viscous length and velocity scales, while the outer layer is described by Coles' law of the wake \citep{Coles1}, modified to include mean density variations. 
The corresponding thermodynamic properties -- e.g. $\rho$, $T$, $\mu$, $\kappa$, and $Pr$ -- are computed using the Van der Waals EoS, under the assumption of constant thermodynamic pressure, and the JST model, with both expressed as functions of the enthalpy obtained via Van Driest's theory. Note that two additional assumptions are made: (i) $Pr_\mathit{m} \approx \overline{Pr}$ (compare  figures~\ref{fig_Tw_Pr}a,c), and (ii) the shear distribution follows \citet{VanDriest2}, i.e. $\tau/\tau_\mathit{w} \approx 1 - \exp[-\kappa/\sqrt{C_\mathit{f}/2}(1-\overline{u}_r)]$, with $\kappa=0.41$ as the von Kármán constant. For further details on the analytical solver, refer to the source code available on GitHub (\url{https://github.com/pcboldini/DragAndHeatTransferEstimation_NonIdealFluids}). 

Figure~\ref{fig_profiles_DNS_esti} presents the predicted temperature, density, viscosity, and velocity profiles from the analytical solver (`Estimation') compared with the DNS profiles for cases Tw095-LA and Tw110-LA. In the subcritical case, the estimated profiles are in very good agreement with the DNS solution. In the transcritical case, the temperature profile in figure~\ref{fig_profiles_DNS_esti}(b) shows minimal deviations in the buffer layer but successfully approximates the wall heat flux $\overline{q}_\mathit{w}$. The estimated profiles for density (figure~\ref{fig_profiles_DNS_esti}d) and viscosity (figure~\ref{fig_profiles_DNS_esti}f) differ significantly only in the buffer layer, where the largest discrepancies in $\overline{h}$ are observed. Note that Jensen's inequality, i.e. $\overline{\chi} \neq \chi(\overline{h},\overline{p})$, applies to these thermophysical quantities \citep{Nemati2}, due to their sharp curvature around the pseudo-critical point, i.e. $\max\{c_\mathit{p}\}$, and their large fluctuations present specifically in the buffer layer. Conversely, the estimated velocity profile agrees very well with the DNS solution, as it results from integration across both inner and outer layers.

\begin{figure}
\centering
\includegraphics[angle=-0,trim=0 0 0 0, clip,width=0.60\textwidth]{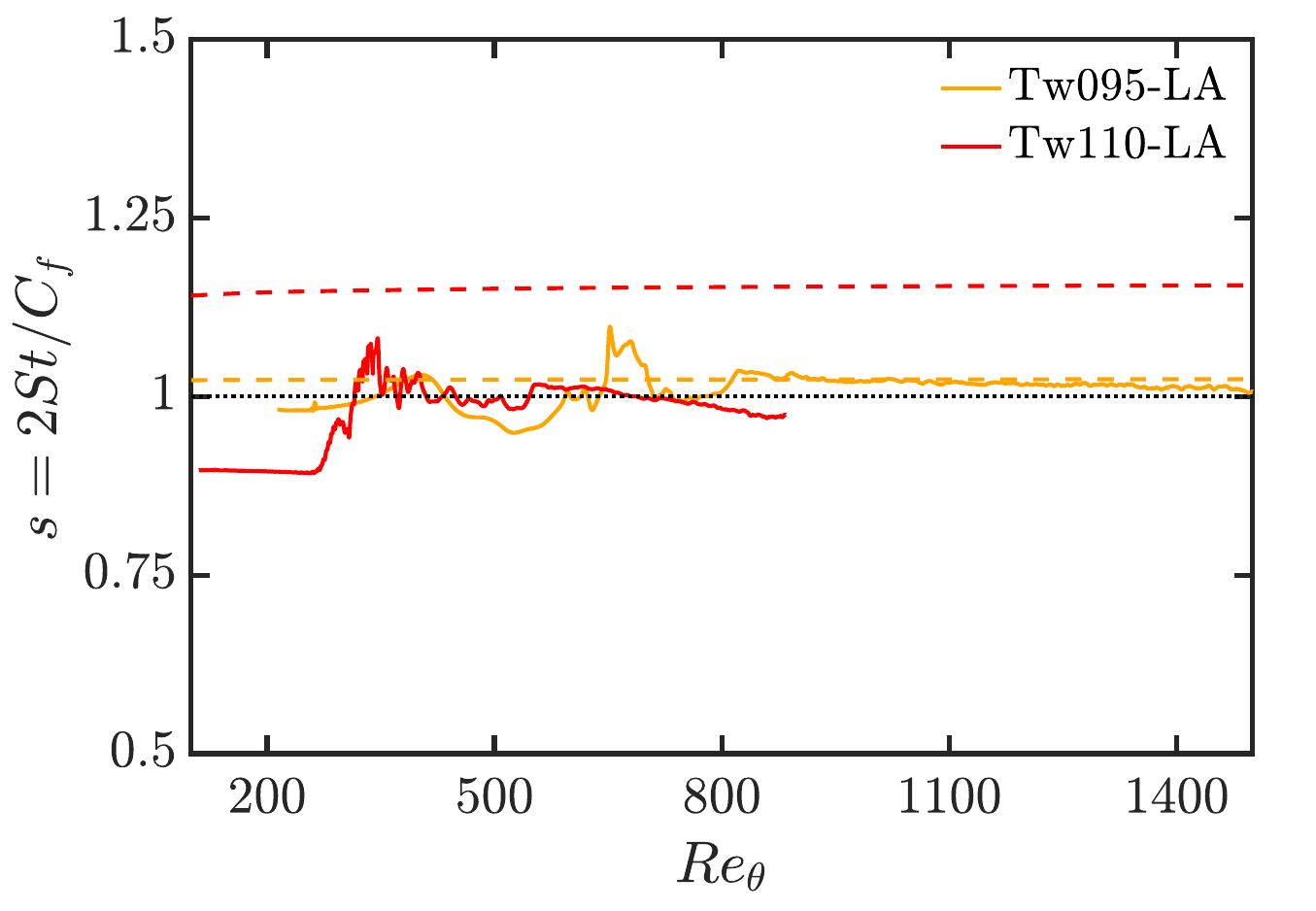}
\captionsetup{justification=justified} 
\caption{\label{fig_DNS_sfact}Reynolds analogy factor $s=2St/C_\mathit{f}$ as a function of the momentum-thickness Reynolds number $Re_\mathit{\theta}$: case Tw095-LA (orange) and Tw110-LA (red). Solid lines correspond to the DNS results, while dashed lines represent the turbulent Reynolds analogy factor, $s=\mathcal{S}_\infty$, according to \citet{VanDriest2}. The Reynolds analogy is indicated by a black dotted line.}
\end{figure}
Before addressing the prediction of the skin-friction coefficient and Stanton number, we first examine the Reynolds analogy, which relates skin friction and heat transfer. Using the Reynolds analogy factor, $s=2St/C_\mathit{f}=\overline{q}^*_\mathit{w}u^*_\infty/(\overline{\tau}^*_\mathit{w}(h^*_\mathit{aw}-h^*_\mathit{w}))$, we investigate how non-ideal gas effects may disrupt the classical coupling between momentum and thermal transport. The impact of the Reynolds analogy is shown in figure~\ref{fig_DNS_sfact}. In the laminar boundary layer, strong property variations in the transcritical regime reduce the Reynolds analogy factor below unity ($s \approx 0.89$). In the transitional region, the value of $s$ rises above unity in both thermodynamic regimes due to the gradual development of turbulent structures. Interestingly, $s$ appears largely insensitive to the degree of fluid non-ideality in the turbulent boundary layer. Here, the Reynolds analogy holds in both thermodynamic regimes, with $s \approx 1$. This behaviour aligns with trends observed in ideal-gas turbulent boundary layers, where $s$ was found to be unaffected by variations in wall temperature and Mach number \citep{Wenzel1}. Nevertheless, a more comprehensive investigation over a broader Reynolds-number range would be needed to extend the applicability of these results to other non-ideal fluid flows. Note that the turbulent Reynolds analogy factor, calculated according to the theory of \citet{VanDriest2} in \eqref{eq:VanDriest1}, with $s=\mathcal{S_\infty}$, is following the DNS results only in the subcritical regime. In the transcritical regime, larger deviations in the estimation of $\tau$ are observed.

\begin{figure}
\centering
\includegraphics[angle=-0,trim=0 0 0 0, clip,width=1.0\textwidth]{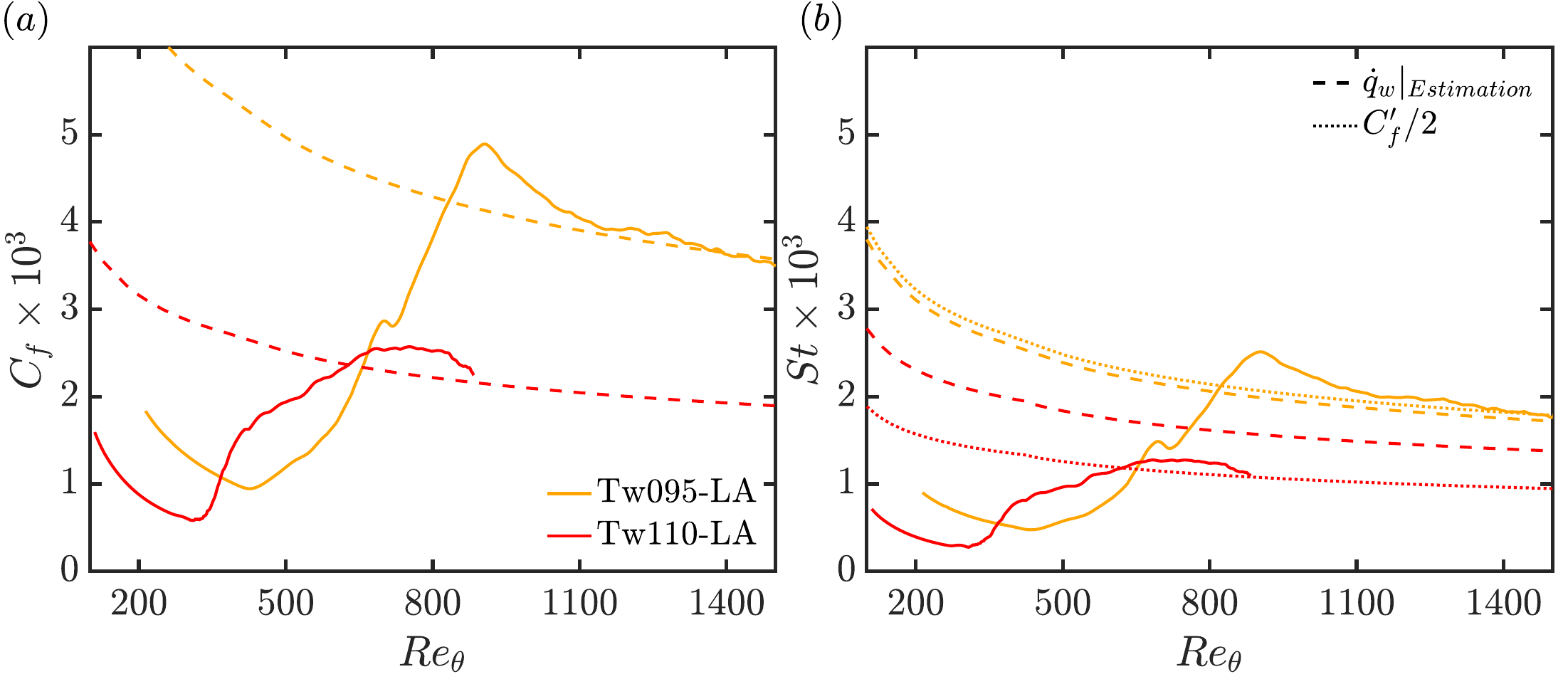}
\captionsetup{justification=justified} 
\caption{\label{fig_DNS_esti}Case Tw095-LA (orange) and Tw110-LA (red): (a) skin-friction coefficient $C_\mathit{f}$ and (b) Stanton number $St$ as functions of the momentum-thickness Reynolds number $Re_\mathit{\theta}$. In (a,b), solid lines correspond to the DNS results, while dashed lines denote the analytical estimations. In (b), dotted lines represent $St=C^{\prime}_\mathit{f}/2$ ($Pr_\infty=1$) according to the Reynolds analogy, where $C^{\prime}_\mathit{f}$ is the analytical prediction from (a).}
\end{figure}

In figures~\ref{fig_DNS_esti}(a,b), the DNS and estimated values of the skin-friction coefficient, $C_\mathit{f}$, and Stanton number, $St$, are plotted as functions of $Re_\mathit{\theta}$.
The predictions of $C_\mathit{f}$ show good agreement with the DNS curves beyond the respective overshoots. Moreover, the less pronounced overshoot in $C_\mathit{f}$ for case Tw110-LA is confirmed. In contrast, the $St$ prediction reveals good accuracy only for the subcritical case Tw095-LA. For the highly non-ideal case Tw110-LA, the $St$ number is over-predicted by $\sim30\%$ at $Re_\mathit{\theta}=881$ due to a slightly higher estimated wall heat flux in figure~\ref{fig_profiles_DNS_esti}(b). Note that the error in the Stanton number prediction using the relation of \citet{Walz1} reaches $\sim200\%$ at $Re_\mathit{\theta}=881$. Interestingly, when applying the Reynolds analogy $St=C^{\prime}_\mathit{f}/2$ for $Pr_\infty=1$, verified in figure~\ref{fig_DNS_sfact}, and using $C^{\prime}_\mathit{f}$ from the analytical prediction in figure~\ref{fig_DNS_esti}(a), the estimated $St$ curve closely agrees with the DNS results in the turbulent regime. In conclusion, this behaviour suggests that even under the examined transcritical conditions with variable mixed Prandtl number, the Reynolds analogy may remain a robust tool for predicting the turbulent heat transfer in ZPG flat-plate boundary layers.

\section{Conclusions}\label{sec:conclusions}
Direct numerical simulations (DNS) are performed to investigate the laminar-to-turbulent transition of zero-pressure-gradient flat-plate boundary layers at supercritical fluid pressure with wall heating to trigger Mode-II instability. Two flow cases are defined based on the pseudo-critical temperature $T^*_\mathit{pc}$, both featuring a liquid-like free stream ($T^*_\mathit{\infty} < T^*_\mathit{pc}$): one in the subcritical, liquid-like regime with $T^*_\mathit{w} < T^*_\mathit{pc}$, and one in the transcritical (pseudo-boiling) regime with a vapour-like near-wall region ($T^*_\mathit{w} > T^*_\mathit{pc}$).

First, only a single fundamental two-dimensional (2-D) wave is excited. Under linear forcing, the Mode-II instability in boundary layers is shown to result from the combination of shear and baroclinic effects, producing two out-of-phase vorticity waves around the critical layer. This confirms the model proposed by \citet{Bugeat2} in plane Couette flow. As the 2-D wave forcing is increased, nonlinearity saturates the unstable 2-D wave in the subcritical regime, case Tw095. Conversely, in the transcritical-heating regime, case Tw110, nonlinear excitation of higher harmonics emerges in the vicinity of the forcing strip. The first higher harmonic $(2,0)$ outgrows the primary mode $(1,0)$ further downstream. A strong modal interaction follows, leading to the development of subharmonic resonance of mode $(1,0)$ relative to the now dominant $(2,0)$ mode. This resonant interaction, similar to the vortex pairing in mixing layers, accelerates the nonlinear breakdown and induces the billowing motion of the Widom line within the boundary layer. The resulting behaviour closely resembles the classical Kelvin-Helmholtz instability in shear layers, generating a sequence of billow structures that grow within the boundary layer in streamwise direction. In addition, large velocity perturbations near the wall cause periodical, localised flow reversal, with a generalised inflection point and a density-weighted vorticity maximum, indicative of inviscid instability. The formation of localised separation zones -- absent under weakly non-ideal gas conditions -- is analogous to that observed in the ideal-gas regime with a prescribed (strong) adverse pressure gradient.

Subsequently, the 3-D breakdown to turbulence is investigated by building upon the 2-D nonlinear analyses of both subcritical and transcritical cases. In addition to the large amplitude 2-D wave, a pair of subharmonic oblique waves at half the frequency of the primary wave is introduced. The amplitude of oblique waves is either infinitesimally small (`IA') or finitely large (`LA').

In the subcritical regime, case Tw095-IA reveals that infinitesimal 3-D perturbations fail to trigger transition under the considered Reynolds number range. Conversely, in Tw095-LA, the classical H-type breakdown is triggered with a sharp skin-friction overshoot, resembling the ideal-gas reference case of \citet{Sayadi1}. However, the onset of the staggered, streamwise-elongated $\Lambda$-vortices is delayed compared to the ideal-gas case. 

In the transcritical-heated regime, despite stronger primary-wave growth, transition to turbulence is delayed compared to the subcritical regime, given the same forcing parameters. In analogy with the 2-D nonlinear analysis, spanwise-oriented billows initially dominate the early transitional stage, with downstream-convected near-wall separation zones causing a moderate rise in $C_f$ and $St$, neither of which become negative. In Tw110-LA, subharmonic resonance emerges only after vortex pairing between mode $(1,0)$ and its first higher harmonic $(2,0)$. Once the 3-D subharmonic reaches finite amplitude, all 3-D modes undergo abrupt nonlinear amplification. This leads to the formation of alternating high-shear layers at spanwise `peak' and `co-peak' (half a spanwise wavelength apart) position, similar to the K-type breakdown with APG \citep{Kloker1, Kloker2}. The `co-peak' high-shear layer, induced by near-wall separation, breaks up first, triggering near-wall longitudinal structures ahead of the outer-region hairpin-like vortices.

In contrast, case Tw110-IA -- with infinitesimally low subharmonic 3-D forcing -- reveals that no oblique-wave forcing is needed to trigger transition, despite its onset shifts slightly downstream compared to Tw110-LA. As a result, the primary 2-D wave is amplified to higher amplitude levels, initiating a rapid fill-up of the spectrum of all 3-D modes from the low numerical background noise, with fundamental-resonance/breakdown mechanism emerging. In other words, subharmonic resonance is no longer the dominant secondary instability mechanism under low-noise conditions.
Three-dimensionality within each spanwise-oriented billow develops gradually, with aligned $\Lambda$-vortices progressively forming and breaking up similar to case Tw110-LA.

The $C_\mathit{f}$ and $St$ curves of both transcritical cases are characterised by (i) the absence of a sharp skin-friction overshoot, due to the lack of strong hairpin-like structures; (ii) lower transitional momentum-thickness Reynolds numbers; and (iii) significantly higher shape factors, $H_{12}$, compared to the subcritical case.

In the turbulent regime under transcritical conditions, the \citet{Patel1} scaling collapses well the velocity profiles. For predicting the mean enthalpy profile under transcritical conditions, the classical enthalpy-velocity relations fail to reproduce the rapid inner-layer enthalpy rise due to strong variation of the molecular Prandtl number. Instead, the theory of \citet{VanDriest2} for variable Prandtl number agrees well with DNS. Based on these results, a predictive model for the turbulent skin-friction coefficient and Stanton number with non-ideal fluid flows is developed, demonstrating good agreement with the DNS results.

The findings in this work highlight the sensitivity of the laminar-to-turbulent transition to both the amplitude of the 3-D perturbations and thermodynamic state. Due to the presence of the Widom line (pseudo-boiling effect), 2-D waves can be strongly amplified -- similar to the ideal-gas mixing layer -- and may trigger transition to turbulence only from 3-D numerical background noise alone. The resulting K-type breakdown features resemble those observed in the ideal-gas APG case -- a scenario absent under ideal-gas and weakly non-ideal gas conditions without a streamwise pressure gradient. These DNS investigations of controlled laminar-to-turbulent transition pave the way for future studies of free-stream turbulence-induced transition that may be encountered in experimental facilities or industrial applications under pseudo-boiling conditions. 

\backsection[Supplementary data]{\label{SupMat}Supplementary material is available.}

\backsection[Acknowledgements]{The authors acknowledge the use of computational resources of the Dutch National Supercomputer Snellius (SURF) (grant no.~2024/ENW/01704792). P.~C.~Boldini acknowledges helpful discussion on the turbulent boundary layer with A.~M.~Hasan (TU Delft).}

\backsection[Funding]{This work was funded by the European Research Council grant no. ERC-2019-CoG-864660, Critical.}

\backsection[Declaration of interests]{The authors report no conflict of interest.}

\backsection[Data availability statement]{The data that support the findings of this study are available on demand. The source code for the estimation of turbulent skin-friction coefficient and Stanton number is available at \url{https://github.com/pcboldini/DragAndHeatTransferEstimation_NonIdealFluids}.}

\backsection[Author ORCIDs]{\\
P.~C.~Boldini, \url{https://orcid.org/0000-0003-0868-5895}; \\
B.~Bugeat, \url{https://orcid.org/0000-0001-5420-7531}; \\
J.~W.~R.~Peeters, \url{https://orcid.org/0000-0003-1265-9040}; \\
M.~Kloker, \url{https://orcid.org/0000-0002-5352-7442}; \\
R.~Pecnik, \url{https://orcid.org/0000-0001-6352-6323}}

\appendix

\section{DNS set-up and grid-resolution analysis} \label{sec:appA}
The grid resolution, with $N_\mathit{x} \times N_\mathit{y} \times N_\mathit{z}$ grid points, varies for each case. The grid is uniform in the streamwise ($x$) and spanwise ($z$) directions, and stretched in the wall-normal ($y$) direction according to $y=y_\mathit{e}[ K_1\eta +(1-K_1)( 1 + \tanh(0.5\sigma(\eta-1)/\tanh(0.5\sigma) )]$, where $\eta=0,...,1$ and $K_1=0.6(N_\mathit{y}-1)/(Re_{\tau,0} y_\mathit{e})$. The stretching factor $\sigma$ and the inlet friction Reynolds number $Re_{\tau,0}=\delta_{99,0}/\delta_{99,v}$ -- with $\delta_{99,v}=\overline{\mu}_\mathit{w}/(\overline{\rho}_\mathit{w}u_\tau)$, and $u_\tau=\sqrt{\overline{\tau}_\mathit{w}/\overline{\rho}_\mathit{w}}$ and $\overline{\tau}_\mathit{w}$ being the friction velocity and the wall shear stress, respectively -- are chosen such that the first grid cell in wall-normal direction, $\Delta y^{+}_\mathit{w}$, (superscript $(\cdot)^{+}$ denotes viscous units), remains below unity throughout the domain. At the inlet $x_0$, the laminar boundary layer is resolved with approximately $140$ grid points in the wall-normal direction for both supercritical cases. Table~\ref{tab:numerical_setup} summarises the relevant grid parameters.
\begin{table}
  \begin{center}
\def~{\hphantom{0}}
  \begin{tabular}{l@{\hspace{2mm}}c@{\hspace{4mm}}c@{\hspace{4mm}}c@{\hspace{4mm}}ccccccccc}
       Case & $L_\mathit{x}$ & $L_\mathit{y}$ & $L_\mathit{z}$  & $N_\mathit{x} \times N_\mathit{y} \times N_\mathit{z}$ & $Re_{x,0}/10^5$ & $\Delta x^+_\mathit{max}$ & $\Delta y^+_\mathit{w,max}$ & $\Delta z^+_\mathit{max}$ & $Re_\mathit{\theta,max}$ \\[0.4em]
       Tw095  & $724.3$ & $40.0$ & $9.63$ & $13550 \times 600 \times 180$ & $1.0$ & $4.48$ & $0.63$ & $4.48$ & $1509$ \\
       Tw110  & $1078.0$ & $40.0$ & $9.63$ & $20150 \times 900 \times 180$ & $0.58$ & $3.49$ & $0.49$ & $3.49$ & $885$ \\
       TadIG  & $515.0$ & $20.0$ & $9.63$ & $4000 \times 600 \times 150$ & $1.0$ & $10.0$ & $0.59$ & $4.9$ & $1229$ \\
  \end{tabular}
  \captionsetup{justification=justified}
  \caption{Numerical parameters for the cases listed in table~\ref{tab:tableBF}: $L_\mathit{x}$, $L_\mathit{y}$, and $L_\mathit{z}$ are the sizes of the computational domain in the streamwise, wall-normal, and spanwise directions, respectively; $N_\mathit{x}$, $N_\mathit{y}$, and $N_\mathit{z}$ denote the number of grid points in the corresponding directions; $\Rey_{x,0}$ is the inlet Reynolds number; $\Delta x^+_\mathit{max}$, $\Delta y^+_\mathit{w,max}$, and $\Delta z^+_\mathit{max}$ are the maximum grid sizes in the $x$-, $y$-, and $z$-directions relative to the maximum viscous length scale in the domain, $\overline{\mu}_\mathit{w}/(\overline{\rho}_\mathit{w} u_\tau)$. In addition, the momentum Reynolds number is defined as $Re_\mathit{\theta} = \rho^*_\infty u^*_\infty \theta^*/\mu^*_\infty$, based on the local momentum thickness $\theta^*$ and free-stream properties. Note that case Tw110 here corresponds to case Tw110-LA in table~\ref{tab:numerical_setup2}, whereas case Tw110-IA differs in $\Delta y^+_\mathit{w,max} \approx 0.48$, $\Delta x^+\approx\Delta z^+\approx 3.39$, and $Re_\mathit{\theta,max}=837$.    }
  \label{tab:numerical_setup}
  \end{center}
\end{table}

One-dimensional non-reflecting boundary conditions for non-ideal flows \citep{Okongo1} are applied at:~(i) the subsonic inlet ($x=x_0$), (ii) the outlet ($x=x_\mathit{e}$), with the incoming wave amplitude set to zero, (iii) the top boundary ($y=y_\mathit{e}$), with constant pressure $p_{r,\infty}$, and (iv) the wall ($y=0$), where no-slip and no-penetration conditions are imposed, except in the disturbance-strip region (see \S\,\ref{sec:disturbance_strip}). In addition, sponge zones are applied at the inlet, outlet, and top boundaries to minimise spurious acoustic reflections \citep{Mani1}, with the local solution gradually dampened towards the laminar boundary-layer profile. The inflow sponge length ($x_0<x<x_0+20.0$) and damping strength ($\sigma_\mathit{p}=0.5$) are the same for all simulations. For case Tw110, both the outlet ($x_\mathit{e}-50.0<x<x_\mathit{e}$, $\sigma_\mathit{p}=0.5$) and top sponge zones ($y_\mathit{e}-13.3<y<y_\mathit{e}$, $\sigma_\mathit{p}=1.0$) are extended to account for the strong fluctuations intensity, e.g. $\rho^{\prime}/\overline{\rho} \approx O(1)$ \citep{Kawai1}. For cases Tw095 and TadIG, sponge zones are active in the regions $x_\mathit{e}-20.0<x<x_\mathit{e}$ ($\sigma_\mathit{p}=0.5$) and $y_\mathit{e}-1.0<y<y_\mathit{e}$ ($\sigma_\mathit{p}=0.5$). A sensitivity analysis of the grid resolution for transcritical case Tw110 (the most computationally expensive case) is performed in the following.

\begin{figure}
    \centering
    \includegraphics[angle=-0,trim=0 0 0 0, clip,width=1.0\textwidth]{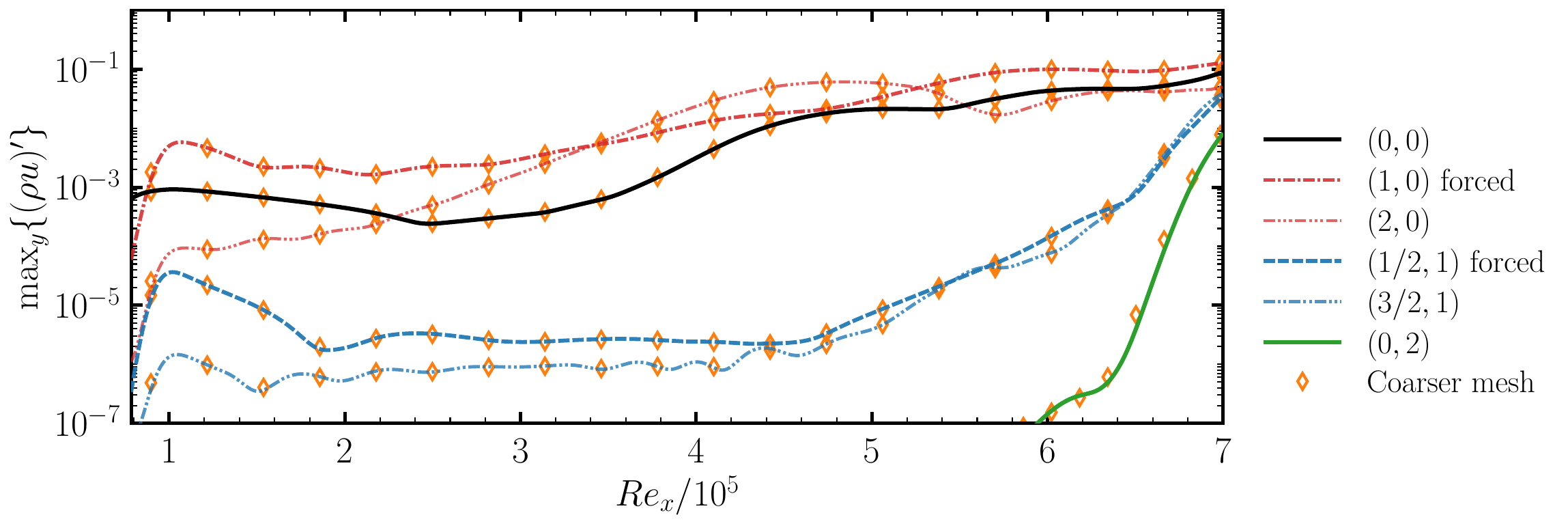}
    \captionsetup{justification=justified}    \caption{\label{fig:fig_sens_analysis_fft}Case Tw110-LA: streamwise development of the $y$-maximum $(\rho u)^\prime$ disturbance amplitudes of the most relevant modes $( \omega / \omega_{\text{2-D}}, \beta / \beta_0)$. Coarser-mesh results are marked with \textcolor{mycolor2}{$\lozenge$}.}
\end{figure}

\begin{figure}
    \centering
    \includegraphics[angle=-0,trim=0 0 0 0, clip,width=1.0\textwidth]{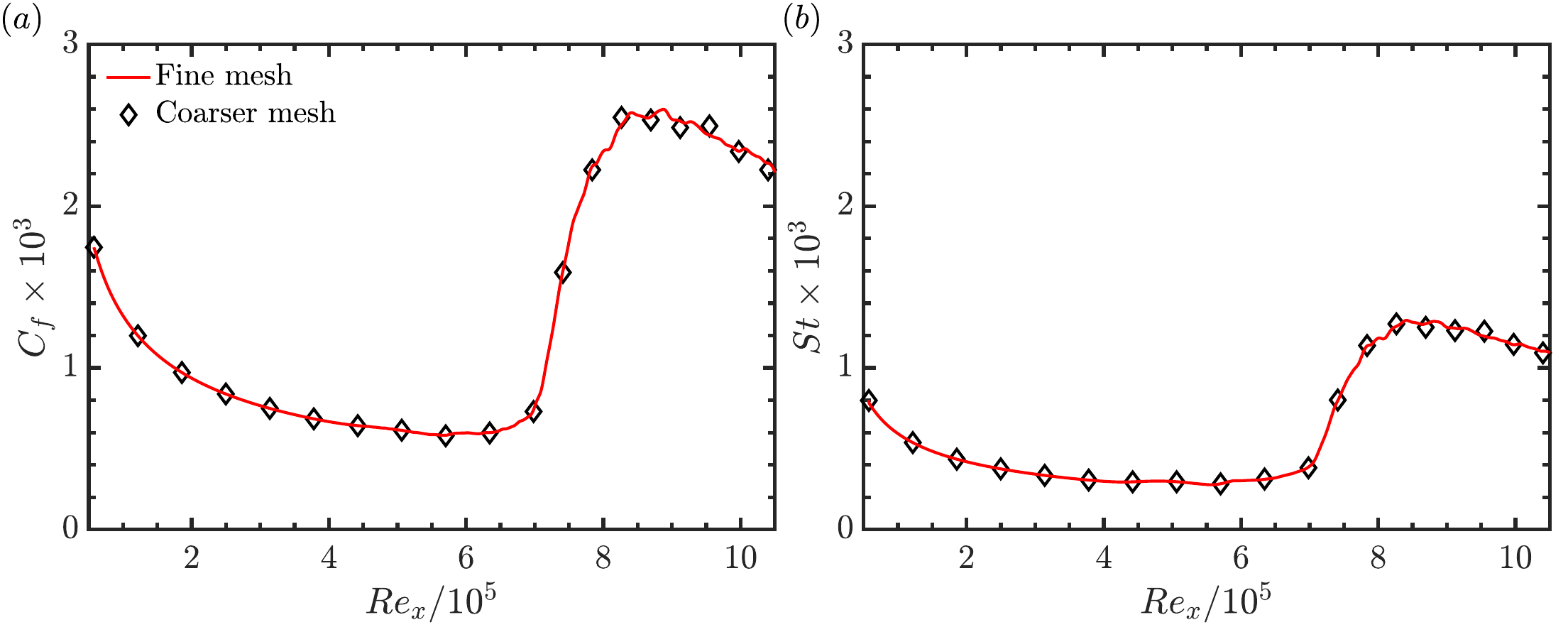}
    \captionsetup{justification=justified}     \caption{\label{fig:fig_sens_analysis_cf}Case Tw110-LA: time- and spanwise-averaged (a) skin-friction coefficient and (b) Stanton number. Coarser-mesh results are marked with \textcolor{black}{$\lozenge$}.}
\end{figure}

The baseline computational grid of case Tw110 (hereafter referred to as `fine grid'), reported in table~\ref{tab:numerical_setup}, is coarsened by factors of approximately $1.44$, $1.2$, and $1.2$ in the streamwise, wall-normal and spanwise directions, respectively, resulting in a grid of $14000 \times 750 \times 150$ points, hereafter referred to as `coarser mesh'. Uniform spacing is applied in the streamwise and spanwise directions, while the same wall-normal grid-clustering and $\Delta y^+_\mathit{w,max}$ as the fine grid are retained. Figures~\ref{fig:fig_sens_analysis_fft} and \ref{fig:fig_sens_analysis_cf} show that the modal evolution, skin-friction coefficient, and Stanton number remain largely insensitive to changes in grid resolution, confirming the robustness of the 3-D simulations.

\section{Laminar boundary layer} \label{sec:appB}
\begin{figure}
    \centering
    \includegraphics[angle=-0,trim=0 0 0 0, clip,width=1.0\textwidth]{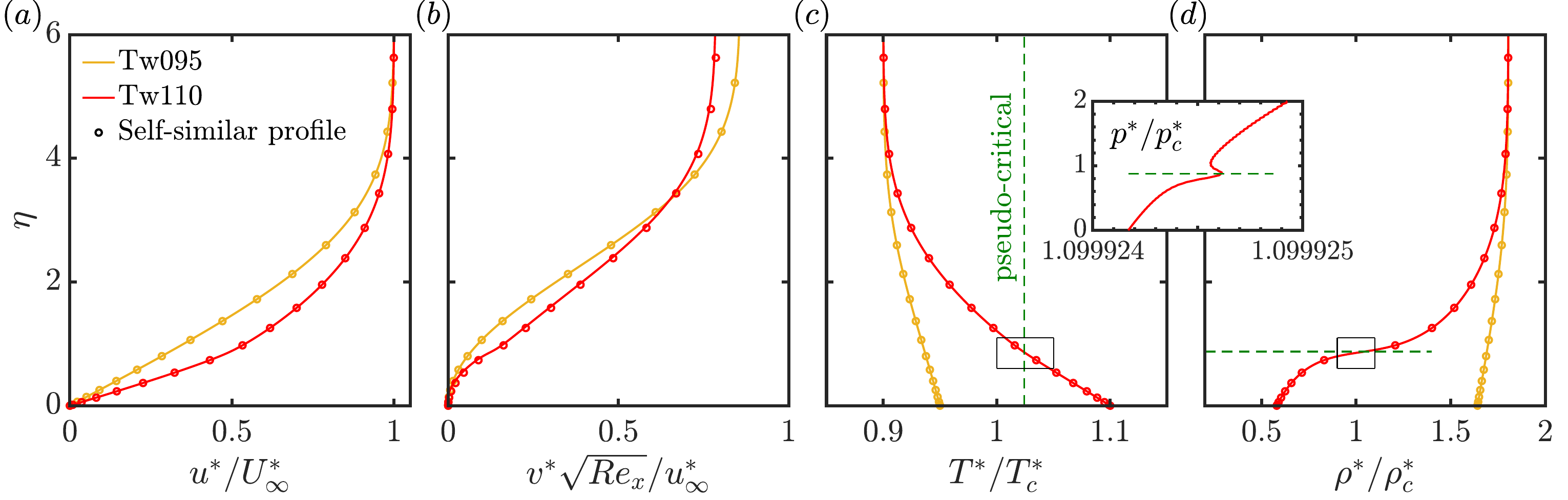}
    \captionsetup{justification=justified}                
    \caption{\label{fig:fig_bl_validation}2-D DNS laminar profiles for cases Tw095 and Tw110: (a) streamwise velocity, (b) wall-normal velocity, (c) reduced temperature, and (d) reduced density plotted against the self-similar wall-normal coordinate $\eta$. The laminar self-similar solutions are indicated by circles (\textcolor{black}{$\circ$}). The DNS reduced pressure $p^*/p^*_\mathit{c}$ is plotted in the inset. The dashed green line (\textcolor{matplotlibgreen}{\rule[0.5ex]{0.2cm}{1pt}} \textcolor{matplotlibgreen}{\rule[0.5ex]{0.2cm}{1pt}}) indicates the pseudo-critical point, i.e. where $T^*=T^*_\mathit{pc}$.}
\end{figure}

For supercritical cases Tw095 and Tw110 in table~\ref{tab:tableBF}, a comparison is made between the initial solution described in \S\,\ref{sec:initial_conditions} and the steady, fully developed 2-D DNS. Note that the disturbance strip is not active in these simulations.
The DNS boundary-layer profiles in figure~\ref{fig:fig_bl_validation} exhibit strong agreement with the self-similar profiles for streamwise velocity, temperature, and density in both regimes. Minor deviations in wall-normal velocity emerge upon crossing the Widom line \citep{Boldini2}, caused by the non-zero DNS wall-normal pressure gradient retained in the numerical integration of \eqref{eq:ns}. In fact, the pressure boundary-layer profile shown in the inset of figure~\ref{fig:fig_bl_validation} reveals a bump at the height of the pseudo-critical point. This phenomenon arises due to the dependency of $p$ on both $\rho$ and $T$. Thus, the resulting pressure gradient is expressed as $\partial p/\partial y \propto \partial p/\partial \rho \, \partial \rho/ \partial y + \partial p/\partial T \, \partial T/\partial y$. Both terms reach their maximum at the pseudo-critical point and nearly cancel each other out. As a result, the deviation from the conventional boundary-layer assumption of $\partial p/ \partial y=0$ remains minimal, thereby justifying the validity of the self-similar boundary-layer solution used in \S\,\ref{sec:initial_conditions}.

\section{Linear disturbance evolution:~LST vs.~DNS} \label{sec:appC}

\begin{figure}
\centering
\includegraphics[angle=-0,trim=0 0 0 0, clip,width=1.0\textwidth]{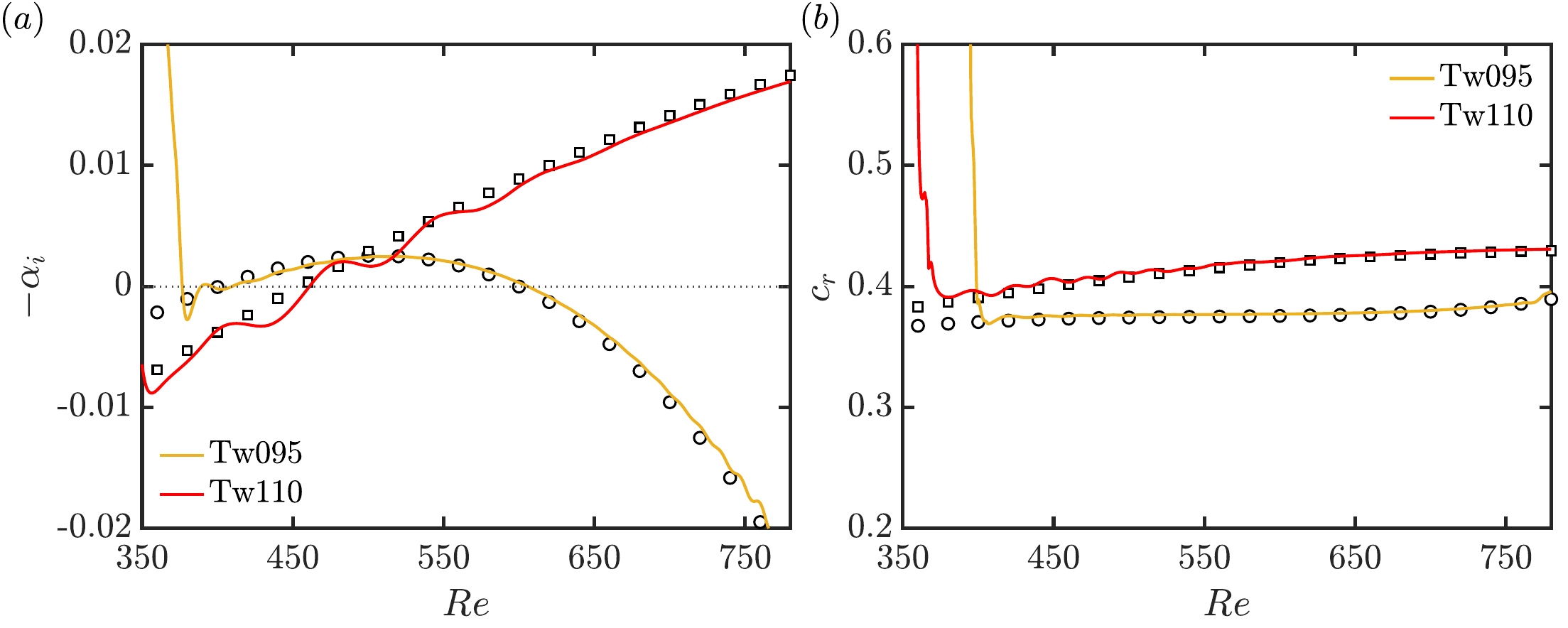}
\captionsetup{justification=justified} \caption{\label{fig_linear_growth}Comparison between low-amplitude DNS (lines) and LST (symbols) for a 2-D wave at $F_{\text{2-D}}=124 \times 10^{-6}$: (a) growth rate ($-\alpha_\mathit{i}$) and (b) phase speed $c_\mathit{r}$. Case Tw095 and Tw110 (Mode II) in blue and red, respectively.}
\end{figure}

To compare DNS with LST results in \S\,\ref{sec:linear_regime}, the disturbances are Fourier transformed in time. The normalised disturbance growth rate and phase speed of the fundamental wave are calculated as
\begin{equation}
    \alpha_\mathit{i}(x)=-\dfrac{Re}{Re_{0}}\dfrac{1}{\hat{u}^\mathit{max}}\dfrac{\partial \hat{u}^\mathit{max}}{\partial x}, \quad c_\mathit{r}(x)=\dfrac{\omega_{\text{2-D}} Re_{0}}{Re}\left( \dfrac{\partial \hat{\phi}}{\partial x}\right)^{-1}, \tag{C1$a{,}b$}
    \label{eq:growth_rate}
\end{equation}
with $\hat{u}^\mathit{max}(x)=\max{\{ |\hat{u}(x=\mathrm{const.},y) | \}}$ and $\hat{\phi}$ the phase angle $\arg(\hat{p}_{1,w})$, with $\hat{p}_\mathit{w}$ being the wall pressure. Figure~\ref{fig_linear_growth} compares the spatial amplification rate $-\alpha_\mathit{i}$ and phase speed $c_\mathit{r}$ for case Tw095 and Tw110. The results reveal very good agreement between DNS and LST, with the phase speed being less sensitive to the criterion used in \eqref{eq:growth_rate}. However, for $\alpha_\mathit{i}$ in figure~\ref{fig_linear_growth}(a), a moderate modulation near the disturbance strip is observed, caused by the excitation of damped waves before the most unstable mode dominates. This behaviour is more pronounced for case Tw110. A smaller disturbance strip and a further upstream disturbance-strip location reduce the modulation (not shown). Thus, for case Tw110, $Re_\mathit{x,mid}$ is shifted upstream relative to $Re_\mathit{x,mid}$ in case Tw095 (see table~\ref{tab:numerical_setup}).

\bibliographystyle{jfm}
\bibliography{main}

\end{document}